\documentclass[notitlepage,apj,numberedappendix]{emulateapj}

\usepackage{verbatim}
\usepackage{amsmath}
\usepackage{subfigure}
\usepackage{hyperref}

\newcommand{\simgt}{\,\hbox{\lower0.6ex\hbox{$\sim$}\llap{\raise0.6ex\hbox{$>$}}}\,}
\newcommand{\simlt}{\,\hbox{\lower0.6ex\hbox{$\sim$}\llap{\raise0.6ex\hbox{$<$}}}\,}
\newcommand{\degree}{\ensuremath{^\circ}}
\newcommand{\code}[1]{\texttt{#1}}

\begin{document}

\title{The Effect of Anisotropic Viscosity on Cold Fronts in Galaxy Clusters}

\author{J. A. ZuHone\altaffilmark{1,2}, M. W. Kunz\altaffilmark{3}, M. Markevitch\altaffilmark{1}, J. M. Stone\altaffilmark{3}, V. Biffi\altaffilmark{4}}

\begin{abstract}
Cold fronts -- contact discontinuities in the intracluster medium
(ICM) of galaxy clusters -- should be disrupted by Kelvin-Helmholtz
(K-H) instabilities due to the associated shear velocity. However,
many observed cold fronts appear stable. This opens the possibility to
place constraints on microphysical mechanisms that stabilize them,
such as the ICM viscosity and/or magnetic fields. We performed
exploratory high-resolution simulations of cold fronts arising from
subsonic gas sloshing in cluster cores using the grid-based
\code{Athena} MHD code, comparing the effects of isotropic Spitzer and
anisotropic Braginskii viscosity (expected in a magnetized plasma).
Magnetized simulations with full Braginskii viscosity or isotropic Spitzer viscosity 
reduced by a factor $f \sim 0.1$ are both in qualitative agreement with observations in 
terms of suppressing K-H instabilities. The RMS velocity of turbulence within the 
sloshing region is only modestly reduced by Braginskii viscosity. We also performed 
unmagnetized simulations with and without viscosity and find that magnetic fields have 
a substantial effect on the appearance of the
cold fronts, even if the initial field is weak and the viscosity is
the same. This suggests that determining the dominant suppression
mechanism of a given cold front from X-ray observations (e.g. viscosity or magnetic 
fields) by comparison with simulations is not straightforward. Finally, we performed 
simulations including anisotropic thermal conduction, and find that including Braginskii 
viscosity in these simulations does not significant affect the evolution of cold
fronts; they are rapidly smeared out by thermal conduction, as in the inviscid
case.
\end{abstract}

\altaffiltext{1}{Astrophysics Science Division, Laboratory for High
 Energy Astrophysics, Code 662, NASA/Goddard Space Flight Center,
 Greenbelt, MD 20771}
\altaffiltext{2}{Department of Astronomy, University of Maryland, College Park, MD, 20742-2421, USA}
\altaffiltext{3}{Department of Astrophysical Sciences, 4 Ivy Lane, Peyton Hall, Princeton University, Princeton, NJ 08544, U. S. A.}
\altaffiltext{4}{SISSA - Scuola Internazionale Superiore di Studi Avanzati, Via Bonomea 265, 34136 Trieste, Italy}

\keywords{conduction --- galaxies: clusters: general --- X-rays: galaxies: clusters --- methods: hydrodynamic simulations}

\nopagebreak[4]

\section{Introduction}\label{sec:intro}

X-ray observations of the intracluster medium (ICM) of galaxy clusters often show sharp 
surface brightness discontinuities. Spectral analysis of these regions have
revealed that in most cases the brighter (and therefore denser) side
of the edge is the colder side, and hence these jumps in gas density
have been dubbed ``cold fronts'' \citep[for a detailed review
see][]{MV07}. Cluster cold fronts are generally classified into
two classes. The first are those which occur in clusters undergoing major
mergers, such as 1E 0657-558 \citep[the ``Bullet Cluster''][]{mar02}, A3667
\citep{vik01,vik01b,vik02}, and also the elliptical galaxy NGC 1404
\citep{mac05}. In those cases, the cold front is
formed by the action of ram pressure on a cluster core moving at high
velocity through another cluster's ICM \citep[cf. Figure 23
of][]{MV07}. The second class of cold fronts occur in ``cool-core''
systems, exhibited as edges in X-ray surface brightness approximately
concentric with respect to the brightness peak of the cluster
\citep[e.g.,][]{maz01,mar01,MVF03}; these are observed
in a majority of cool-core systems \citep[][]{ghi10}. Prominent
examples include A2142 \citep{mar00}, A1644 \citep{sri01,joh10}, and A1795
\citep{mar01}. These primarily spiral-shaped cold fronts are believed to arise from gas
sloshing in the deep dark-matter--dominated potential well. These
motions are initiated when the low-entropy, cool gas of the core is
displaced from the bottom of the potential well, either by
gravitational perturbations from infalling subclusters
\citep[][]{AM06} or by an interaction with a shock front \citep[][]{chu03}. 

Most \citep[but not all, see][for an example]{rod12b} observed cold
fronts have smooth, arc-like shapes. As cold fronts move through the surrounding ICM, significant (but subsonic) shear flows develop. These flows
can develop Kelvin-Helmholtz (K-H) instabilities. This is certainly the case if the ICM were an inviscid, unmagnetized
plasma, as was demonstrated by \citet[e.g.][hereafter ZMJ10]{zuh10} and \citet{rod12}.

However, Faraday rotation measurements and radio
observations of clusters indicate that the ICM is magnetized
\citep[see][for recent reviews]{car02,fer08}. Magnetic fields, oriented
parallel to a shearing surface and contributing a pressure comparable to the kinetic energy per unit volume of the shear flow, 
will suppress the growth of the K-H instability \citep{lan60,cha61}. For a cold front in the galaxy cluster A3667, \citet{vik01} and \citet{vik02} determined that the magnetic-field strength required to stabilize the front is $B \sim 10~\mu$G,
roughly an order of magnitude higher than the field strengths usually
inferred from radio observations and rotation measure
estimates. However, \citet{lyu06} pointed out that the shear flows
around cold fronts lead to ``magnetic draping.'' Provided the motion
of the front is super-Alfv\'{e}nic, a weak, tangled magnetic field in the surrounding medium will be ``swept up" and stretched by this shear flow to produce a layer parallel to the the front surface. The energy of the magnetic field in this layer would be
increased due to shear amplification, possibly becoming strong enough to
stabilize the front against K-H instabilities
\citep{kes10}. A number of numerical simulations 
\citep[e.g.][]{asa04,asa07,dur07,dur08} have demonstrated this stabilizing effect in simplified situations.

In the case of sloshing cold fronts, there is a nearly identical effect, since at the cold front surface there is a strong velocity shear. In this case, the fast-moving gas of the sloshing motions underneath the front amplify the magnetic field {\it inside} the front surface, in the denser, colder gas, instead of sweeping up magnetic fields in the hotter, more rarified outer medium. The inviscid MHD simulations presented in \citet[][hereafter ZML11]{zuh11} showed that gas sloshing in cool cores produces relatively strong magnetic fields by shear amplification along the cold front surfaces, resulting in layers of strong magnetic field stretched tangentially to the fronts that suppress K-H instabilities along them. The suppression depended on the assumed strength of the initial field; for a range of field strengths close to those observed in cluster cores, the suppression was only partial, indicating that in general the magnetic field cannot always be relied upon to stabilize cold fronts. In these cases, K-H instabilities not only disrupt the cold fronts but also the magnetized layers, re-tangling the field lines. Similar simulations including anisotropic thermal conduction were performed by \citet[][]{zuh13a}. These demonstrated that the tangling of field lines along the magnetized layers as well as hot gas that is unshielded from the cold gas of the front results in a significant flow of heat to the cold front, smearing the fronts out to the extent that they would be unobservable in X-rays. 

Viscosity may also play a role in stabilizing cold fronts against K-H
instabilities. Recently, \citet{rod13a} investigated the effects of isotropic
viscosity on cold fronts in a hydrodynamic simulation of the Virgo cluster. Those authors 
showed that suppression of these instabilities at a level consistent with 
observational constraints could be obtained by assuming an isotropic Spitzer
viscosity with a suppression factor $f = 0.1$. One limitation in this approach is that it neglects the effects of the magnetic field in the ICM. Typical conditions prevailing in the ICM indicate that the mean free path is $\sim 10^{11}-10^{13}$ times larger than the Larmor radius $r_{g,i}$ \citep{sch06}, so that the transport of momentum is strongly anisotropic. The consequent restriction of viscous dissipation to the motions and
gradients parallel to the magnetic field is known as the
``Braginskii'' viscosity \citep{bra65}. For an isotropically tangled
magnetic field, the effective viscosity on large scales is 1/5 the Spitzer value, so that 
in some situations it may be possible to model the viscous flux as
isotropic with a suppression factor. However, as mentioned
above, the magnetic-field lines at the cold front are
likely to be (mostly) draped parallel to the front by shear
amplification, preventing viscous damping of the velocity shear
perpendicular to the front surface. Along these lines, \citet{suz13} recently simulated the propagation of a cold, dense mass of gas in an idealized magnetized ICM with
isotropic and anisotropic viscosity. They showed that anisotropic
viscosity is far less effective than isotropic viscosity at
suppressing K-H instabilities at the cold front that
develops. However, their study was limited since it involved a
simplified setup, including an initially uniform magnetic field
geometry.

In this work, we build upon earlier work by ZMJ10, ZML11, and Z13 to include the effects 
of viscosity in simulations of gas sloshing in cool-core galaxy clusters. We try various 
forms and strengths of the viscosity, and construct synthetic X-ray observations in
order to see if the effects can be detectable in X-ray images of real clusters. We show 
that Braginskii viscosity, in combination with magnetic fields, may provide an explanation 
for the observed smoothness of cold fronts in sloshing cool cores without completely 
suppressing K-H instabilities. We also find that an isotropic Spitzer viscosity with $f = 
0.1$, in combination with magnetic fields, produces smooth cold fronts and turbulence 
similar to the Braginskii case. However, using unmagnetized simulations for comparison, we 
also show that even if the ICM is viscous, the presence or the absence of the magnetic 
field can still have a substantial effect on the appearance of the cold fronts, 
complicating the use of these simulations to discern the dominant mechanism for 
suppression of K-H instabilities. Finally, we show that the additional stability imparted 
to cold fronts and their associated magnetized layers by viscosity does not prevent anisotropic thermal conduction (at its full Spitzer strength) from eliminating the strong temperature gradient of the cold front jumps, thus rendering them unobservable. 

%
%
\begin{figure*}
\begin{center}
\includegraphics[width=0.95\textwidth]{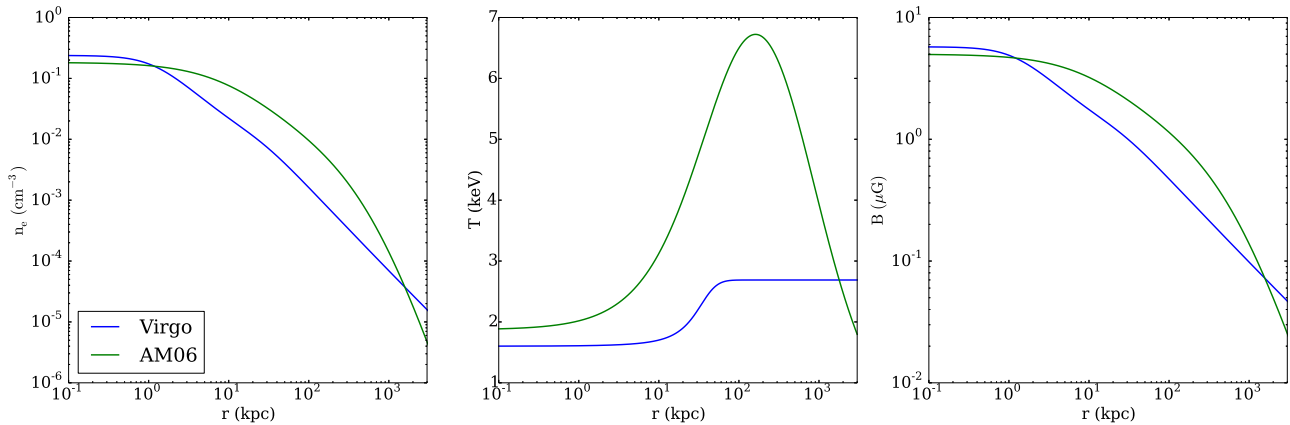}
\caption{Initial profiles of the density, temperature, and magnetic
 field strength for both cluster models.\label{fig:profiles}}
\end{center}
\end{figure*}

This paper is organized as follows. In Section \ref{sec:method} we
describe the simulations and the code. In Section \ref{sec:results} we
describe the effects of anisotropic viscosity. Finally, in Section \ref{sec:discussion} we 
summarize our results and discuss future developments of this work. Throughout, we assume 
a flat $\Lambda$CDM cosmology with $h = 0.71$ and $\Omega_{\rm m} = 0.27$.

%
%
\section{Method of Solution}\label{sec:method}

%
%
\subsection{Equations}\label{sec:physics}

Our simulations solve the following set of MHD equations:
\begin{eqnarray}
\frac{\partial{\rho}}{\partial{t}} + \nabla \cdot (\rho{\bf v}) &=& 0, \label{eqn:density}\\
\frac{\partial{(\rho{\bf v})}}{\partial{t}} + \nabla \cdot
\left(\rho{\bf vv} - \frac{\bf BB}{4\pi} + p{\sf I}\right) &=&
\rho{\bf g} - \nabla \cdot {\sf \Pi} \label{eqn:momentum}, \\
\frac{\partial{E}}{\partial{t}} + \nabla \cdot \left[{\bf v}(E+p) - \frac{{\bf
 B}({\bf v \cdot B})}{4\pi}\right] &=& \rho{\bf g \cdot v} - \nabla
\cdot {\bf Q}\nonumber  \\
&-& \nabla \cdot ({\sf \Pi} \cdot {\bf v}), \\
\frac{\partial{\bf B}}{\partial{t}} + \nabla \cdot ({\bf vB} - {\bf
 Bv}) &=& 0, \label{eqn:bfield}
\end{eqnarray}
where
\begin{equation}
p = p_{\rm th} + \frac{B^2}{8\pi}
\end{equation}
is the total pressure,
\begin{equation}
E = \frac{\rho{v^2}}{2} + \epsilon + \frac{B^2}{8\pi}
\end{equation}
is the total energy per unit volume, $\rho$ is the gas density,
$\bf{v}$ is the fluid velocity, $p_{\rm th}$ is the gas pressure, $\epsilon$ is the gas internal energy per unit volume, $\bf{B}$ is the magnetic field vector, and $\sf{I}$ is the unit dyad. We assume an ideal equation of state with $\gamma = 5/3$, equal electron and ion temperatures, and primordial abundances with molecular weight $\bar{A} = 0.6$. 

We model the viscous transport of momentum as
either isotropic or anisotropic. The viscous flux in the isotropic case is given by
\begin{equation}
{\sf \Pi}_{\rm iso} = -\mu\nabla{\bf v},
\label{eqn:isotropic_visc}
\end{equation}
where $\mu$ is the dynamic viscosity coefficient. To illustrate the physics behind anisotropic (Braginskii) viscosity,
we will follow a more detailed derivation. Equation (\ref{eqn:momentum}) may be written in an alternative form:
\begin{equation}
\frac{\partial{(\rho{\bf v})}}{\partial{t}} + \nabla \cdot
\left(\rho{\bf vv} + {\sf P}\right) = \rho{\bf g},
\label{eqn:momentum_alt}
\end{equation}
where ${\sf P}$ is the total (thermal + magnetic) pressure tensor:
\begin{equation}
{\sf P} = \left(p_\perp+\frac{B^2}{8\pi}\right){\sf
 I}-\left(p_\perp-p_\parallel+\frac{B^2}{4\pi}\right)\hat{\textbf{b}}\hat{\textbf{b}}, 
\end{equation}
$p_\perp$ ($p_\parallel$) is the thermal pressure perpendicular
(parallel) to the magnetic field, and $\hat{\textbf{b}} = {\bf B}/B$ is
the unit vector in the direction of the magnetic field. The total thermal
pressure satisfies
\begin{equation}
p = \frac{2}{3} \,p_\perp+\frac{1}{3} \,p_\parallel.
\end{equation}
Differences in these two components of the thermal pressure arise from the
conservation of the first and second adiabatic invariants for each
particle on timescales much greater than the inverse of the
ion gyrofrequency, $\Omega_g^{-1}$ \citep{che56}. When the ion-ion
collision frequency $\nu_{\rm ii}$ is much larger than the rates of
change of all fields, an equation for the pressure anisotropy can be
obtained by balancing its production by adiabatic invariance with its
relaxation via collisions \citep[cf.][]{sch05}:
\begin{equation}
p_\perp-p_\parallel = 0.960\,\frac{p_{\rm i}}{\nu_{\rm ii}}\frac{d}{dt}\ln{\frac{B^3}{\rho^2}},
\label{eqn:pressure_anisotropy}
\end{equation}
where $p_{\rm i}$ is the thermal pressure of the ions. The ion dynamic viscosity 
coefficient for the ICM is given by \citep{spi62,bra65,sar88}
\begin{eqnarray}
\mu             &=& 0.960\,\frac{n_{\rm i}k_{\rm B}T}{\nu_{\rm ii}} \\
\nonumber  &\approx& 2.2 \times 10^{-15}\frac{T^{5/2}}{\ln\Lambda_{\rm i}}~{\rm g~cm^{-1}~s^{-1}},
\end{eqnarray}
where the temperature $T$ is in Kelvin and $\ln\Lambda_{\rm i}$ is the ion Coulomb logarithm, which is a weak
function of $\rho$ and $T$; for simplicity we follow \citet{rod13a}
by approximating $\ln\Lambda_i = 40$, appropriate for conditions in
the ICM. The kinematic coefficient of viscosity $\nu = \mu/\rho$. Using Equations (\ref{eqn:density}) and (\ref{eqn:bfield})
to replace the time derivatives of density and magnetic field strength
with velocity gradients, Equation (\ref{eqn:pressure_anisotropy}) may be written as
\begin{equation}
p_\perp-p_\parallel = 3\mu\left(\hat{\textbf{b}}\hat{\textbf{b}}-\frac{1}{3}{\sf
I}\right):\nabla{\bf v} .
\end{equation}
It can then be shown that Equations (\ref{eqn:momentum}) and
(\ref{eqn:momentum_alt}) are equivalent, with the resulting viscous flux
for the anisotropic (Braginskii) case being
\begin{equation}
{\sf \Pi}_{\rm aniso} = -3\mu\left(\hat{\textbf{b}}\hat{\textbf{b}}-\frac{1}{3}{\sf
I}\right)\left(\hat{\textbf{b}}\hat{\textbf{b}}-\frac{1}{3}{\sf
I}\right):\nabla{\bf v} .
\label{eqn:anisotropic_visc}
\end{equation}
These two different fluxes (isotropic, Eqn. \ref{eqn:isotropic_visc},
and anisotropic, Eqn. \ref{eqn:anisotropic_visc}) are implemented in
our different simulations. For the isotropic cases, we model
viscosities less than the Spitzer value by including a multiplicative suppression factor $f$. 

We also include anisotropic thermal conduction in some of our
runs. The heat flux due to thermal conduction by electrons is given by
\begin{equation}
{\bf Q} = -\kappa\hat{\textbf{b}}\hat{\textbf{b}}  \cdot \nabla{T},
\end{equation}
where the conductivity coefficient \citep{spi62,bra65,sar88}
\begin{eqnarray}
\kappa       &=& 3.2 k_{\rm B} \frac{n_{\rm e} k_{\rm B} T}{m_{\rm e}\nu_{\rm ee}} \\
\nonumber &\approx& 1.84 \times 10^{-5}\frac{T^{5/2}}{\ln\Lambda_{\rm e}}~{\rm erg~cm^{-1}~s^{-1}~K^{-1}},
\end{eqnarray}
where $\nu_{\rm ee}$ is the electron-electron collision frequency and
$\ln\Lambda_{\rm e} = 40$ is the electron Coulomb logarithm. In this model, the conduction proceeds at full Spitzer rate parallel to the field lines, and is zero perpendicular to the lines. The corresponding thermal diffusivity $\chi = \kappa T/p$. Following \citet{cow77}, we include the effect of conduction saturation whenever the characteristic length scale associated with the temperature gradient is smaller than the electron mean free path  (although in the bulk of the ICM this effect is insignificant).

%
%
\begin{figure*}
\begin{center}
\includegraphics[width=0.49\textwidth]{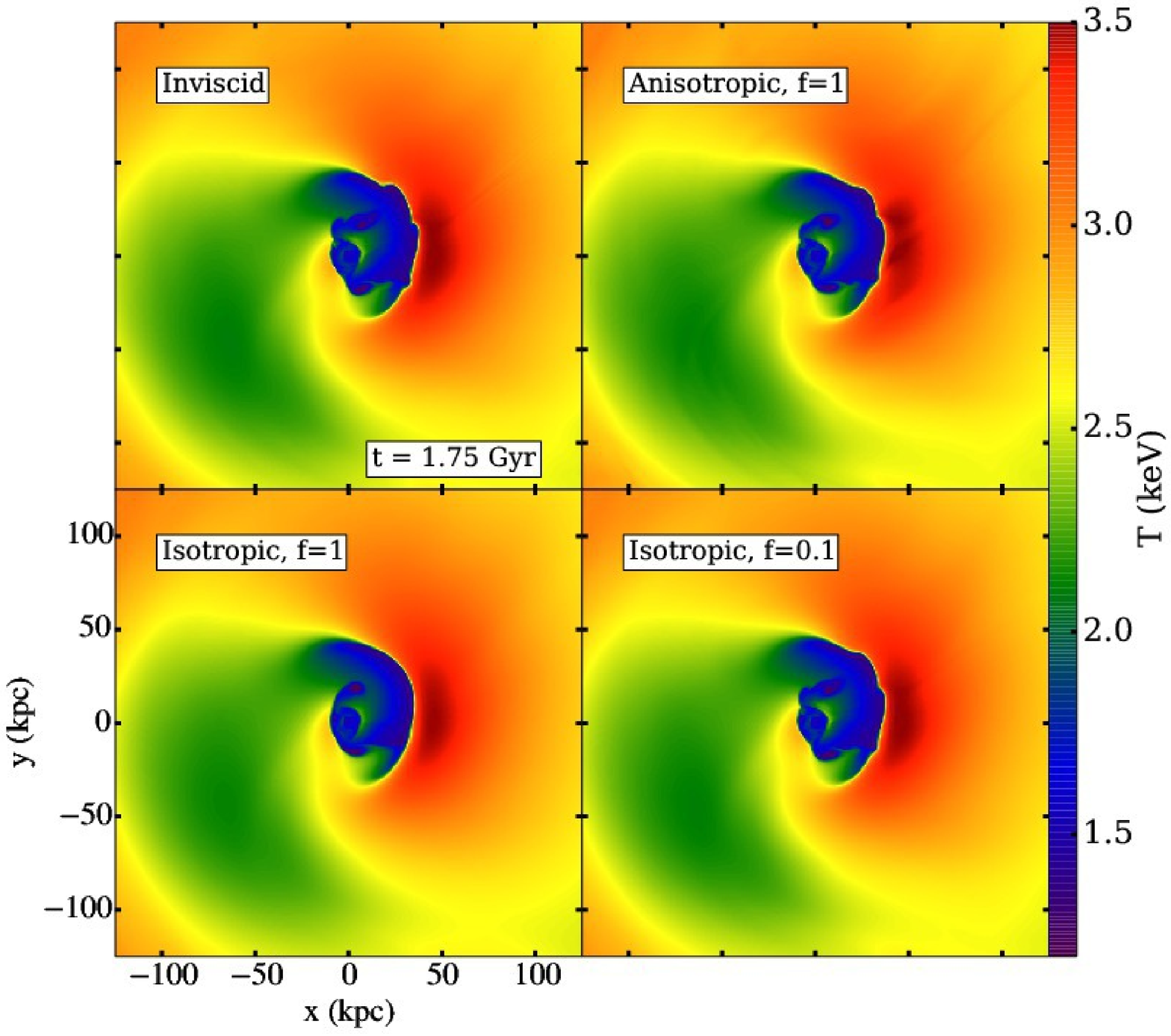}
\includegraphics[width=0.49\textwidth]{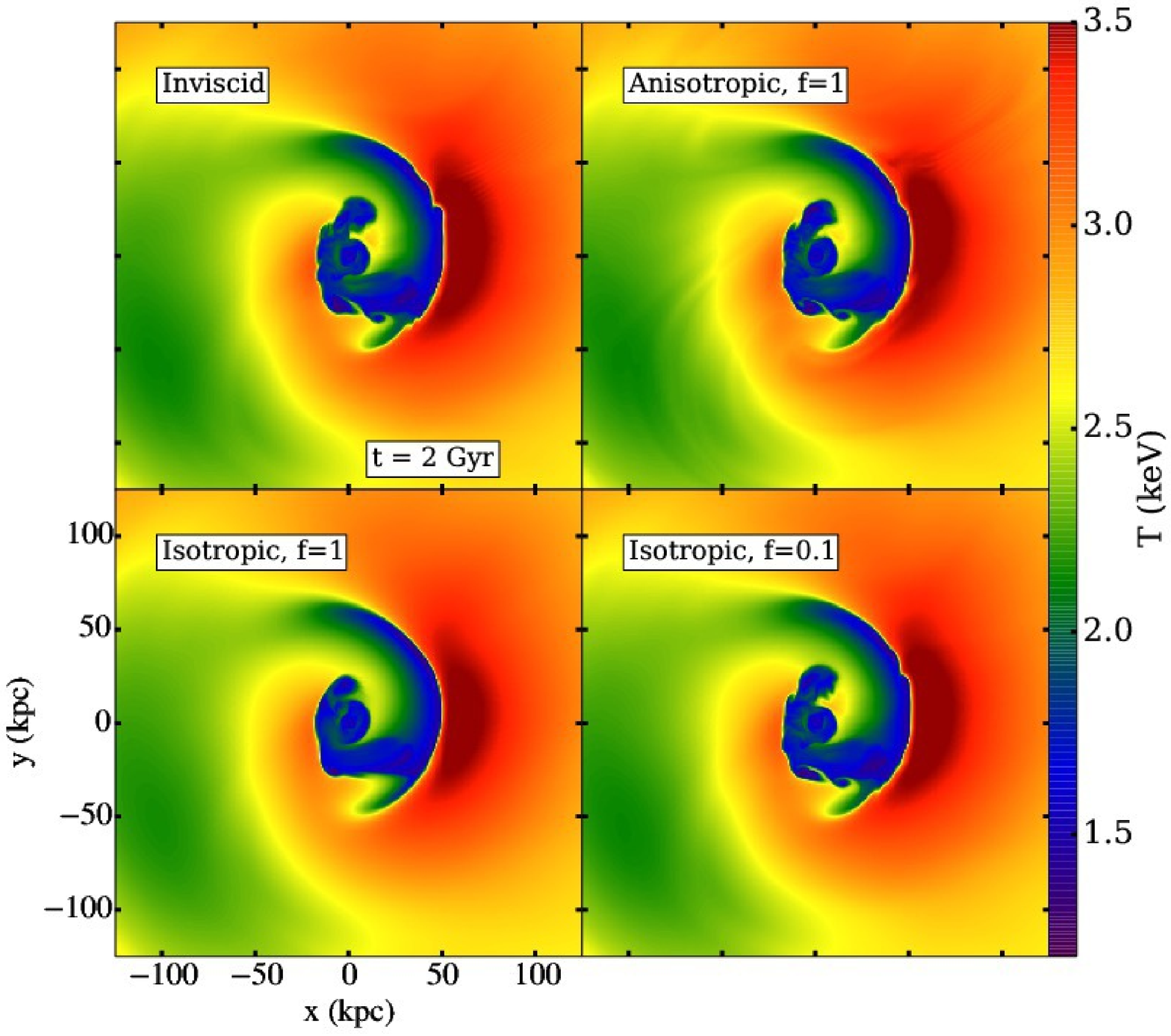}
\includegraphics[width=0.49\textwidth]{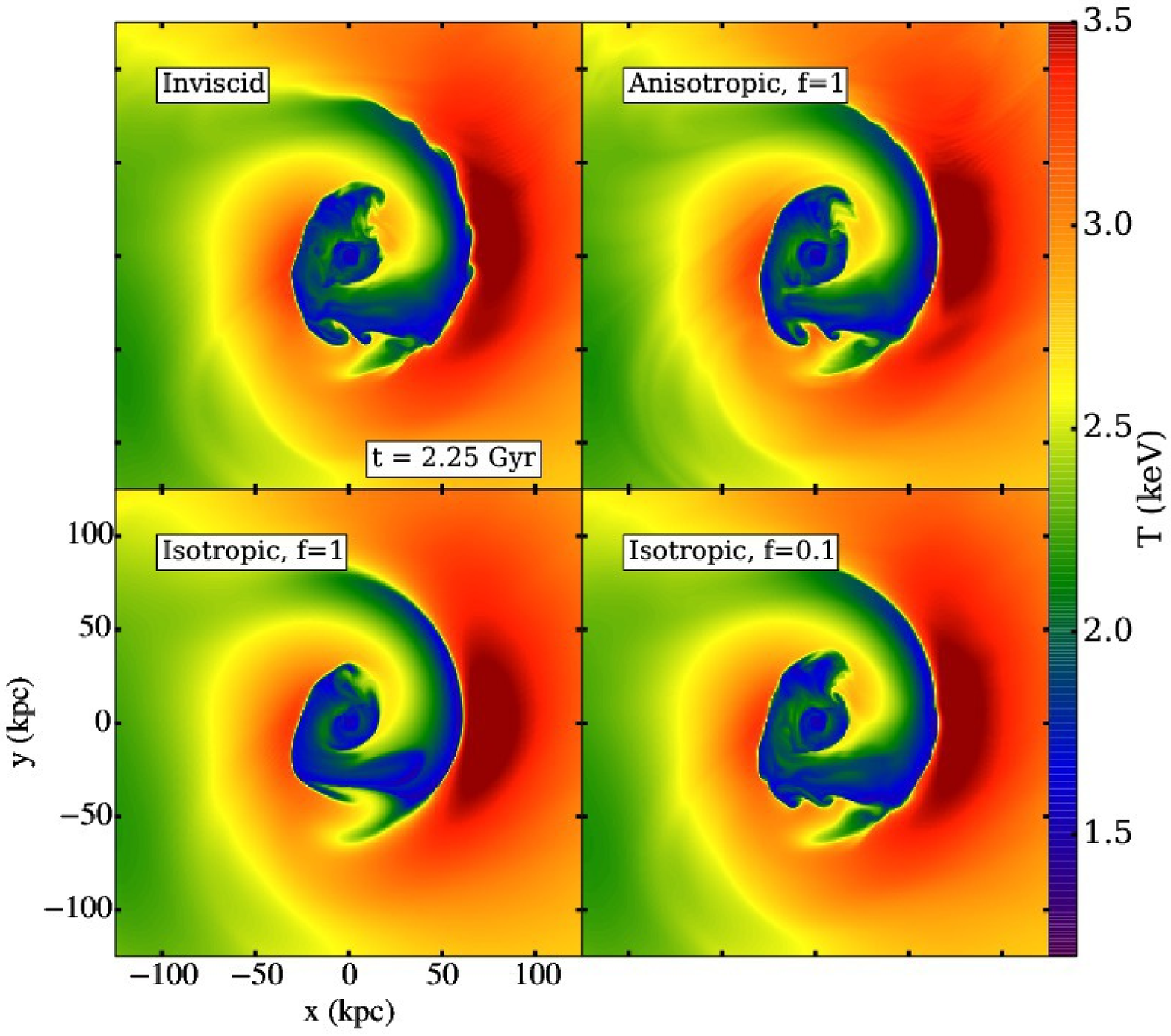}
\includegraphics[width=0.49\textwidth]{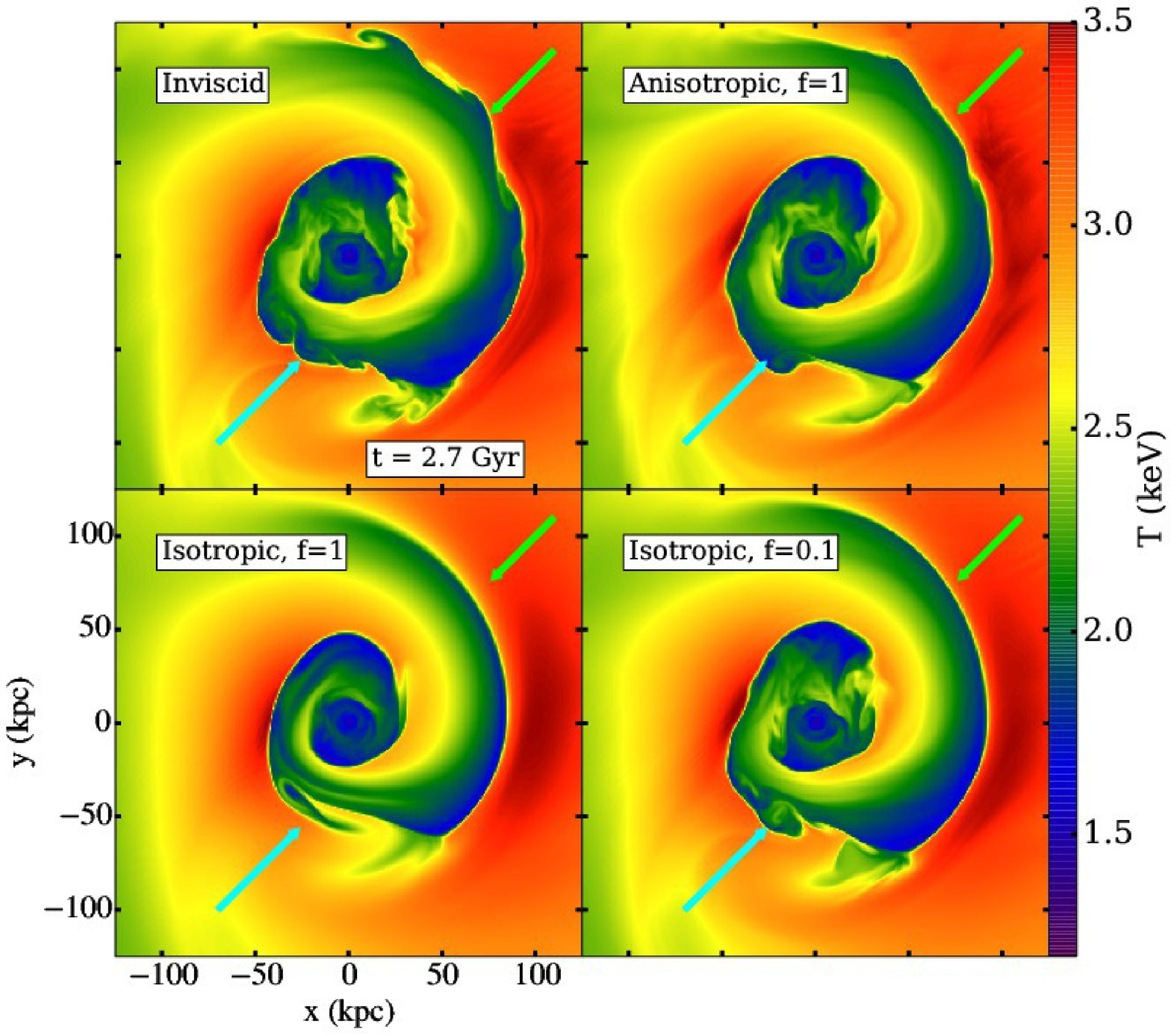}
\caption{Slices of temperature through the center of the ``Virgo'' simulation domain 
at times $t$ = 1.75, 2.0, 2.25, and 2.7~Gyr, the latter of which corresponds to the time identified 
as closely matching the merger state of the Virgo cluster in the simulations of \citet{rod11} and \citet{rod13a}. 
Arrows mark the positions of cold fronts with morphologies that are altered by viscosity.\label{fig:virgo_temp}}
\end{center}
\end{figure*}

%
%
\begin{figure*}
\begin{center}
\includegraphics[width=0.49\textwidth]{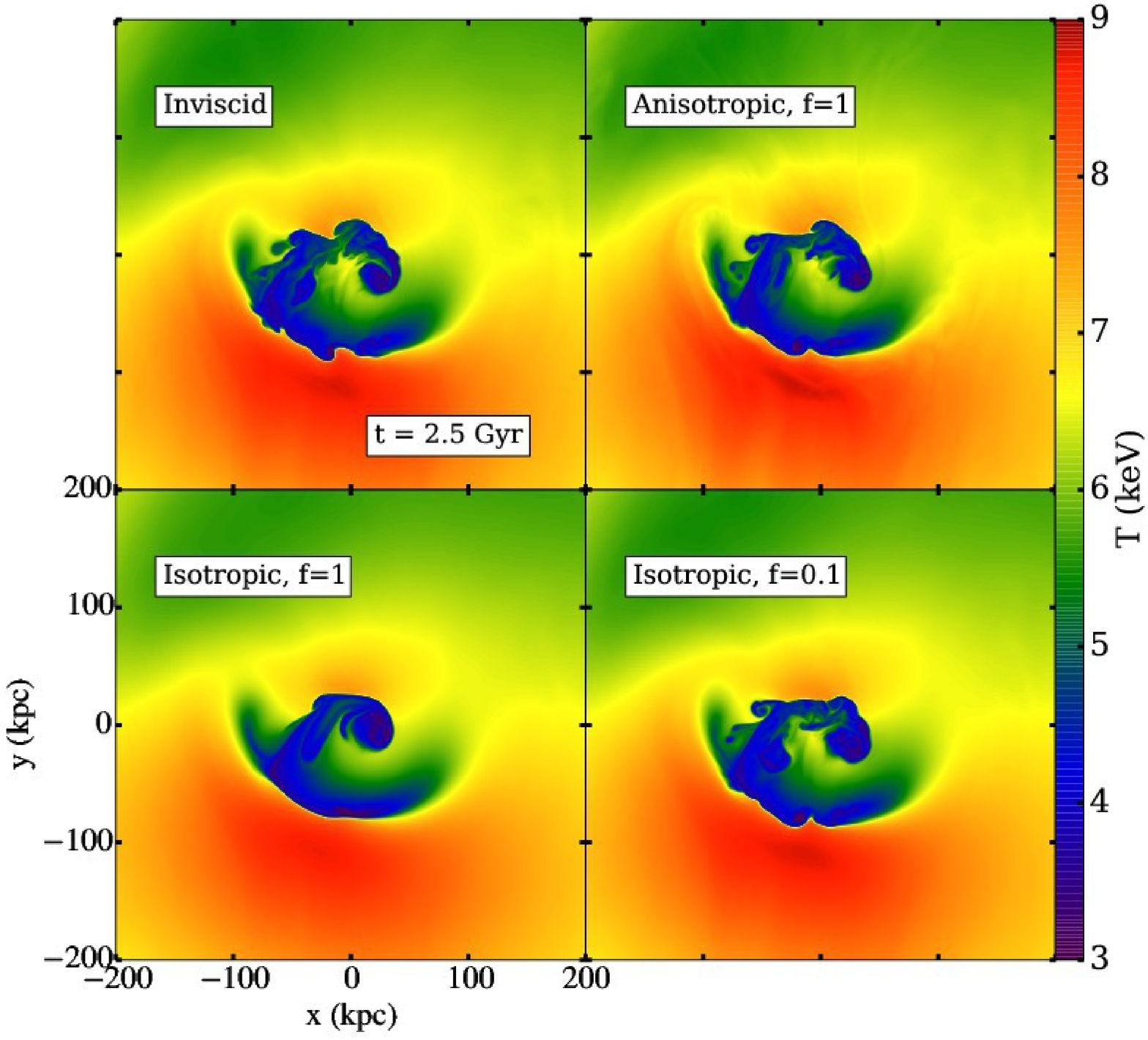}
\includegraphics[width=0.49\textwidth]{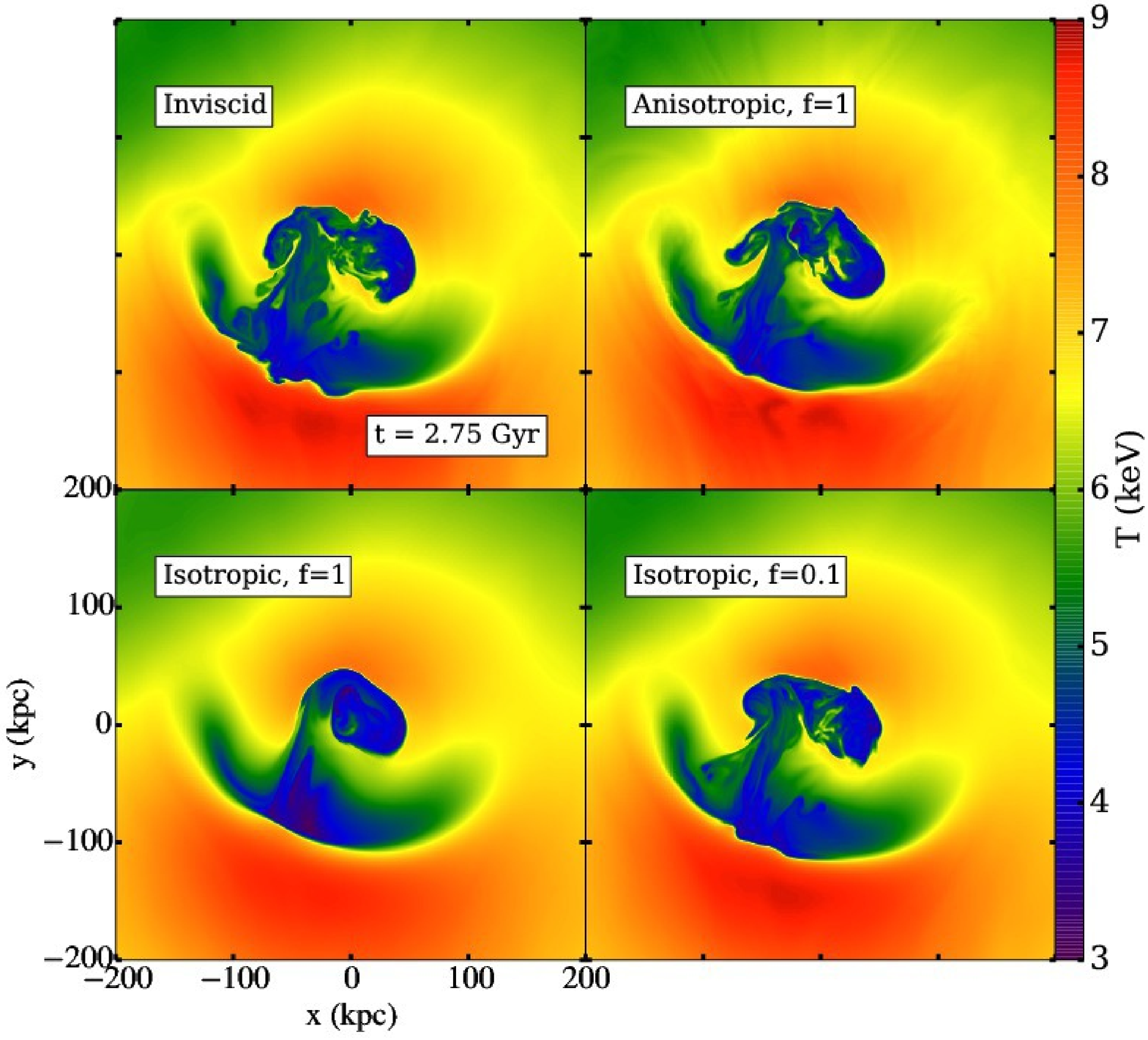}
\includegraphics[width=0.49\textwidth]{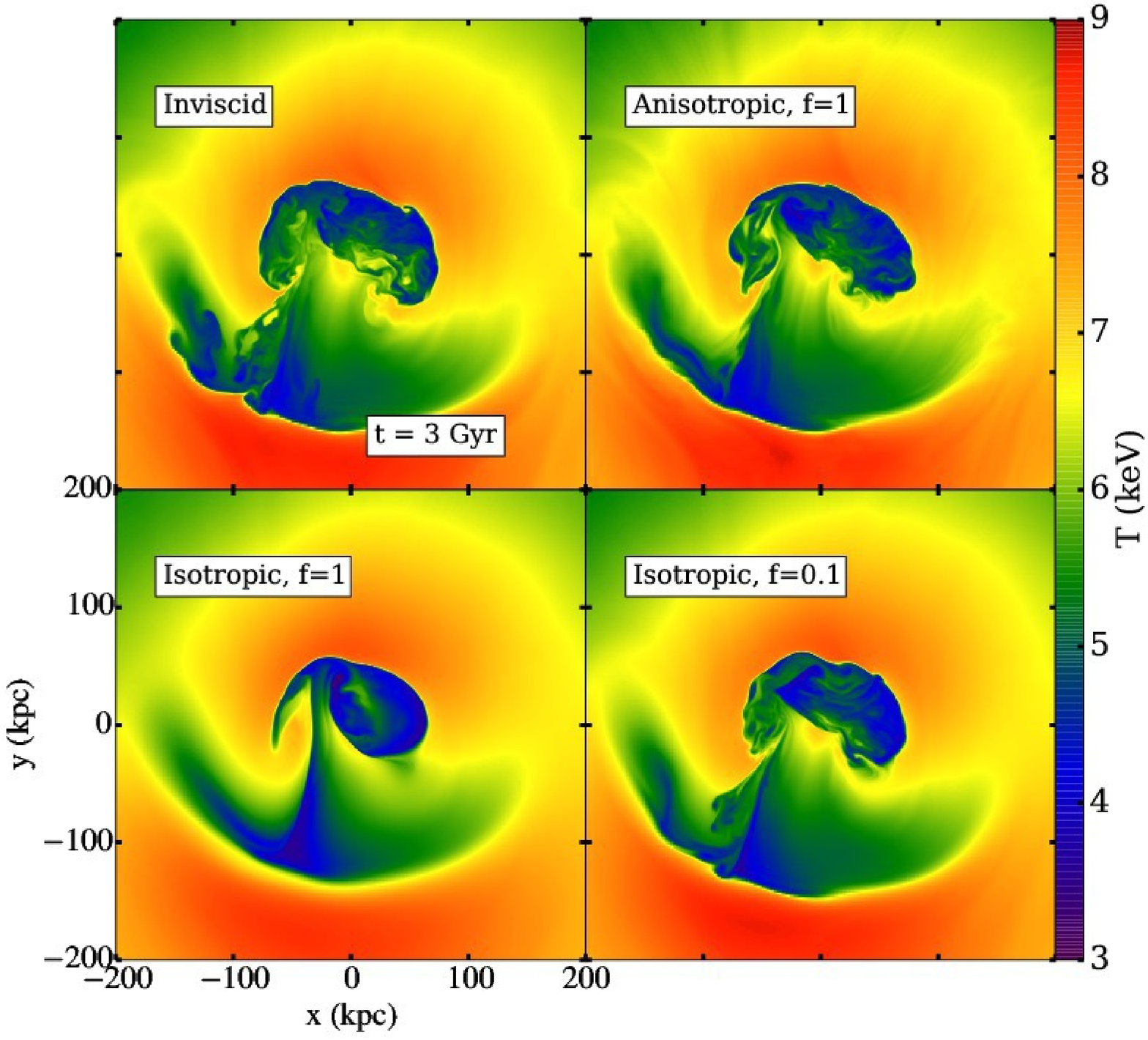}
\includegraphics[width=0.49\textwidth]{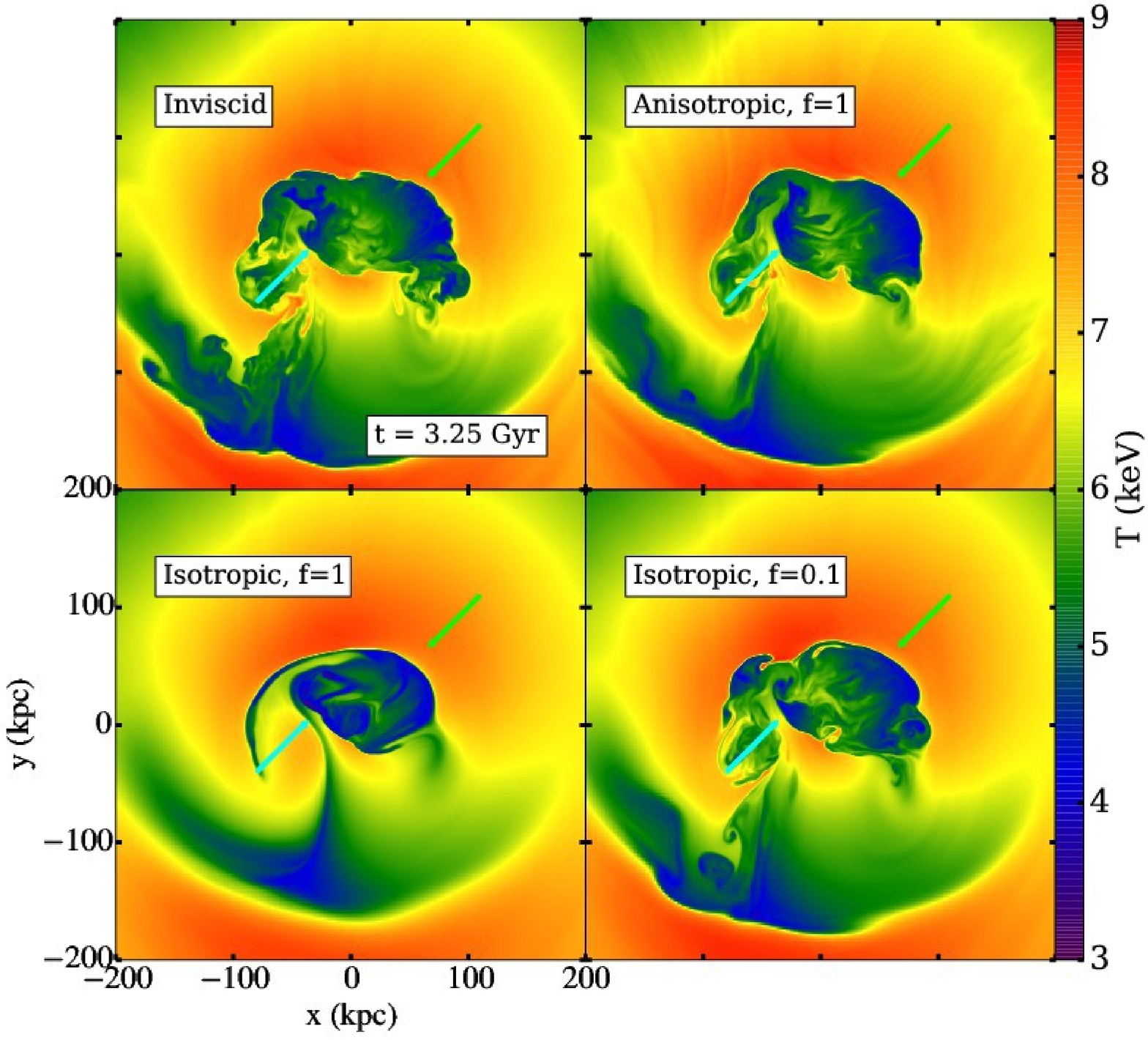}
\caption{Slices of temperature through the center of the ``AM06" simulation domain at times 
$t$ = 2.5, 2.75, 3.0, and 3.25~Gyr. Arrows mark the positions of cold fronts with morphologies
that are altered by viscosity.\label{fig:AM06_temp}}
\end{center}
\end{figure*}

%
%
\subsection{Code}\label{sec:code}

We performed our simulations using \code{Athena} 4.1, a parallel
conservative magnetohydrodynamic (MHD) astrophysical simulation code
\citep{sto08}. The MHD algorithms in \code{Athena} are detailed in
\citet{gar05,gar08}. The directionally unsplit corner transport upwind
(CTU) integration method and the HLLD Riemann solver are used in all
of our simulations, with third-order \citep[piecewise parabolic][]{col84}
reconstruction. 

We included anisotropic conduction and Braginskii viscosity
following the approaches of \citet{sha07}, \citet{don09}, \citet{par12}, and \citet{kun12}. The conductive and
viscous fluxes are implemented via operator splitting, with monotonized central (MC) 
limiters applied to the fluxes to preserve monotonicity. The latter ensures
that unphysical transport of energy and momentum does not occur in the
presence of steep gradients.\footnote{Our implementation differs slightly 
from that presented in the Appendix of \citet{par12}. Those authors applied a slope limiter $\mathcal{L}$ to arithmetically 
averaged transverse velocity gradients, which are located at cell corners (e.g.~at $i-1/2$, $j+1/2$ -- see their eq.~A5 and fig.~A1), in order to obtain the 
face-centered quantities (e.g.~at $i-1/2$, $j$) necessary to compute the viscous fluxes. They claim that this interpolation preserves monotonicity. 
By contrast, we do not arithmetically average at any step. We compute cell-corned transverse velocity gradients by slope limiting
face-centered gradients (e.g.~those at $i-1$, $j+1/2$ and $i$, $j+1/2$), and then slope limit those quantities to compute the viscous fluxes on the cell faces. 
Using our algorithm, their eq.~A4 would instead read
\begin{eqnarray}
\left( \frac{\partial v_x}{\partial y} \right)_{i-\frac{1}{2},j} = \mathcal{L} \left\{ \mathcal{L} \left[ \left( \frac{\partial v_x}{\partial y} \right)_{i,j+\frac{1}{2}} , \left(\frac{\partial v_x}{\partial y} \right)_{i-1,j+\frac{1}{2}} \right] , \right. \nonumber\\*
\left. \mathcal{L} \left[ \left( \frac{\partial v_x}{\partial y}\right)_{i,j-\frac{1}{2}}, \left( \frac{\partial v_x}{\partial y}\right)_{i-1,j-\frac{1}{2}} \right] \right\} .\nonumber
\end{eqnarray}
This interpolation preserves monotonicity and is more in line with the original 
monotonicity-preserving algorithm 
developed for anisotropic conduction by \citet[][their eq.~17]{sha07}.} Since these 
diffusive processes are
modeled by explicit time-stepping methods, they have very restrictive
Courant-limited timesteps $\propto$$(\Delta{x})^2$. For this
reason, we choose to accelerate the calculation via the method of
super-time-stepping \citep[STS,][]{ale96}. STS achieves this speedup by requiring 
stability of the calculation at a timestep that is significantly larger than the 
Courant-restricted timestep (in our case, the timestep associated with the diffusive 
processes of viscosity and conduction). This timestep is typically on the order of the 
hydrodynamical timestep associated with the sound speed of the gas. Finally, since we are 
only concerned in this work with the effects of ICM microphysics on the cold fronts, we do 
not explicitly include the effects of radiative cooling during the simulation, though our 
initial condition implicitly assumes that the cool core was originally formed by such 
cooling.

%
%
\begin{figure*}
\begin{center}
\includegraphics[width=0.49\textwidth]{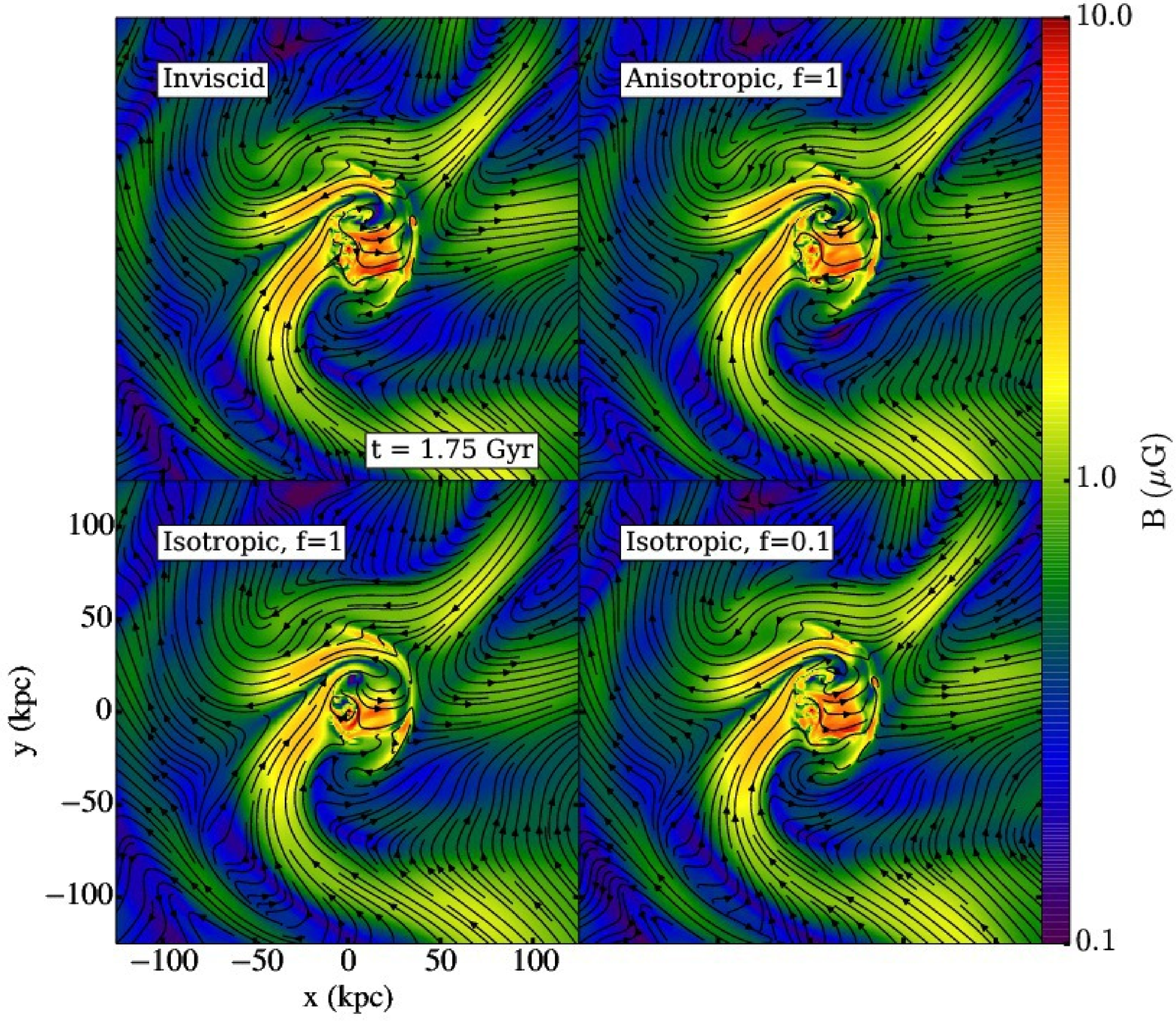}
\includegraphics[width=0.49\textwidth]{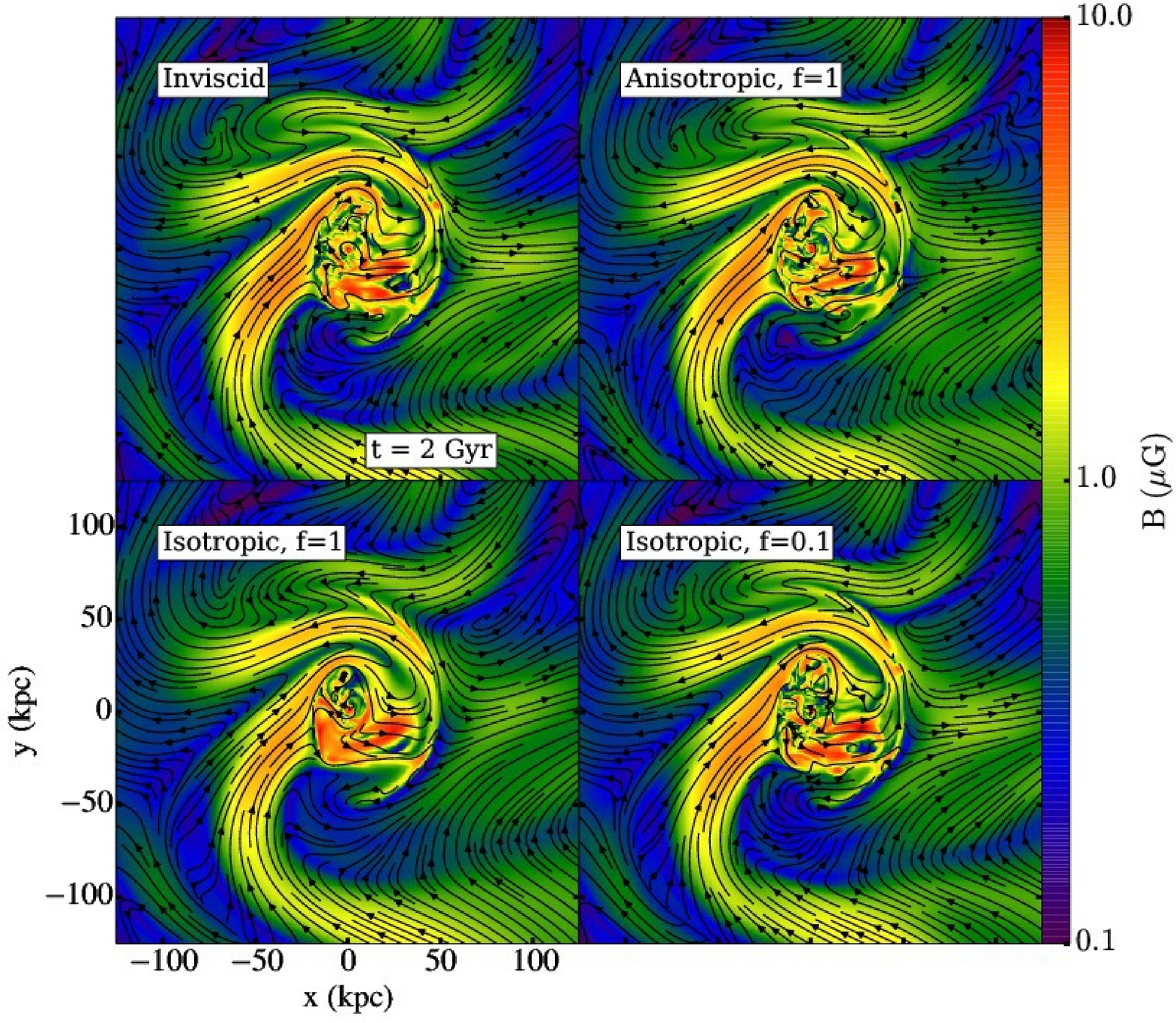}
\includegraphics[width=0.49\textwidth]{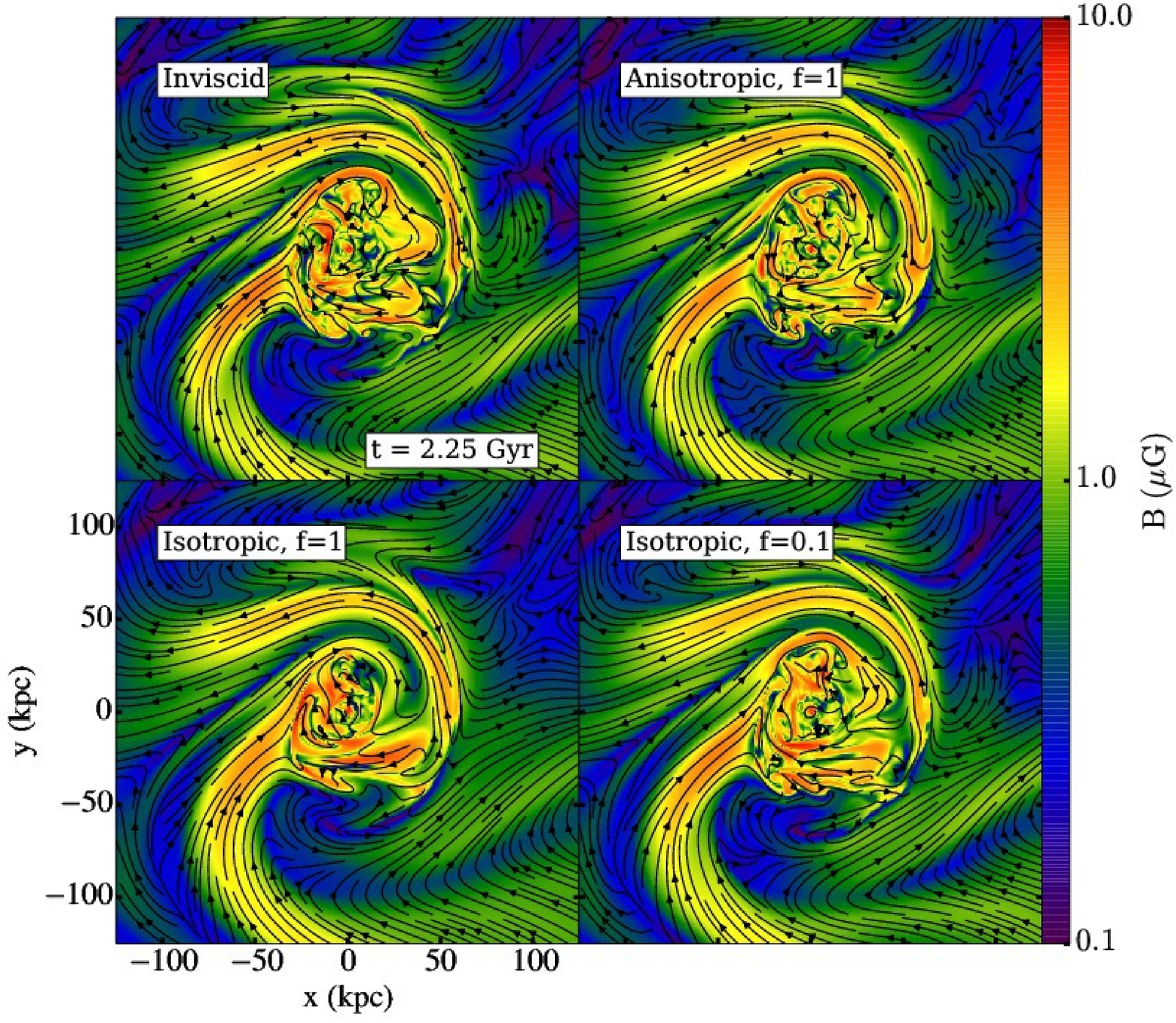}
\includegraphics[width=0.49\textwidth]{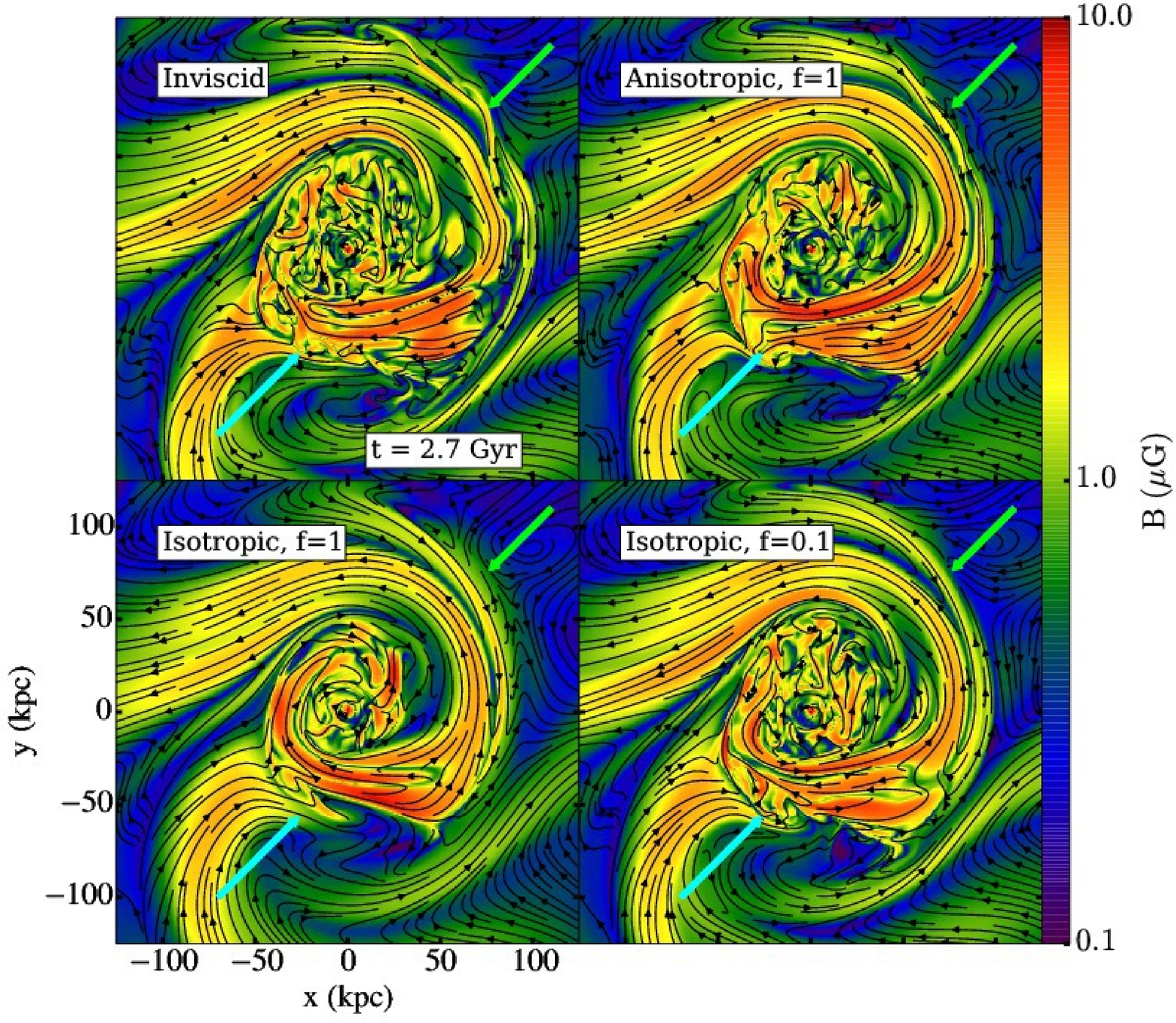}
\caption{Slices of magnetic-field strength through the center of the ``Virgo'' simulation domain 
at times $t$ = 1.75, 2.0, 2.25, and 2.7~Gyr, with magnetic-field lines in the plane overlaid. The
 final time corresponds to the epoch identified as closely matching
 the merger state of the Virgo cluster in the simulations
 of \citet{rod11} and \citet{rod13a}. Arrows mark the positions of
 cold fronts with morphologies that are altered by viscosity.\label{fig:virgo_bmag}}
\end{center}
\end{figure*}

%
%
\subsection{Initial Conditions}\label{sec:ICs}

We model two cluster merger scenarios, both drawn from
previous investigations. The first is the setup from
\citet{rod11} and \citet{rod13a}, which we dub ``Virgo''. Virgo is modeled as a cool-core
cluster with $M \sim 2 \times 10^{14}~{\rm M}_\odot$ and $T_{\rm X} \sim
2$~keV. The gas-less subcluster is modeled as a Hernquist
dark-matter (DM) halo with a mass of $2 \times 10^{13}~{\rm M}_\odot$. The trajectory
results in a closest passage of the subcluster of $\approx 100$~kpc at 
time $t = 1.0$~Gyr in the simulation.\footnote{In
contrast to \citep{rod11} we define $t = 0.0$ at the beginning of
 the simulation instead of at the core passage.} The moment identified by
\citet{rod11} as most consistent with the shape and size of the Virgo cold fronts is $t =
2.7$~Gyr in our simulations. 

%
%
\begin{figure*}
\begin{center}
\includegraphics[width=0.49\textwidth]{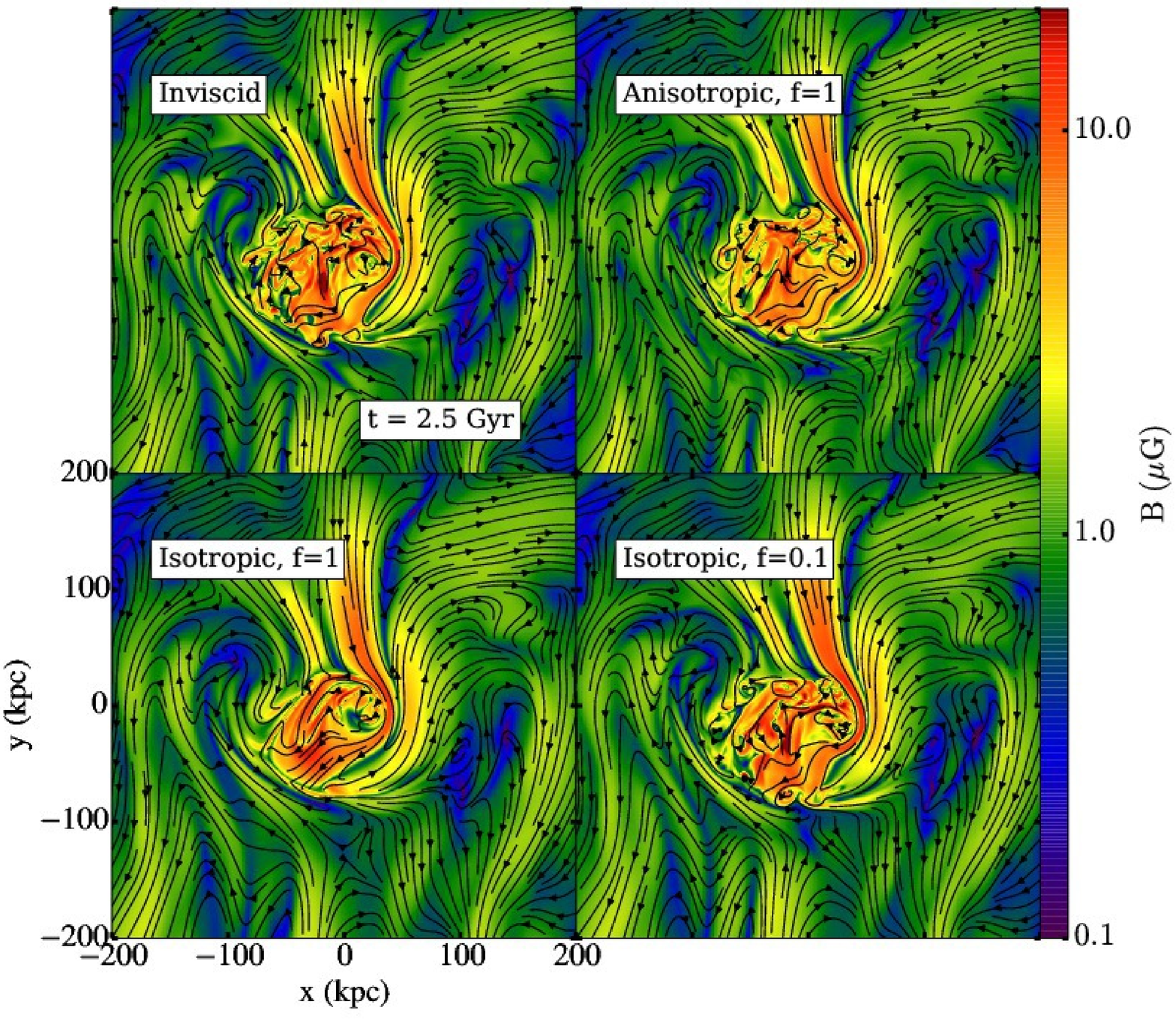}
\includegraphics[width=0.49\textwidth]{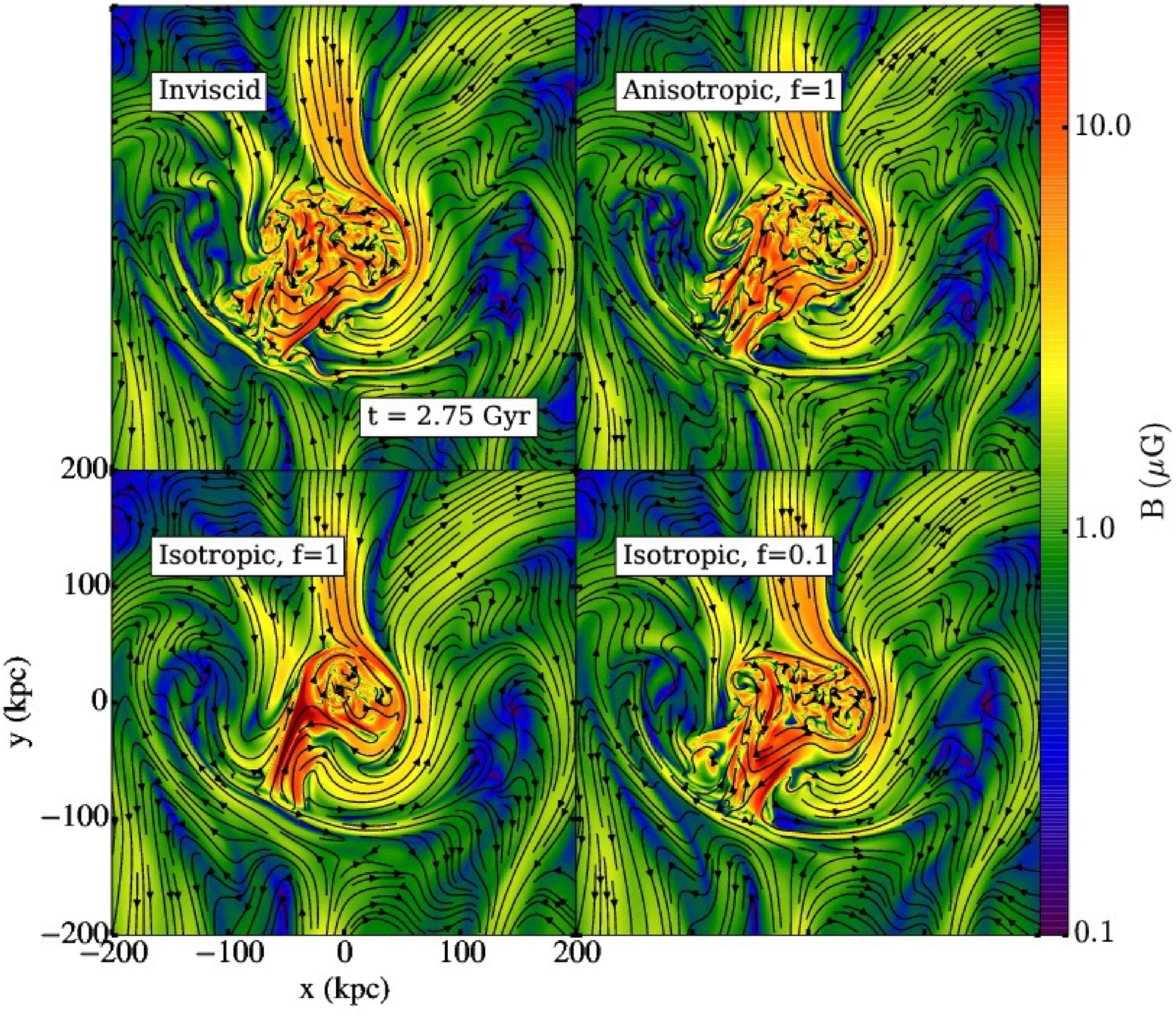}
\includegraphics[width=0.49\textwidth]{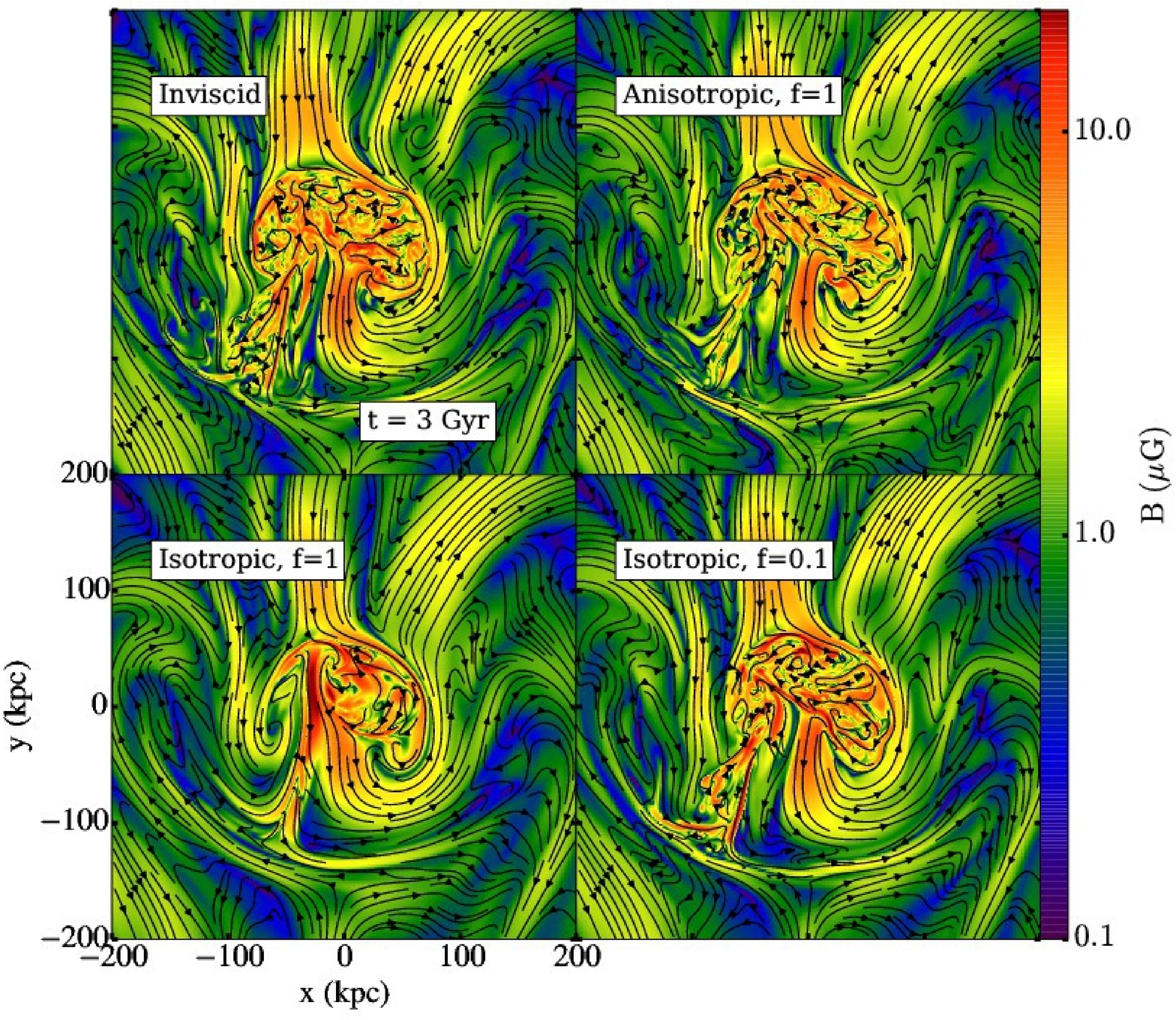}
\includegraphics[width=0.49\textwidth]{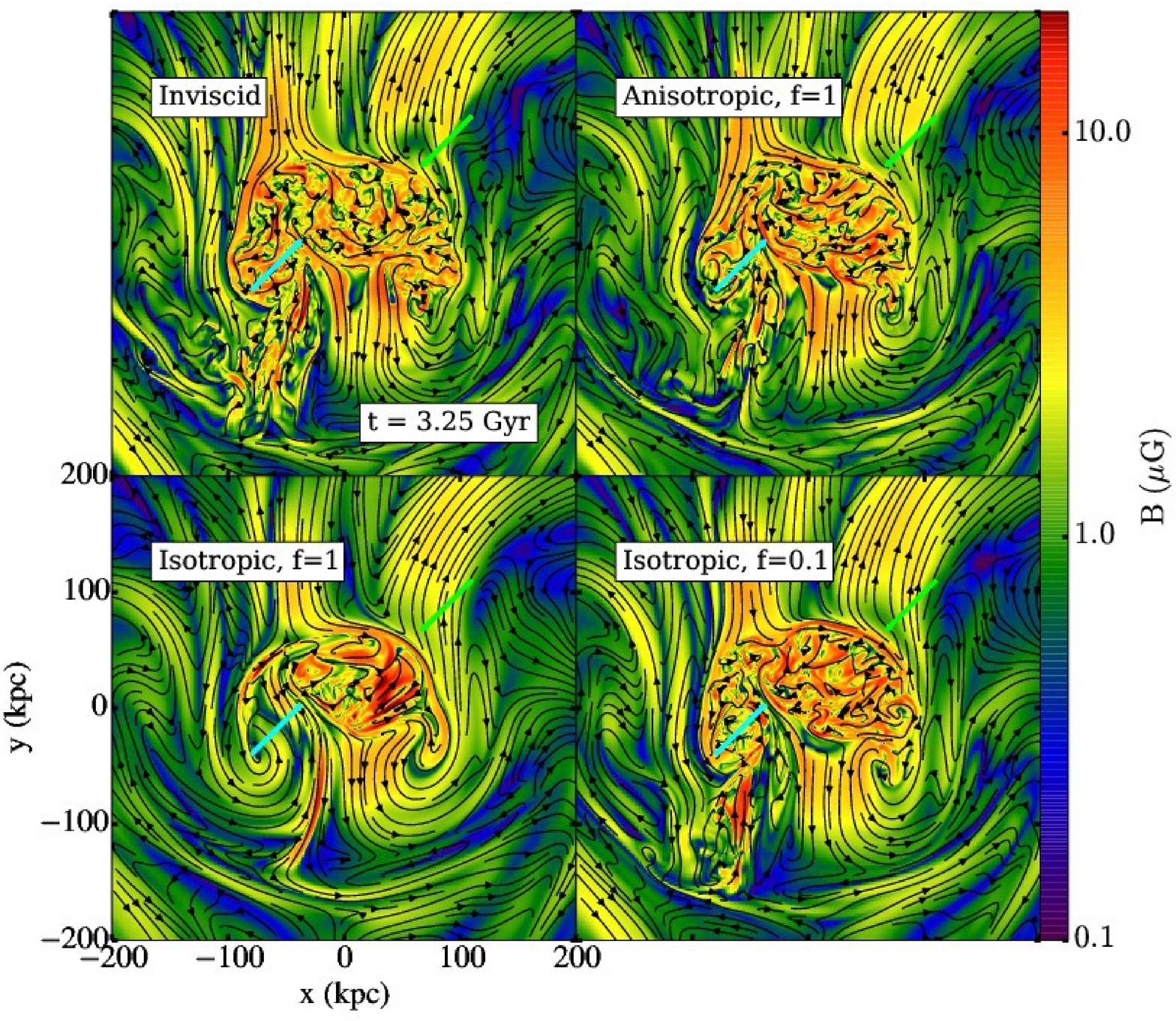}
\caption{Slices of magnetic-field strength through the center of the ``AM06" simulation domain 
at times $t$ = 2.5, 2.75, 3.0, and
 3.25~Gyr, with magnetic-field lines in the plane overlaid. Arrows mark the positions of cold fronts with
 morphologies that are altered by viscosity.\label{fig:AM06_bmag}}
\end{center}
\end{figure*}

The second merger scenario is identical to our previous setup in ZML11 and Z13, which was originally derived from
\citet{AM06}. In this case, the cool-core cluster is more massive ($M
\sim 10^{15}~{\rm M}_\odot$) and hotter ($T \sim 7$~keV), resembling A2029
(though not exactly reproducing it). The gas-less
subcluster is 5 times less massive than the main cluster and
approaches it with an initial impact parameter of $b =
500$~kpc. We will refer to this setup as ``AM06''.

The gravitational potential on the grid is the sum of two
collisionless ``rigid bodies'' corresponding to the contributions to the potential
from both clusters. This approach to modeling the potential is
used for simplicity and speed over solving the Poisson equation for
the matter distribution, and is an adequate approximation for our
purposes. It is the same approach that we used in ZML11, and is
justified and explained in \citet{rod12}. 

The tangled magnetic field of the cluster is set up in an similar way to the
simulations of ZML11 and Z13. We use a radial profile
for the magnetic field strength of $B = B_0 [\rho(r)/\rho_0]^\eta$ with
$\eta = 0.5$ \citep{bon10}. The average magnetic field strength in the core region corresponds to $\beta = p_{\rm th}/_B \sim 1000-1500$, which is on the weaker end of the parameter space from ZML11. This low field strength was chosen to ensure the development of K-H instabilities in order to determine the effect of viscosity on their development. For the three-dimensional tangled structure of the initial magnetic field, we follow the approach of \citet{rus07, rus08} and \citet{rus10}, and we refer the reader to those papers for the details. Figure \ref{fig:profiles} shows the initial radial profiles of density, temperature, and magnetic-field strength for both clusters.

For all of the simulations, we set up the main cluster within a cubical
computational domain of width $L = 4$~Mpc on a side. We employ static
mesh refinement (SMR), with 256 cells on a side on the
top-level domain and five smaller domains at increasing levels of
refinement, each half the size of the domain just above it, centered on the cluster potential minimum. This results in a finest cell size of $\Delta{x}
= 0.98$~kpc within a region of $l^3$ = (250~kpc)$^3$. Appendix A
details the results of a test where we experiment with higher and
lower spatial resolutions. \citet{rod13b} examined the effect of
varying resolution on the development of K-H instabilities in a
simulation of the ICM, and determined that the gross morphologies of
the K-H rolls is captured well down to resolutions of 32 cells per
perturbation wavelength, a wavelength of $\lambda \sim 32$~kpc in our
simulations (for further discussion of the effects of resolution see
Section \ref{sec:res}). 

Similar to the approach taken in Z13 for simulations with diffusive
effects (viscosity and/or conduction), we switch on these effects at
the pericentric passage ($t$ = 1.0~Gyr in the case of the ``Virgo''
simulations and $t \sim 1.5$~Gyr in the case of the ``AM06''
simulations). This is just before the onset of sloshing, and allows us to examine the effects of these processes on the sloshing cold fronts without any significant change
in our initial condition of the main cluster that would occur if these
effects (particularly conduction) were switched on. 

%
%
\begin{figure*}
\begin{center}
\includegraphics[width=0.49\textwidth]{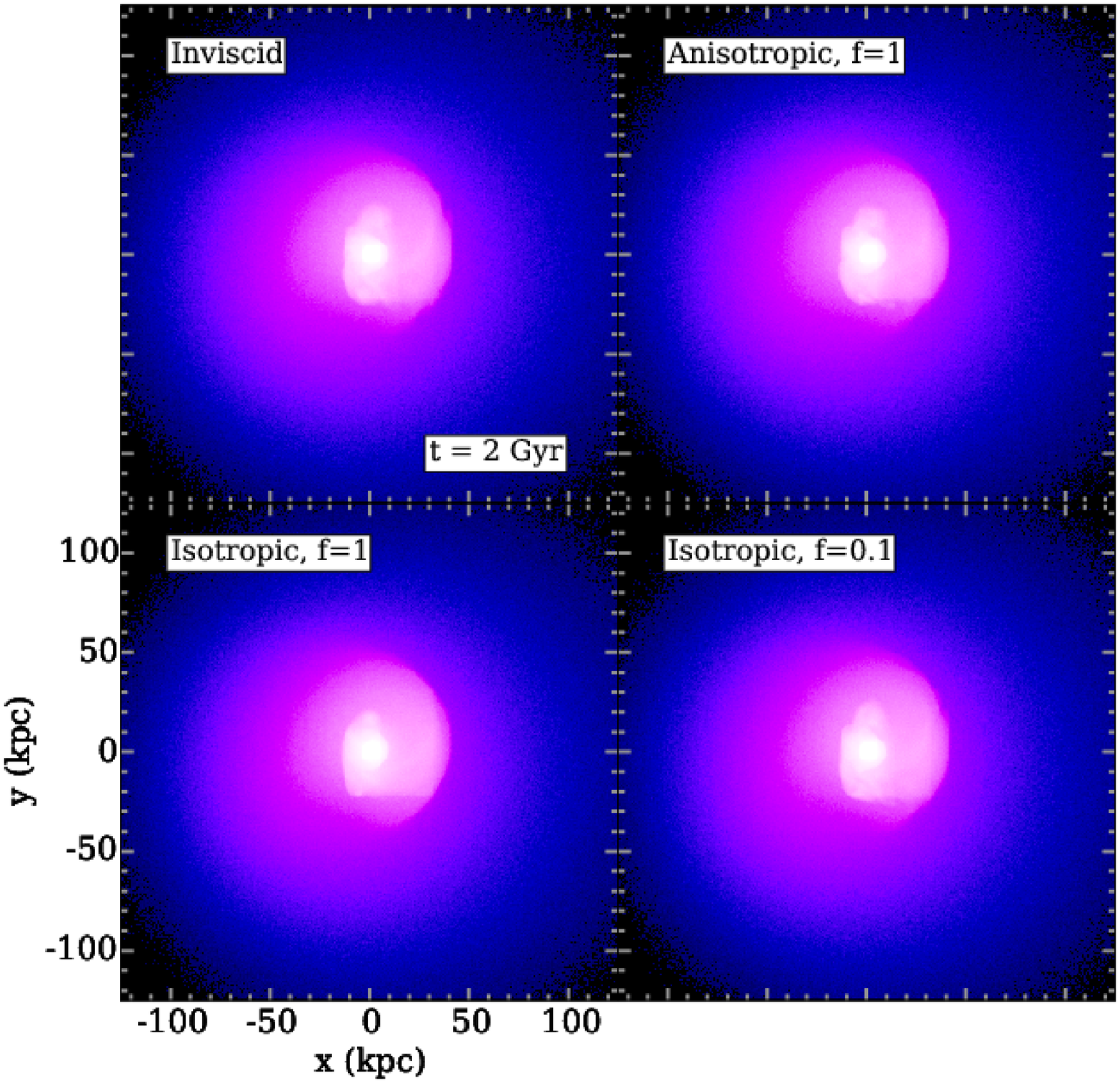}
\includegraphics[width=0.49\textwidth]{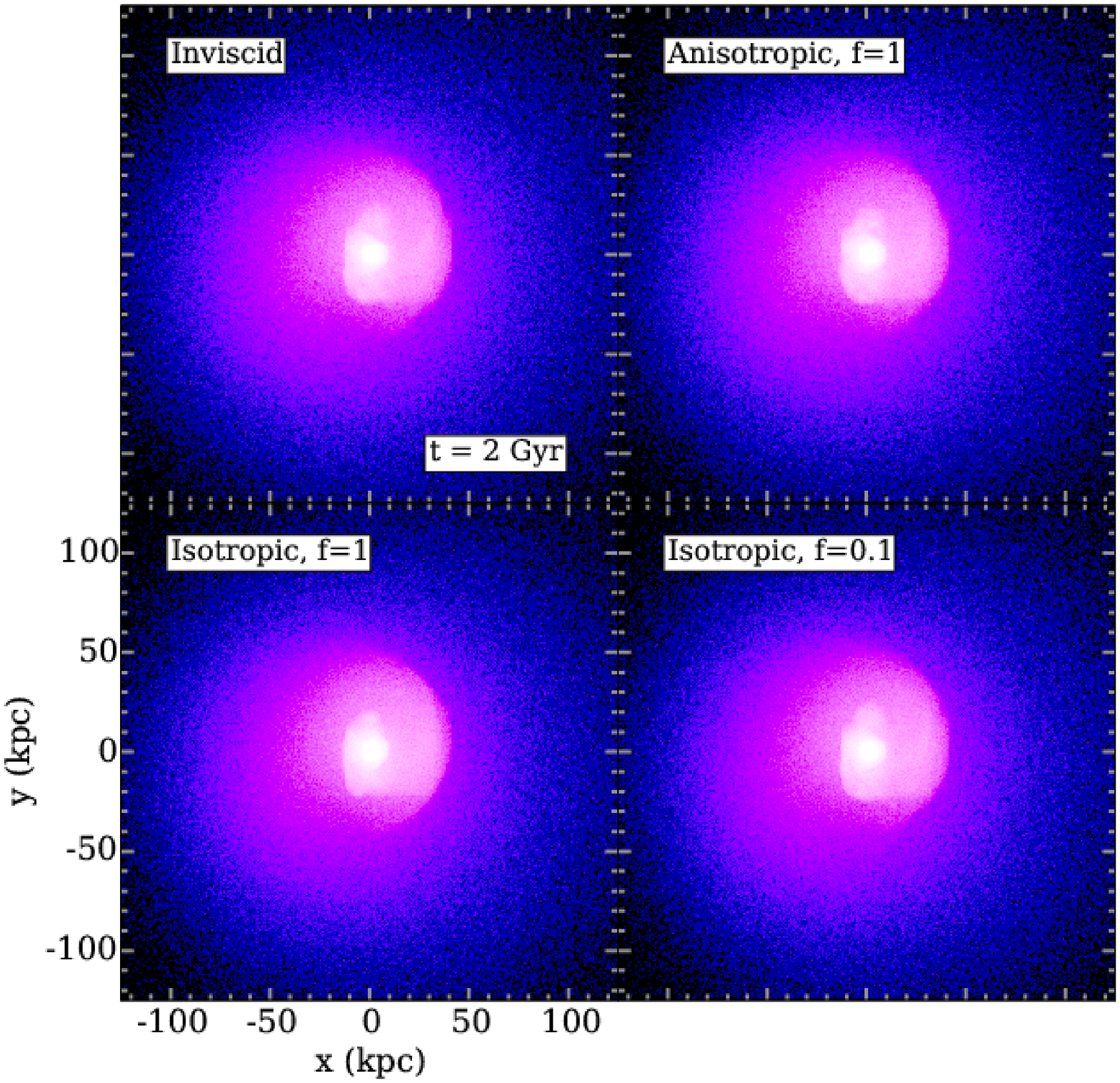}
\includegraphics[width=0.49\textwidth]{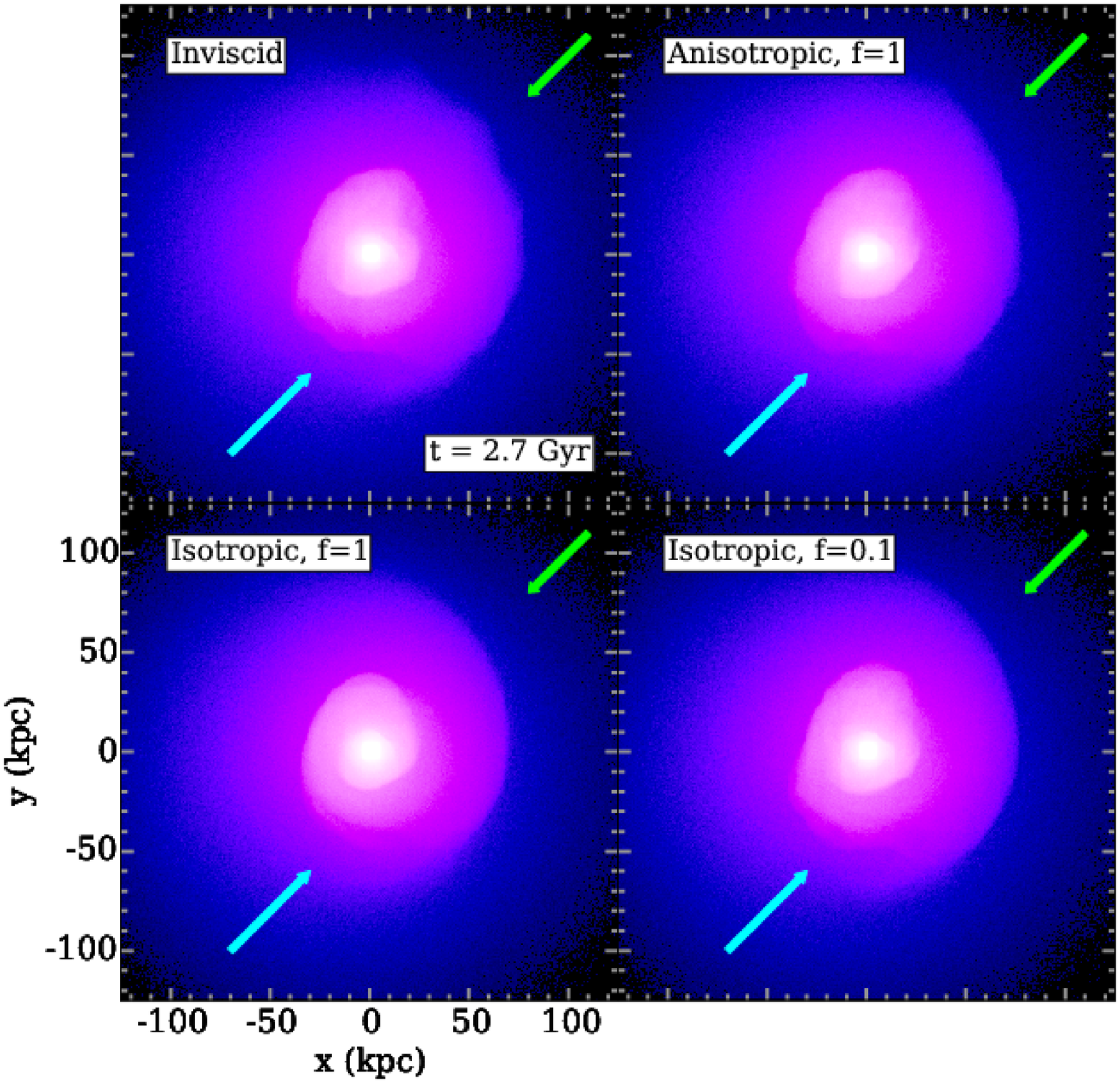}
\includegraphics[width=0.49\textwidth]{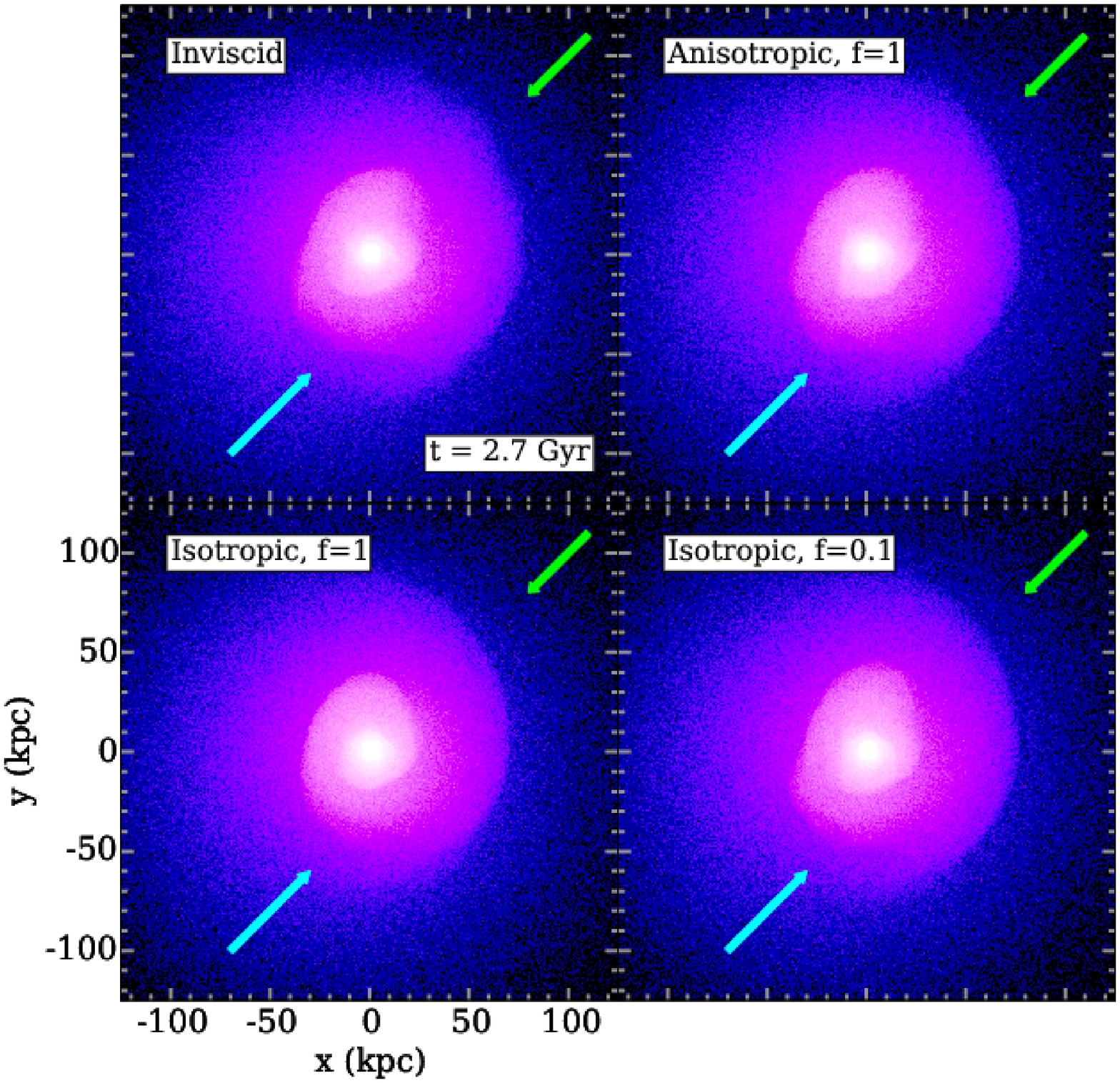}
\caption{Synthetic X-ray counts images of the ``Virgo'' cluster for the
 four different viscosity simulations at times $t = 2.0$ and 2.7~Gyr. Left panels: 300-ks
 exposure. Right panels: 30-ks exposure. Arrows mark the positions of
 cold fronts with morphologies that are altered by viscosity.\label{fig:virgo_counts}}
\end{center}
\end{figure*}

%
%
\begin{figure*}
\begin{center}
\includegraphics[width=0.49\textwidth]{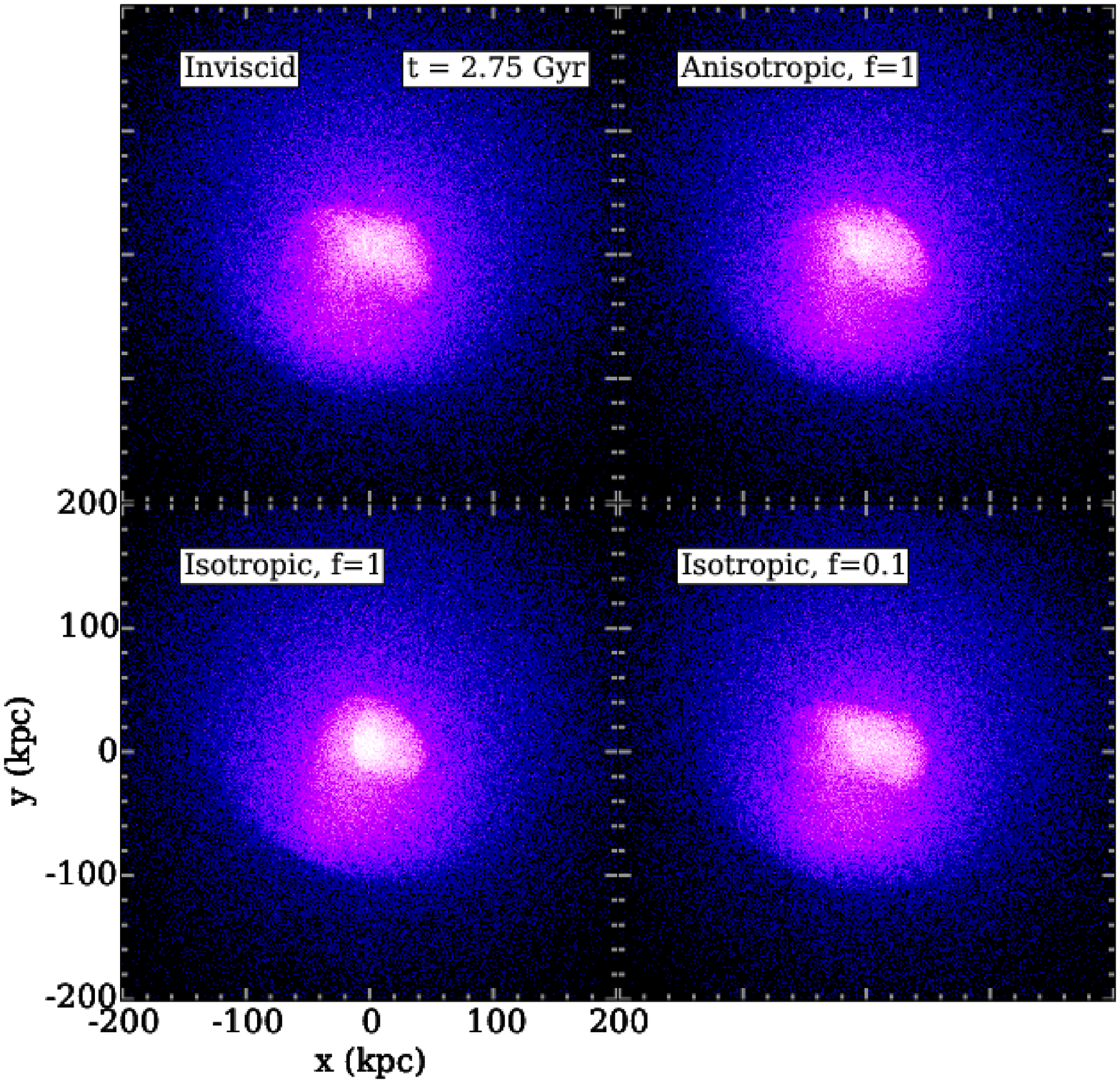}
\includegraphics[width=0.49\textwidth]{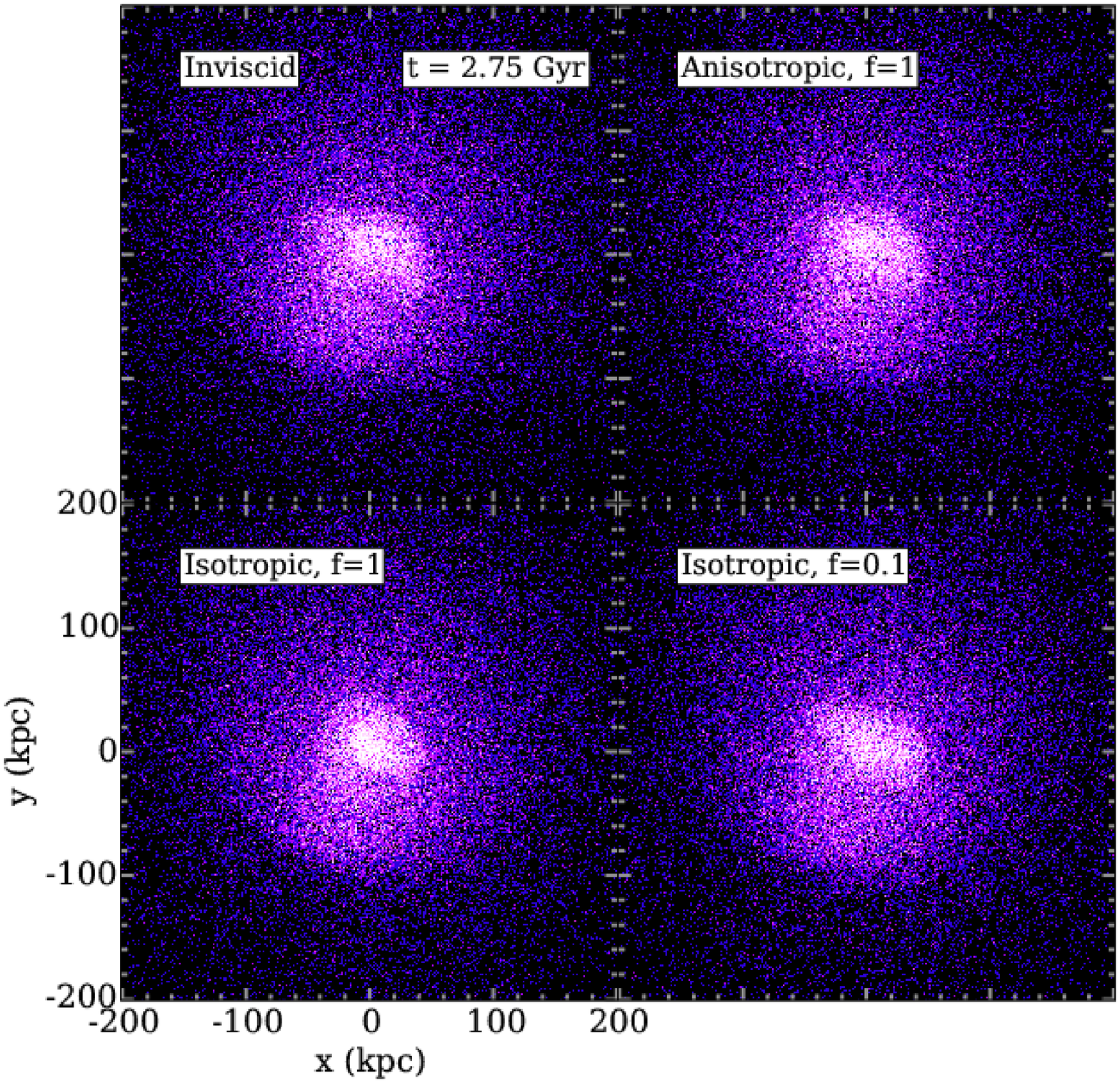}
\includegraphics[width=0.49\textwidth]{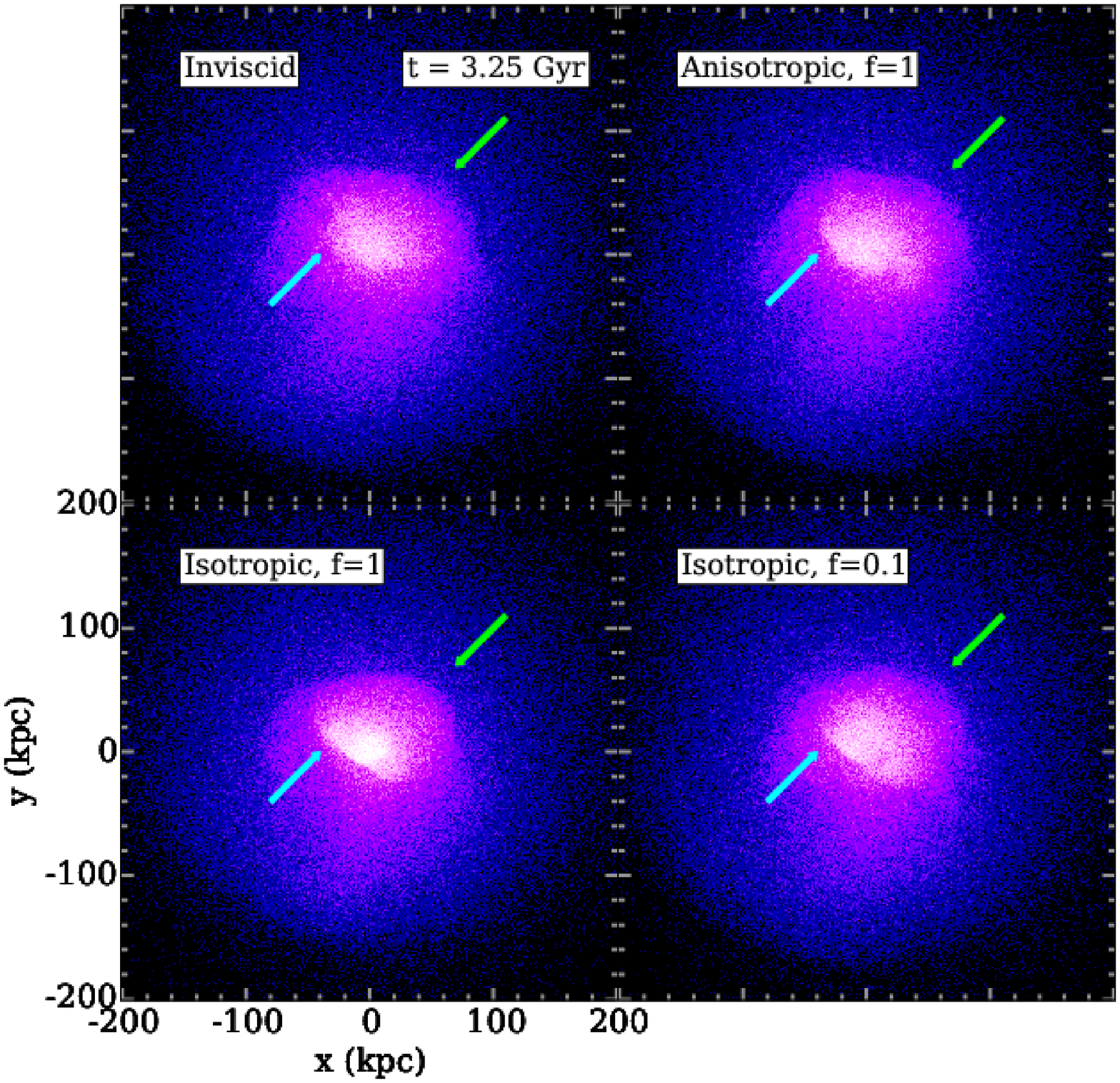}
\includegraphics[width=0.49\textwidth]{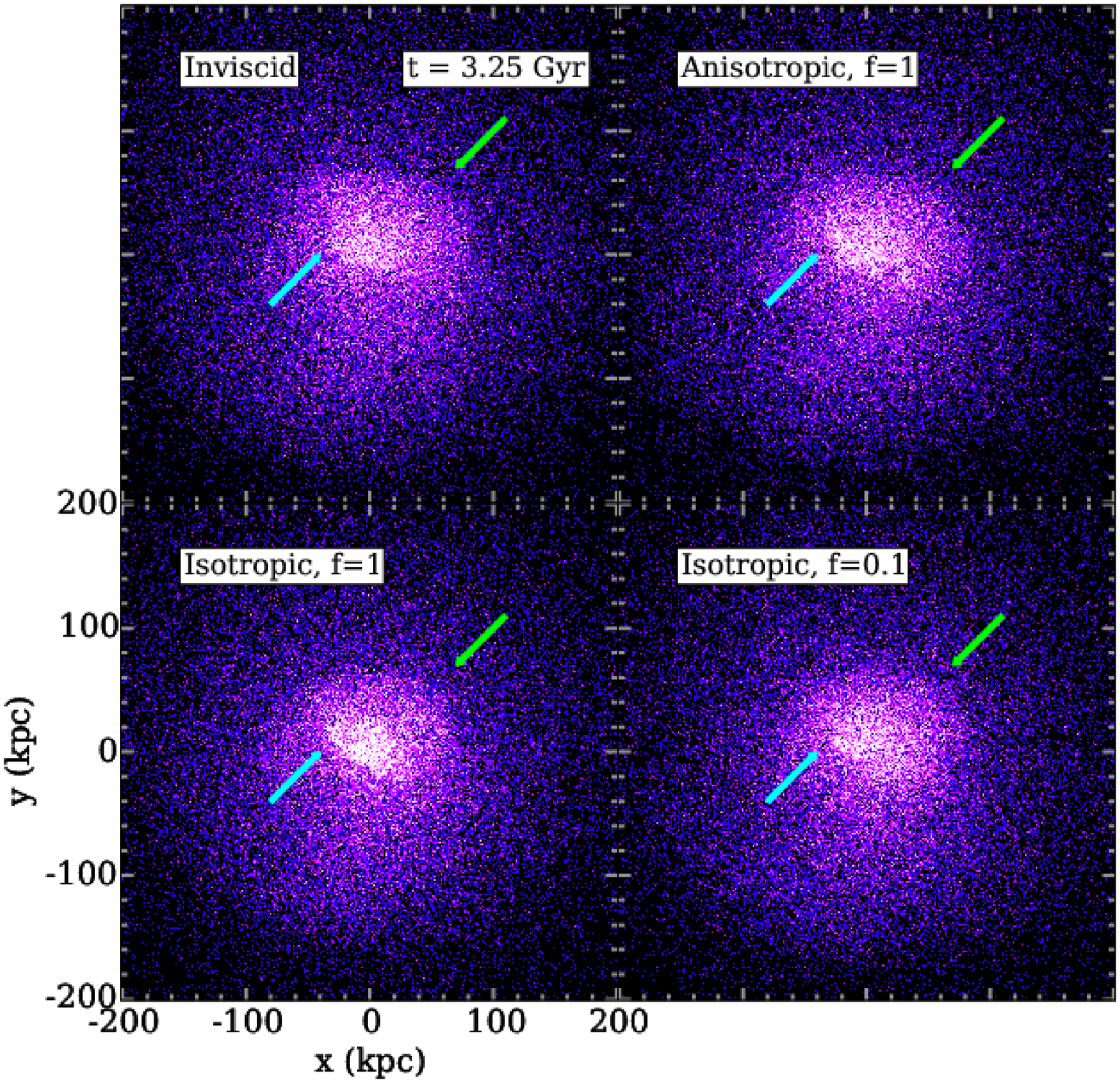}
\caption{Synthetic X-ray counts images of the ``AM06'' cluster for the
 four different viscosity simulations at times $t = 2.75$ and 3.25~Gyr. Left panels: 300-ks
 exposure. Right panels: 30-ks exposure.  Arrows mark the positions of
 cold fronts with morphologies that are altered by viscosity.\label{fig:AM06_counts}}
\end{center}
\end{figure*}

%
%
\begin{figure*}
\begin{center}
\includegraphics[width=0.49\textwidth]{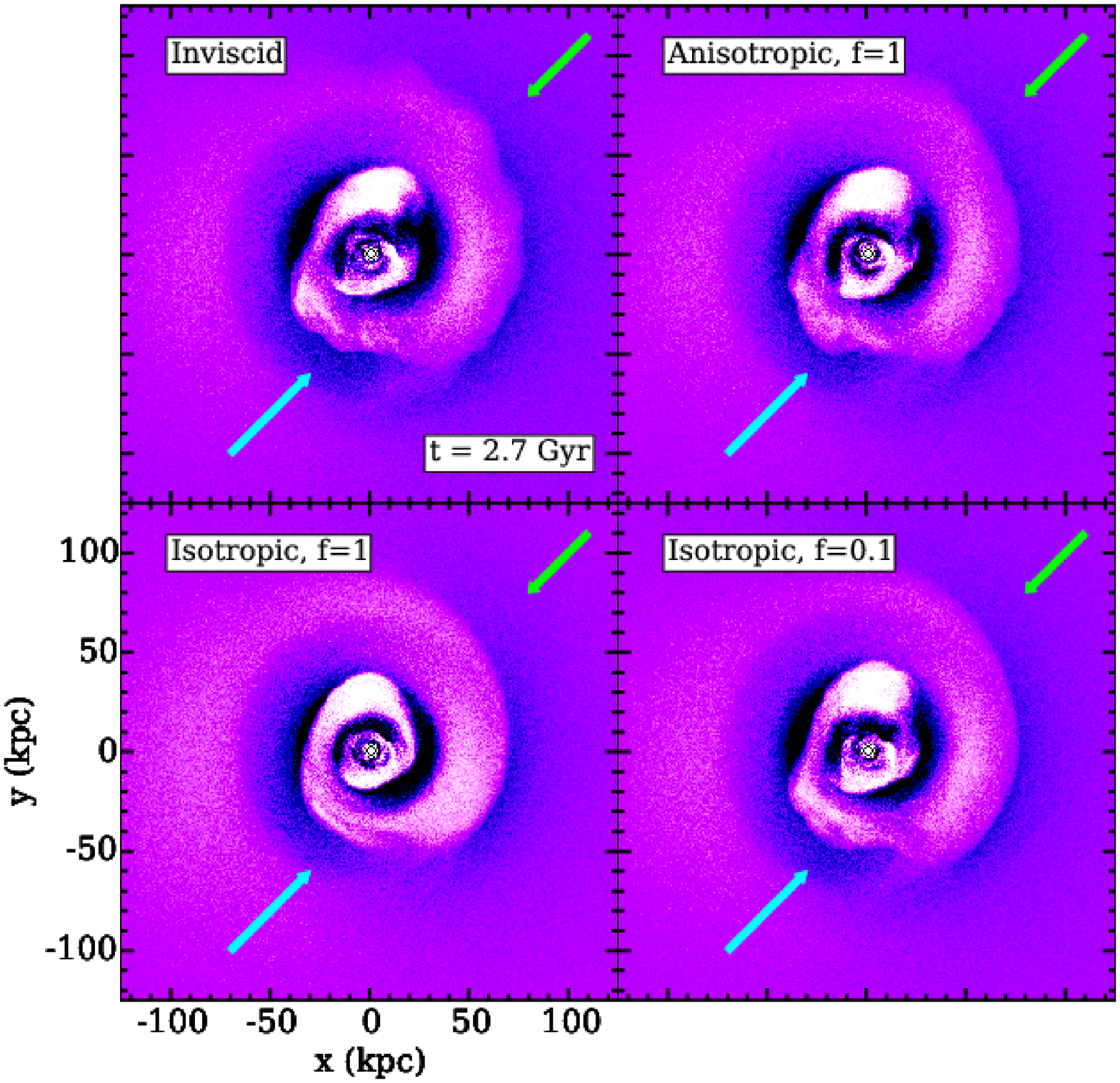}
\includegraphics[width=0.49\textwidth]{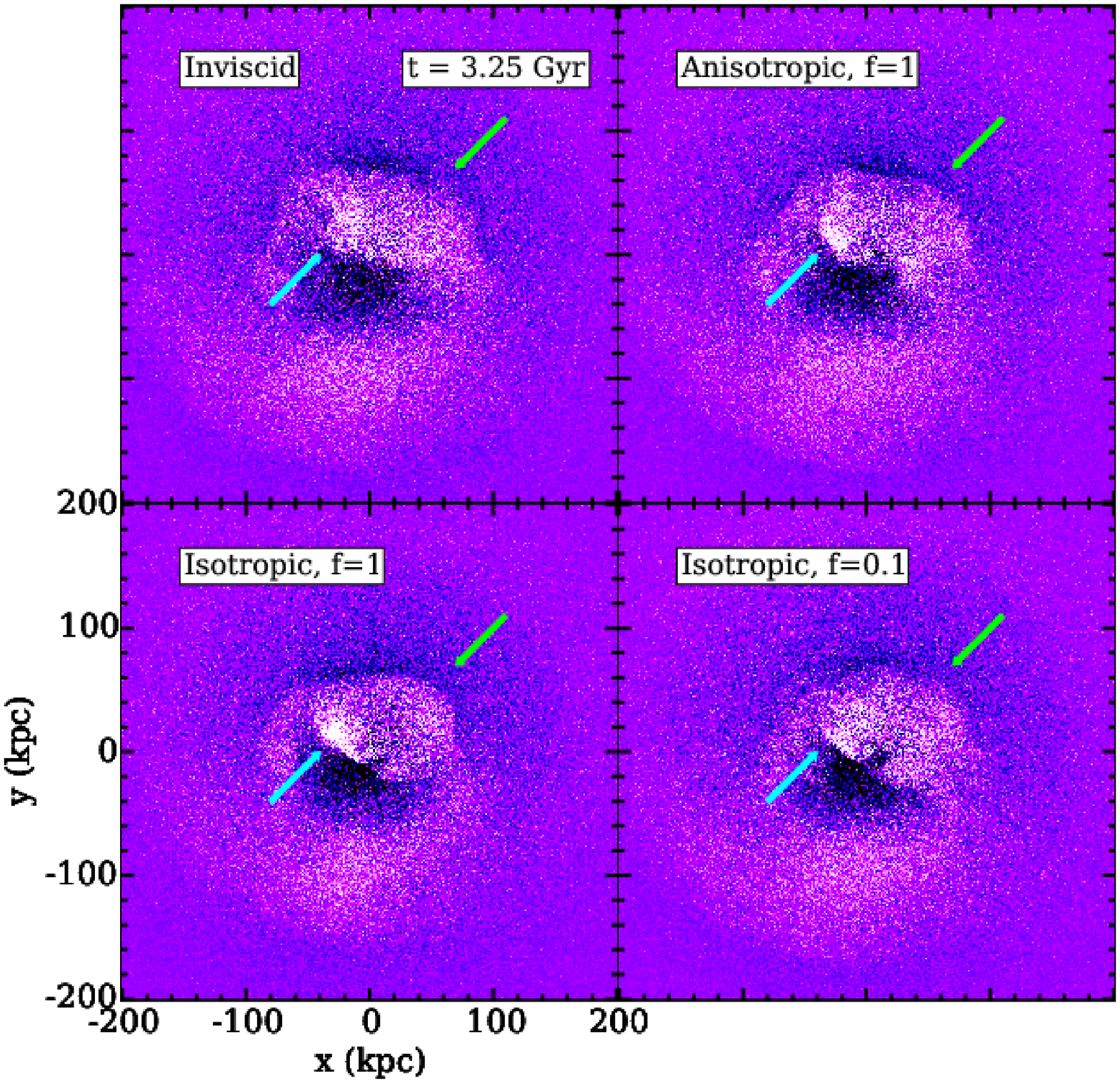}
\caption{Synthetic X-ray residual images (from the 300-ks exposures, after subtraction of the azimuthally averaged profile) of the two clusters for the four different viscosity simulations at late times. Left panels: ``Virgo'' model. Right panels: ``AM06'' model.\label{fig:resid}}
\end{center}
\end{figure*}

%
%
\begin{figure*}
\begin{center}
\includegraphics[width=0.9\textwidth]{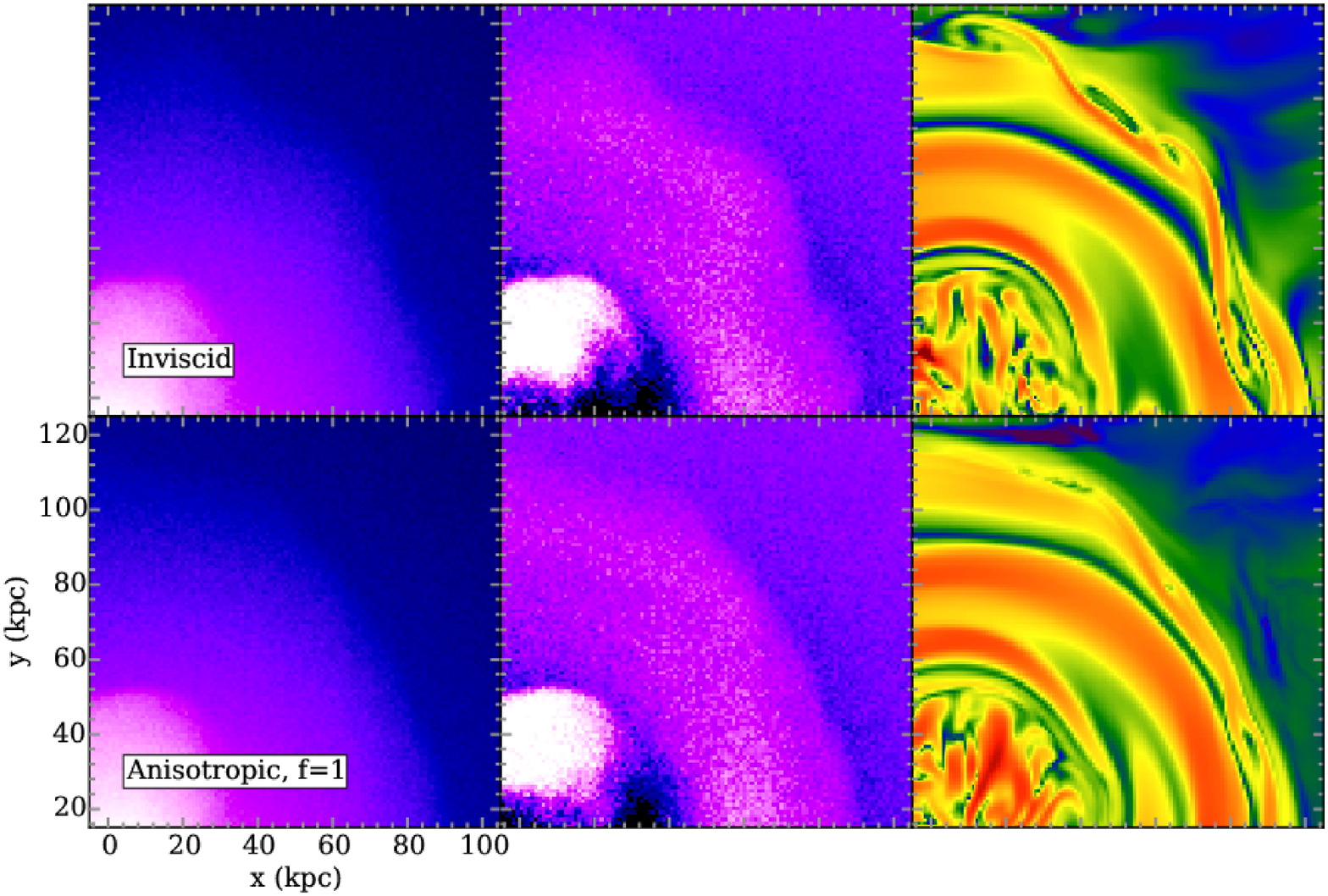}
\caption{Close-up of a cold front in the ``Virgo'' model with an appearance that has been modified by viscosity. The top panels show the {\it inviscid} simulation, whereas the bottom panels show the {\it anisotropic, f=1} simulation. From left to right, the panels show the 300~ks exposure, 300~ks residuals, and slice through the magnetic field strength.\label{fig:virgo_closeup}}
\end{center}
\end{figure*}

%
%
\begin{figure*}
\begin{center}
\includegraphics[width=0.9\textwidth]{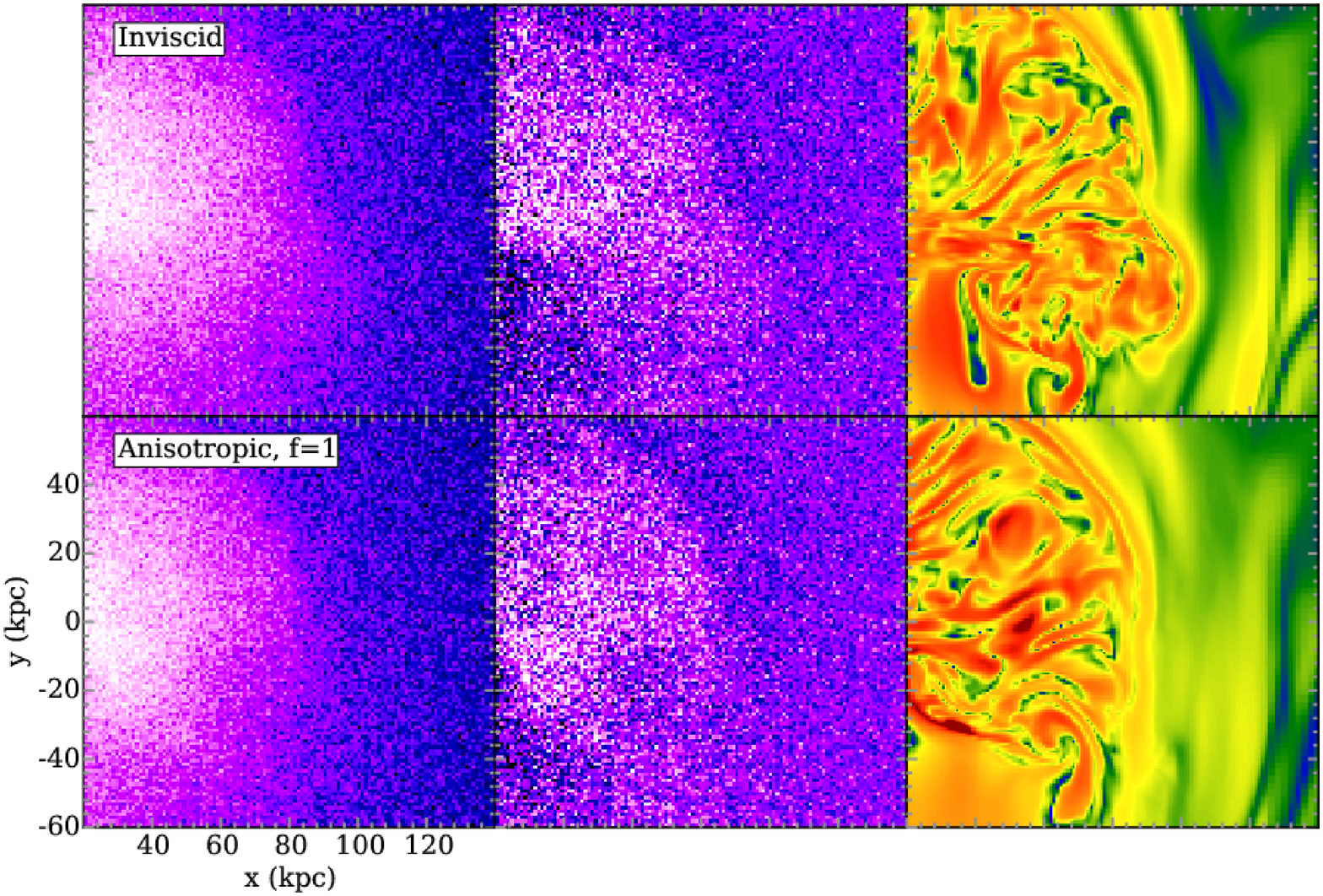}
\caption{Close-up of a cold front in the ``AM06'' model with an appearance that has been modified by viscosity. The top panels show the {\it inviscid} simulation, whereas the bottom panels show the {\it anisotropic, f=1} simulation. From left to right, the panels show the 300~ks exposure, 300~ks residuals, and slice through the magnetic field strength.\label{fig:AM06_closeup}}
\end{center}
\end{figure*}

For these two cluster models, we perform four simulations each with
different types of viscosity. Also, for each model we perform two
additional simulations with anisotropic thermal conduction, one of the
two including Braginskii viscosity. In addition, to see the effect of viscosity separate from that of the magnetic field for the same cluster setup, we also ran unmagnetized runs with and without viscosity of both setups (detailed in Section \ref{sec:nomag}). These are all exploratory runs, not intended to cover the full parameter space of the viscosity and magnetic field values; we will try to reproduce particular observed clusters in a future work.

%
%
\begin{figure*}
\begin{center}
\includegraphics[width=0.49\textwidth]{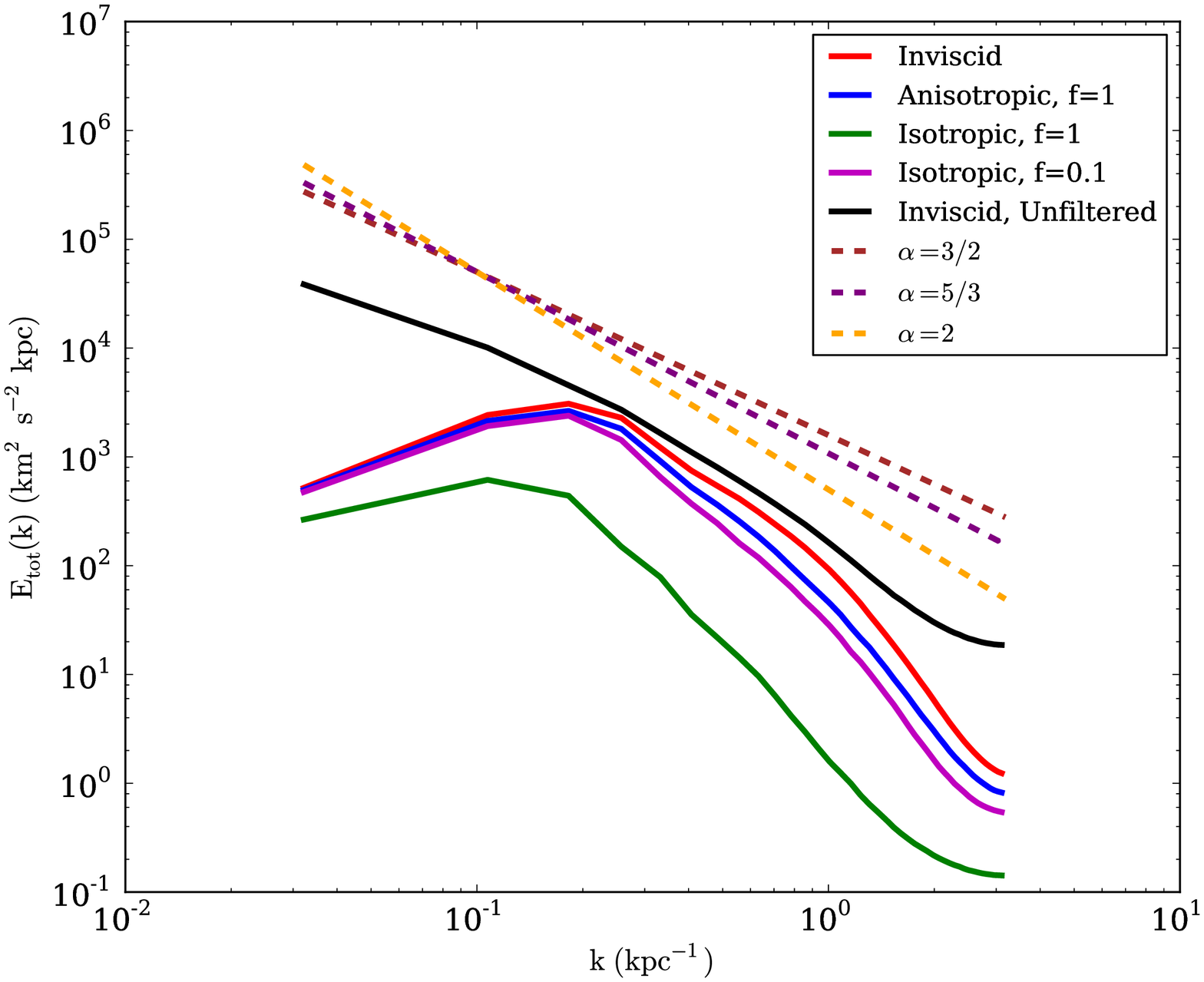}
\includegraphics[width=0.49\textwidth]{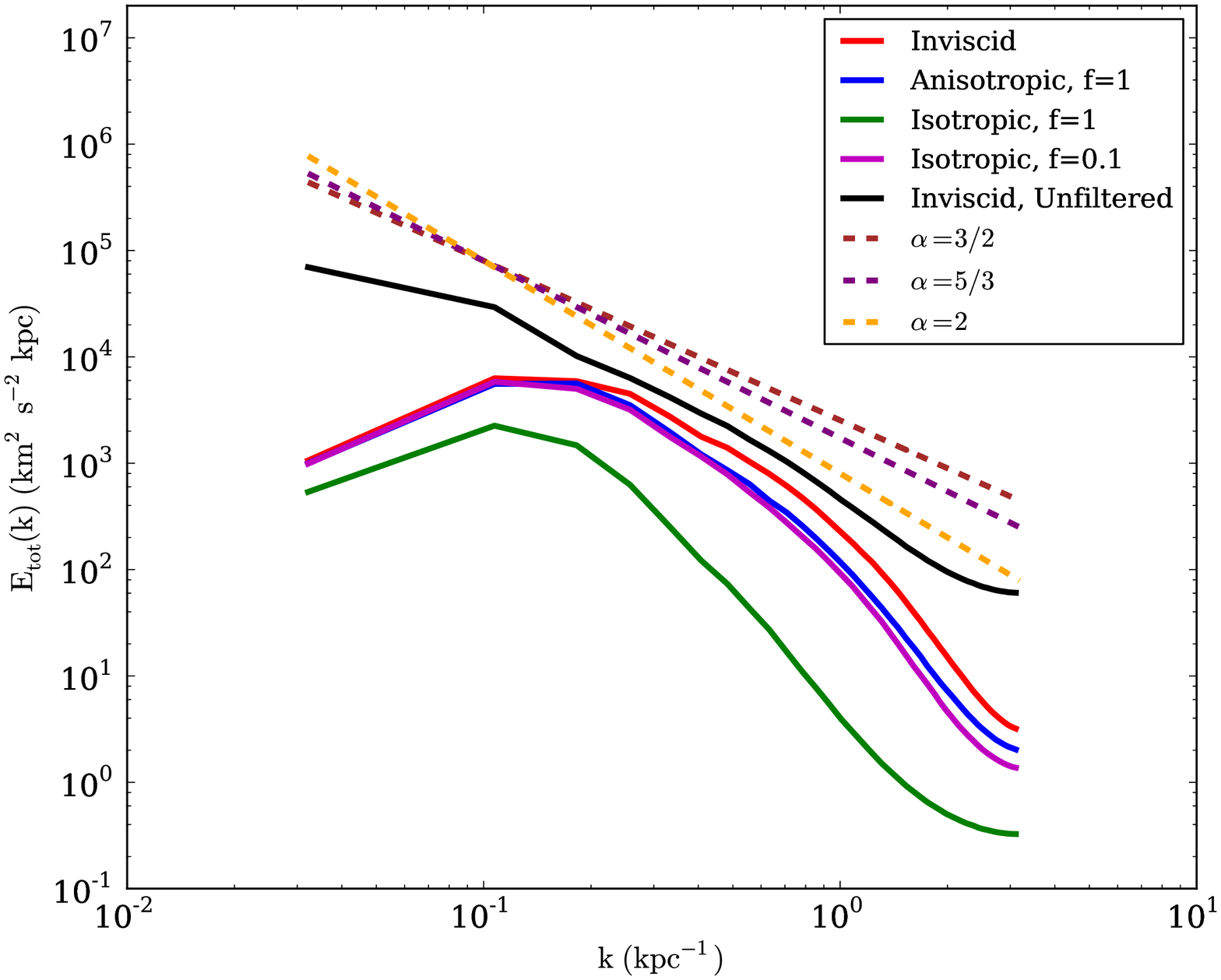}
\caption{Velocity power spectra for the ``Virgo'' simulations at time $t$ = 2.7~Gyr (left panel) and the ``AM06'' simulations at time $t$ = 3.25~Gyr (right panel).\label{fig:power_spectrum}}
\end{center}
\end{figure*}

%
%
\section{Results}\label{sec:results} 

%
%
\subsection{Temperature and Magnetic-Field Maps}\label{sec:temperature_bfield_maps}

We begin by visually examining the development of K-H 
instabilities at cold front surfaces in slices through the
center of the simulation domain in the merger plane. Figures \ref{fig:virgo_temp}
and \ref{fig:AM06_temp} show temperature slices for the two clusters at
various times in the different simulations. The {\it inviscid} simulations
quickly develop K-H instabilities at the cold front surfaces, which
continue to grow and develop as the simulations progress, though the development of K-H instabilities is suppressed to a degree by the weak magnetic field compared to what would obtain in an unmagnetized simulation (ZML11 and Section \ref{sec:nomag}). These
instabilities generate turbulence (to be discussed in Section
\ref{sec:turbulence}) and significant small-scale structure in temperature. In the
{\it anisotropic} and {\it isotropic} $f=0.1$ simulations, the same instabilities develop at early times, but as the cold fronts expand,
these are progressively damped out by viscosity. We mark the
positions of cold fronts with different morphology in the different
simulations with arrows. However, there are still some locations along the fronts where there are indications of instability, as well as some small-scale temperature
fluctuations within the envelope of the cold fronts. In the case of the {\it isotropic, f=1} simulations, no K-H instabilties or small-scale temperature structures develop at all during the course of the simulations.

%
%
\begin{figure*}
\begin{center}
\includegraphics[width=0.49\textwidth]{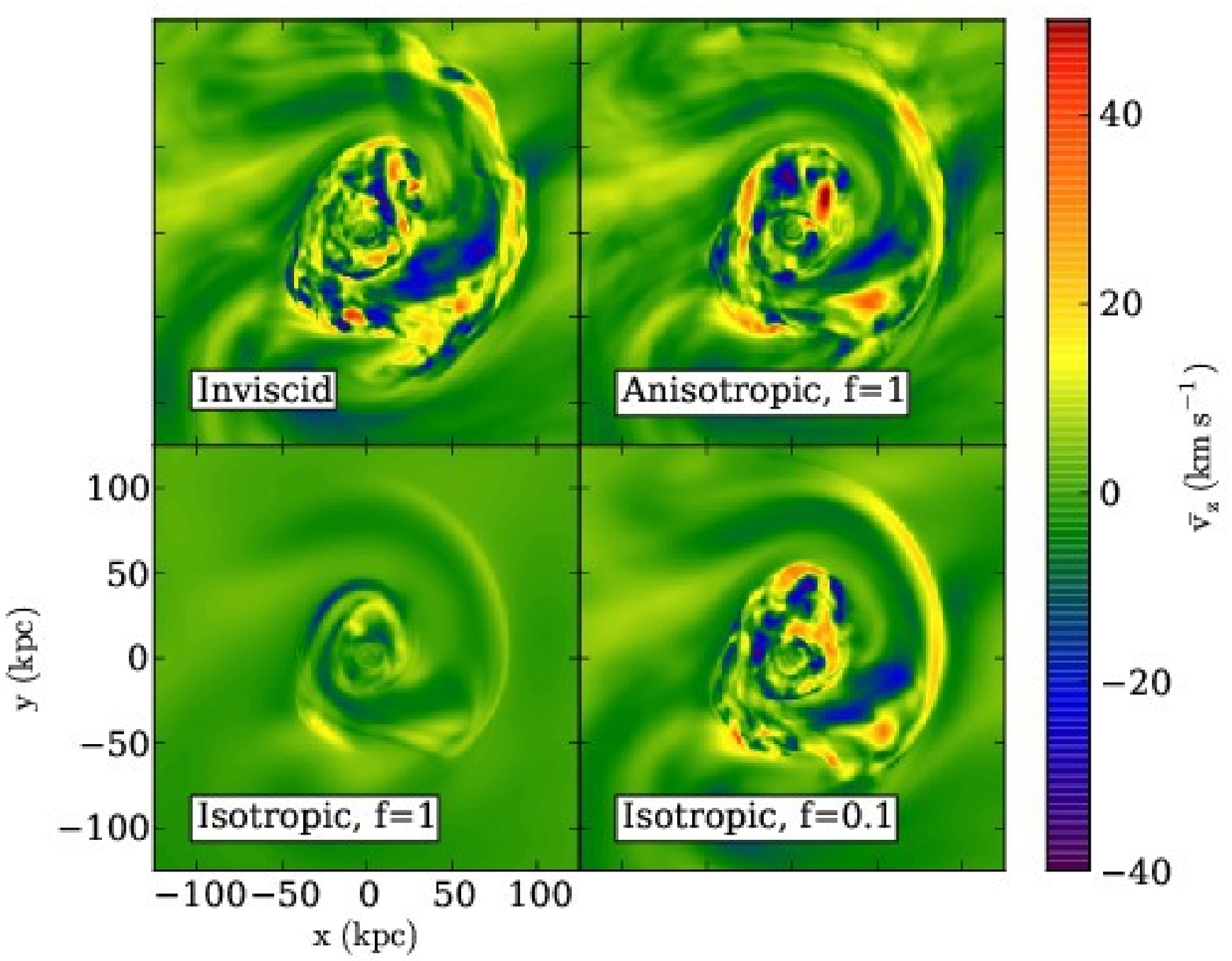}
\includegraphics[width=0.49\textwidth]{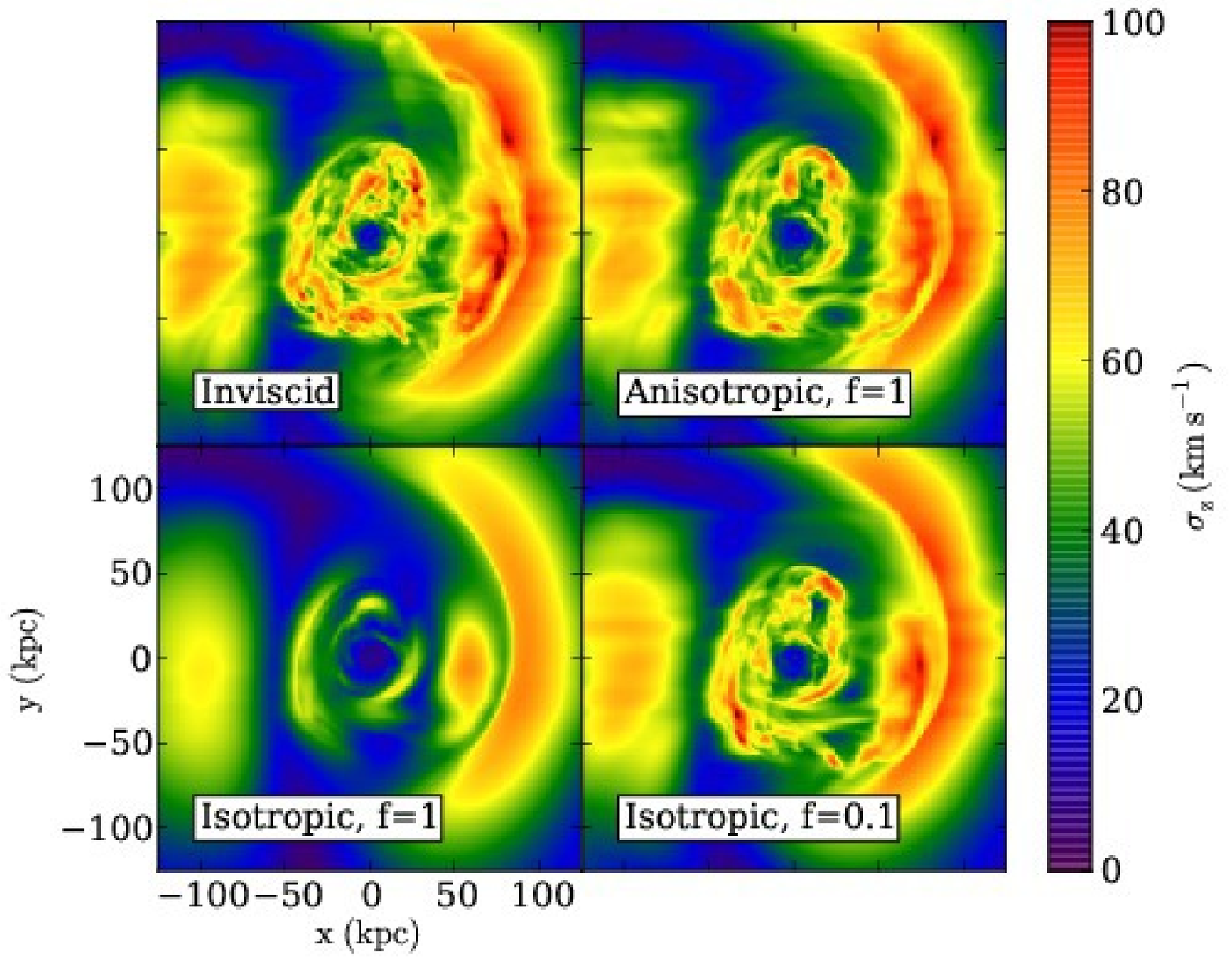}
\caption{Emission-weighted line-of-sight velocity along the $z$-axis for the ``Virgo'' simulations at time $t$ = 2.7~Gyr. 
Left panels: velocity mean, right panels: velocity dispersion, which is a good measure of turbulence \citep{zuh13b}.\label{fig:virgo_vel_z}}
\end{center}
\end{figure*}

Figures \ref{fig:virgo_bmag} and \ref{fig:AM06_bmag} show slices of the magnetic-field
strength for the two clusters at the same times for the different
simulations. The overall evolution of the magnetic field is very
similar in all the simulations, but there are some
differences. The magnetized layers that occur along cold front surfaces are
disrupted by K-H instabilities in the {\it inviscid} simulations,
whereas viscosity results in magnetic-field layers that are more
smooth (see the arrows in the figures), though there is not a perfect wrapping of fields around the
cold fronts in any of the simulations (we will examine the
consequences of this for anisotropic thermal conduction in Section \ref{sec:conduction}). As viscosity is
increased, there is less small-scale structure in the magnetic field. 

%
%
\subsection{Synthetic X-ray Observations}\label{sec:sim_xray}

To make closer comparisons with observations of cold fronts in actual clusters,
we construct synthetic {\it Chandra} observations. We generate X-ray photons
from an \code{APEC} model \citep{smi01} of a thermal gas with
$Z = 0.3Z_\odot$ for each simulation cell within a radius of $R = 250$~kpc from the cluster potential minimum.\footnote{This radius contains the cold fronts at
 all epochs that we simulate; we find that including gas at larger
 radii in the synthetic images does not affect our conclusions.} We
also apply Galactic absorption assuming $N_H = 10^{21}$~cm$^2$. Our
procedure for generating these observations is outlined in detail in Appendix B. To generate an approximate number of counts that would be expected from a mosaic
of {\it Chandra} observations of each of our simulated clusters, we
use the on-axis effective area curve of the ACIS-S3 chip. For the
``Virgo'' simulations, we use the actual redshift of the Virgo
cluster, $z$ = 0.0036, and for the ``AM06'' simulations, we choose a
redshift of $z$ = 0.1. The images have been blocked so that the pixels
correspond to the size of the finest simulation cell size, $\Delta{x}
= 0.98$~kpc.\footnote{In the ``AM06'' case, this happens to 
 correspond roughly to the {\it Chandra} pixel size (0.492'') at $z = 0.1$. In the
 ``Virgo'' case, the {\it Chandra} pixels are $\sim$0.04~kpc, much
 smaller than our cell size.} To determine the effect of photon statistics, we simulate
two different exposures, 300~ks and 30~ks. 

%
%
\begin{figure*}
\begin{center}
\includegraphics[width=0.49\textwidth]{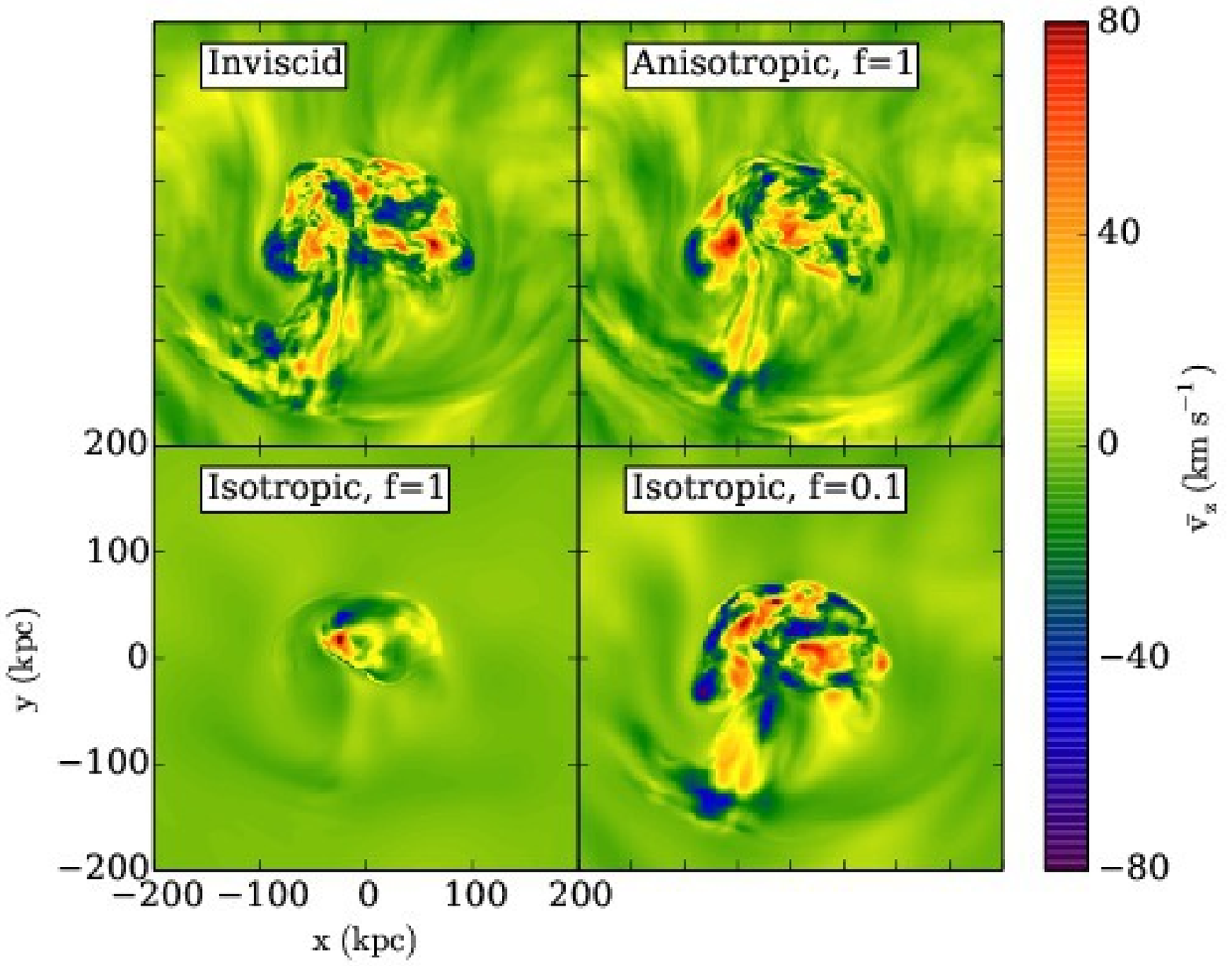}
\includegraphics[width=0.49\textwidth]{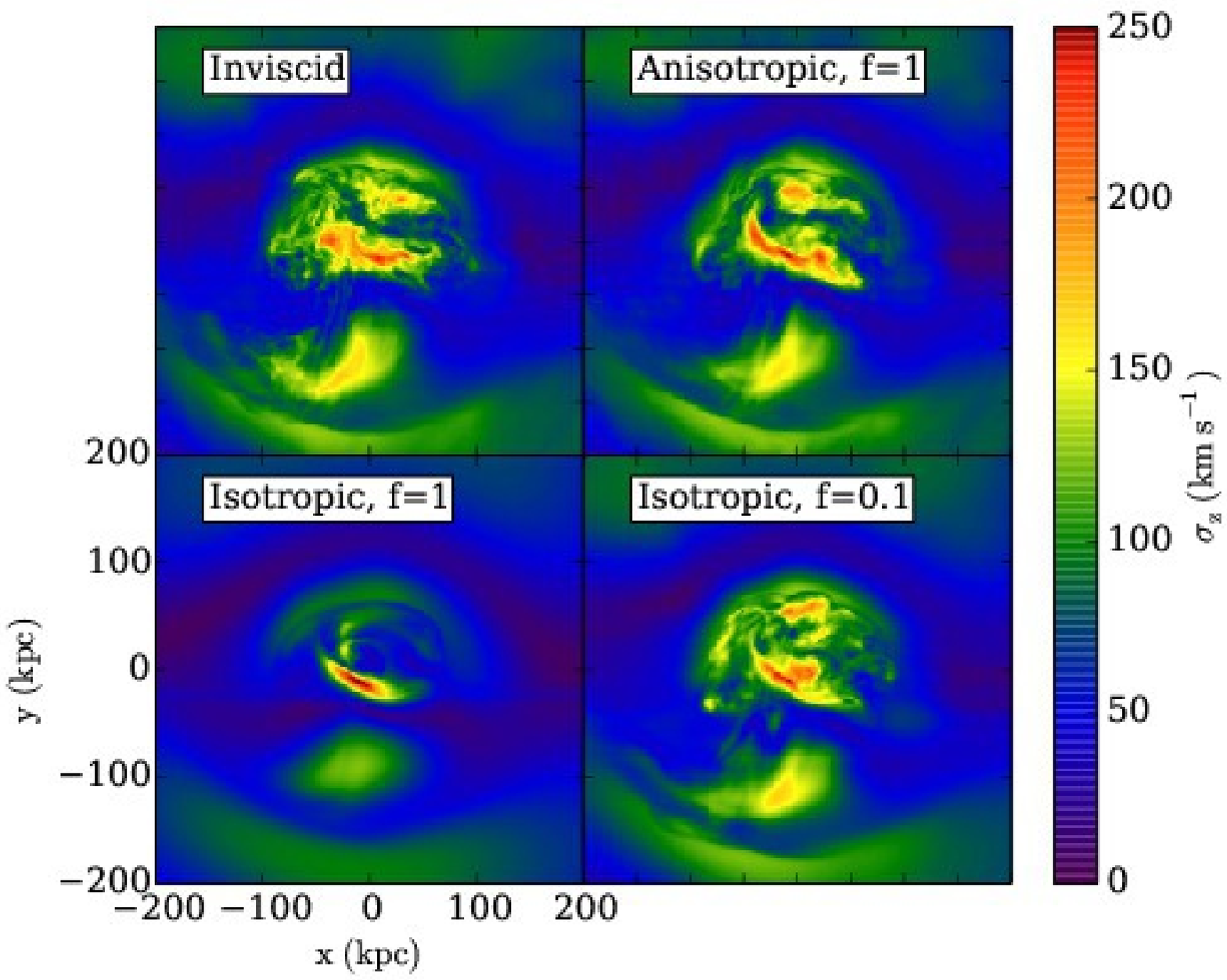}
\caption{Emission-weighted projections of the line-of-sight velocity while looking along the 
$z$-axis for the ``AM06'' simulations at time $t$ = 3.25~Gyr. Left panels: velocity mean, right panels: velocity dispersion.\label{fig:AM06_vel_z}}
\end{center}
\end{figure*}

Figures \ref{fig:virgo_counts} and \ref{fig:AM06_counts} show the X-ray counts images of
the two model clusters at different epochs and for different exposure
times. Projection effects and Poisson noise make the interpretation of
these images more difficult than for the temperature slices. It is clear
that long exposures (in our case, $\sim$300~ks) are needed for
determining the amount of K-H disruption at the cold front
surfaces. In the 30-ks exposures, it is much more difficult to make distinctions
between the cold fronts in the simulations due to poor statistics. For the rest of this section, we will focus on the 300-ks exposures. 

In the {\it inviscid} simulations, the presence of K-H instabilities
along the cold front surfaces is readily apparent from the X-ray
images. By contrast, in the {\it isotropic, f=1} simulations, the cold
front arcs are completely smooth in appearance. In the {\it isotropic}, $f=0.1$ and {\it
  anisotropic} simulations, while the cold front surfaces are mostly smooth,
small indications of instability are visible. Some fronts that
appear in the viscous simulations are so disrupted in the {\it inviscid} simulations that they are no longer technically cold fronts (one example is in the ``AM06'' simulations, indicated by the cyan arrow in Figure \ref{fig:AM06_counts}). Another indicator of cold front instability is the overall shape of the cold fronts. Unstable fronts tend to have a ``boxy'' shape, due to the development of the instability over the curved surface, whereas fronts stabilized by viscosity appear more round \citep[a fact previously pointed out by][]{rod12c}. This is most evident in the ``AM06'' case. 

The characteristics of the cold front structures may be seen more
clearly by looking at the surface brightness residuals instead of the
surface brightness itself. Figure \ref{fig:resid} shows the surface
brightness residuals for both models at a late epoch, 
generated by computing an azimuthally averaged radial profile of the counts from the 300-ks
observation and subtracting the corresponding image from the original
image. As in Figures \ref{fig:virgo_counts} and \ref{fig:AM06_counts}, the arrows mark the positions of cold fronts
with different morphologies in the different simulations. The
difference in the degree of K-H disruption can be more clearly
seen. In particular, it is obvious from the right panel of Figure
\ref{fig:resid} that the cold front at the inner radius (indicated by
the cyan arrow) is essentially impossible to discern even in the residual
image in the {\it inviscid} case. Figures \ref{fig:virgo_closeup} and Figures \ref{fig:AM06_closeup} show two "close-ups" of several quantities side by side in the region of cold fronts where instabilities have been suppressed due to anisotropic viscosity.

%
%
\subsection{The Effect of Viscosity on Sloshing-Driven Turbulence}\label{sec:turbulence}

Viscosity can have an effect on turbulence driven by
sloshing by damping out turbulent motions on the smallest length scales.
Future X-ray telescopes such as {\it Astro-H}\footnote{\url{http://astro-h.isas.jaxa.jp/
}} and {\it Athena+}\footnote{\url{http://www.the-athena-x-ray-observatory.eu/}} will be able to measure turbulence and bulk motions from the line shift and line widths of spectral lines \citep{ino03,sun03}. It is instructive to determine the effect of our different viscosity models on the turbulence driven by sloshing. 

%
%
\begin{figure*}
\begin{center}
\includegraphics[width=\textwidth]{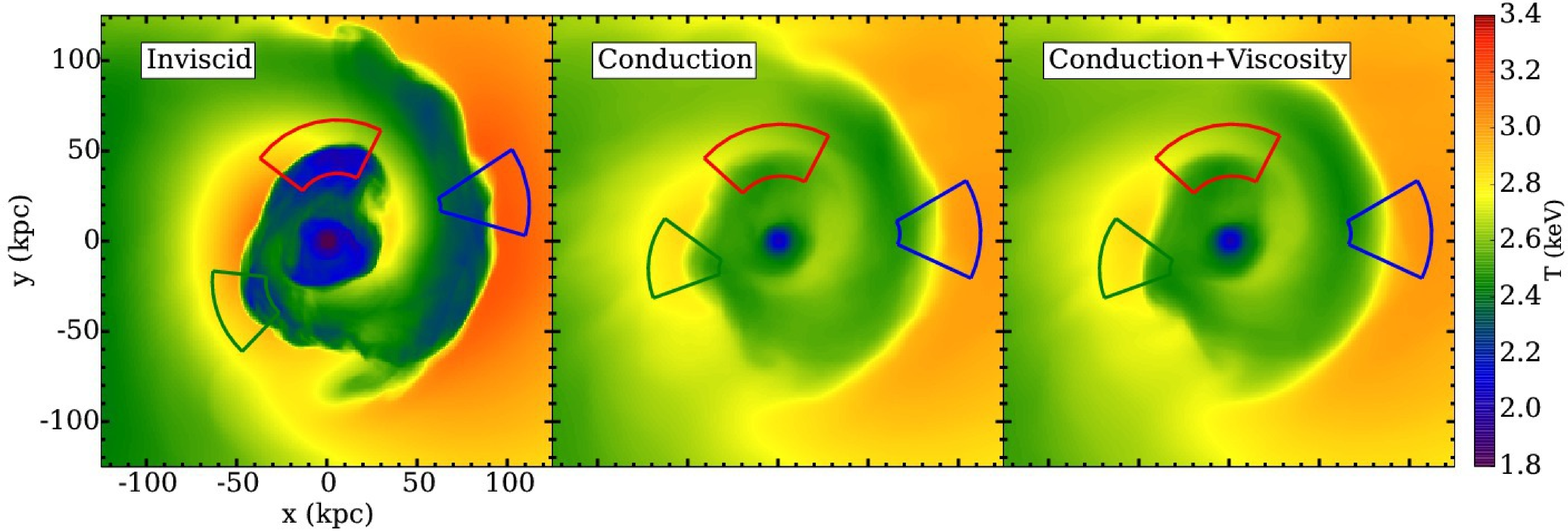}
\includegraphics[width=\textwidth]{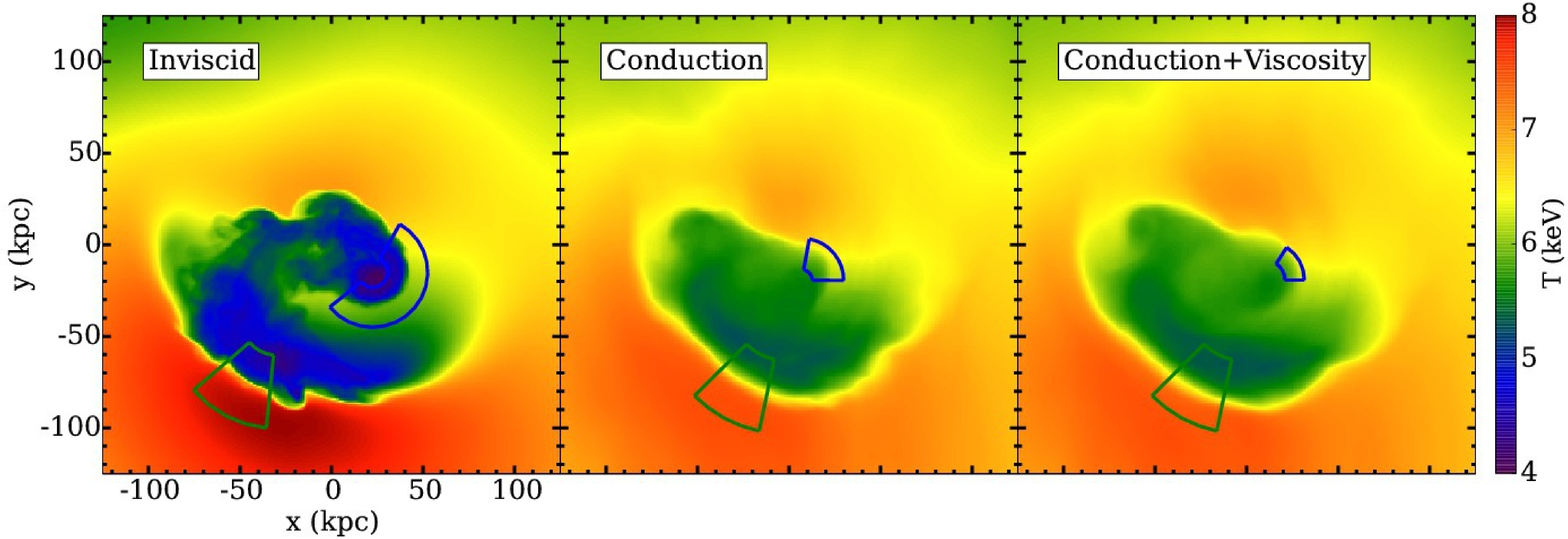}
\caption{Projected ``spectroscopic-like'' temperature images of the
 ``Virgo'' (top) and ``AM06'' (bottom) clusters for the
 three different conduction simulations. Regions correspond to the
 profiles in Figure \ref{fig:cond_profiles}.\label{fig:cond_temp}}
\end{center}
\end{figure*}

A straightforward way to distinguish the turbulent velocity field
between the different models is to compare the velocity power
spectra. The sloshing motions themselves complicate matters, since the
velocity jumps across the cold fronts impose a $\sim$${k^{-2}}$
signature on the power spectrum at all wavenumbers $k$. Since the bulk
of the motion in the simulation is in the sloshing motions, this
signal dominates. This reduces to the more general problem of separating
turbulence from bulk motions in hydrodynamic simulations, a subject of previous investigations \citep{vaz12,zuh13b}. To carry out this decomposition, we use the procedure outlined in \citet{vaz12}, which uses a multi-scale filtering scheme to iteratively converge on the mean velocity field around each cell, which is then subtracted from the
total velocity to get the turbulent component. As noted by
\citet{zuh13b}, carrying out the filtering procedure naively on the
velocity field from sloshing cold fronts will not completely remove
the effect of the bulk motions, because the velocity differences across the
cold fronts will result in the local mean velocity changing
discontinuously across these surfaces. \citet{vaz12} noted the same
difficulty with shock fronts, and proposed filtering on the skew of
the local velocity field as a way to avoid these difficulties. We
adopt the same approach here for our simulations. 

Figure \ref{fig:power_spectrum} shows the filtered (``turbulent'') velocity power spectrum
as a function of wavenumber $k$ for all of the simulations, as well as the unfiltered velocity power spectrum for the {\it inviscid} simulations (for comparison). The {\it
 anisotropic} simulations with the full Braginskii viscosity result
in a very modest reduction of the turbulent power, whereas for the
{\it isotropic} simulations with full Spitzer viscosity the turbulent
power is reduced by more than an order of magnitude. 

Figures \ref{fig:virgo_vel_z} and \ref{fig:AM06_vel_z} show the emission-weighted projected line-of-sight mean velocity and line-of-sight velocity dispersion along the $z$-axis of the simulation domain for all of the
simulations. For this exercise, we have not filtered out the bulk
motions. The projections of the two {\it inviscid} simulations show evidence of
turbulence in small-scale random velocity fluctuations. In
the {\it anisotropic} simulations, turbulence is somewhat reduced, but the
small-scale velocity fluctuations are still present. This is also the
case for the {\it isotropic}, $f=0.1$ simulations. In the {\it
 isotropic}, $f=1$ simulations, essentially all of the small-scale
motions are absent.

%
%
\begin{figure*}
\begin{center}
\includegraphics[width=\textwidth]{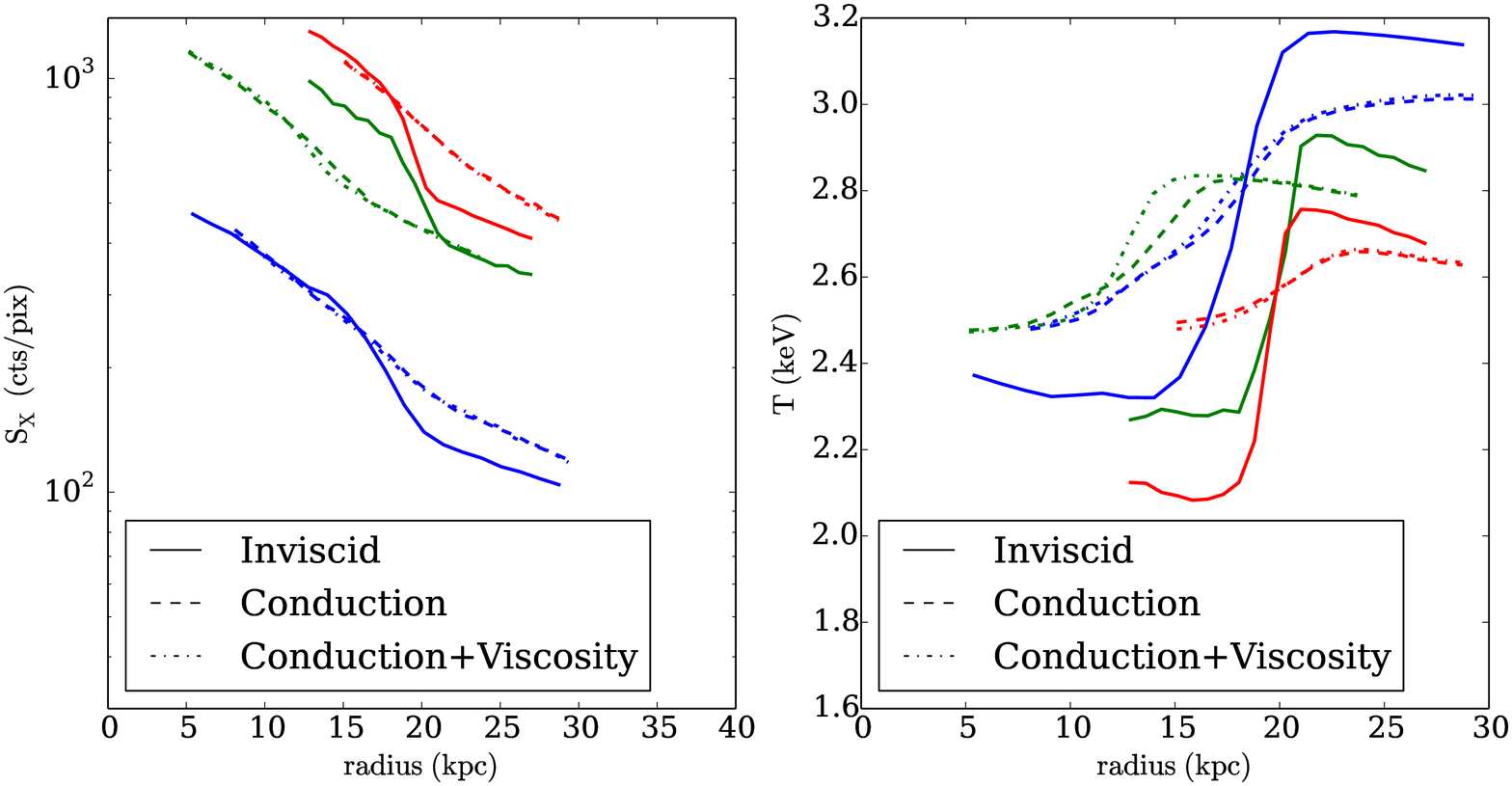}
\includegraphics[width=\textwidth]{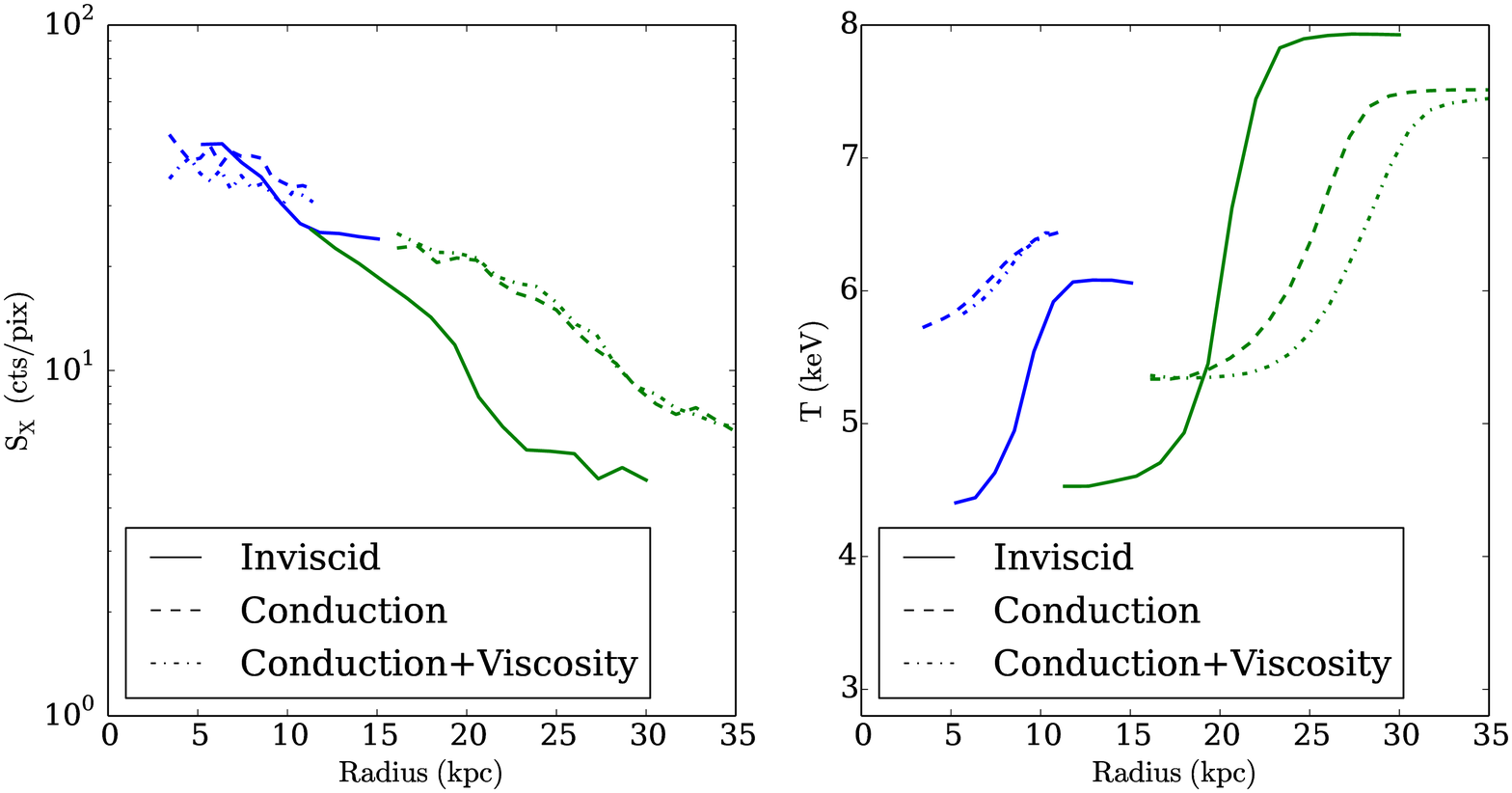}
\caption{Profiles of the surface brightness and projected temperature from the
 ``Virgo'' (top) and ``AM06'' (bottom) clusters for the anisotropic conduction
 simulations, corresponding to the regions in Figure \ref{fig:cond_temp}.\label{fig:cond_profiles}}
\end{center}
\end{figure*}

We also find in all of our simulations that significant velocity
dispersions may be measured along the line of sight that do not
correspond to turbulent motions but instead to large variations in the
bulk motion along the line of sight. Though we have
determined the dispersion here by simply computing it from the
velocity field itself, this effect will also show up in the broadening
of X-ray spectral lines, which is how turbulence is to be measured. We stress that care must be taken in interpreting the results of future measurements of the width of
spectral lines as resulting from turbulence, since it is possible for
bulk motions along the line of sight to produce similar signatures. We
will examine this effect in the context of a sloshing cool core in
more detail in a separate work.

%
%
\subsection{Simulations with Anisotropic Thermal Conduction}\label{sec:conduction}

Z13 found that anisotropic thermal conduction acted to eliminate
the sharp cold front temperature and density jumps despite the presence of strongly magnetized layers stretched along the fronts. This may be the result of imperfections in the magnetic layers due to field-line tangling from K-H instabilities, resulting in some ``leaking'' of heat from the hot plasma above the cold front, or to the presence of other regions of hot plasma that are not magnetically isolated from the colder gas inside the 
fronts (e.g.~the spiral-shaped ``tongues'' of hot gas in between the 
cold fronts, see Figures \ref{fig:virgo_temp} and 
\ref{fig:AM06_temp}). If viscosity acts to prevent the development of K-H 
instabilities, it is possible that the magnetized layers that 
develop will be more aligned to the smooth cold front surfaces and 
hence they may be more effective at suppressing conduction than in the 
inviscid case. In Figures \ref{fig:virgo_temp} and 
\ref{fig:virgo_bmag}, we saw two cold fronts where viscosity has suppressed 
K-H instabilities and produced magnetic fields more aligned to the 
cold front surfaces. This hypothesis motivates our simulations with 
anisotropic thermal conduction and anisotropic viscosity.

We performed two additional simulations for each of our model 
clusters, one with anisotropic thermal conduction (full Spitzer rate along the field lines and zero across the lines, see Section \ref{sec:physics}) and one with both 
anisotropic thermal conduction and Braginskii viscosity. Figure 
\ref{fig:cond_temp} shows images of projected temperature constructed at 
specific epochs for the ``Virgo'' and ``AM06'' simulations. In the simulations with anisotropic thermal conduction, though there are still surface brightness gradients at the 
positions of the cold fronts in the {\it inviscid} simulations, these once-sharp gradients have been reduced. This is demonstrated more quantitatively in Figure 
\ref{fig:cond_profiles}, which shows the extracted surface brightness and projected temperature profiles from the colored annular regions in Figure \ref{fig:cond_temp}. In the simulations with conduction, the gradients are no longer sharp, with 
or without viscosity, and the radial profiles are essentially 
identical. The inclusion of Braginskii viscosity appears to have little 
effect on the reduction of cold front jumps by conduction.

%
%
\section{Discussion}\label{sec:discussion}

%
%
\subsection{The Feasibility of Constraining Viscosity from X-ray
 Observations of Cold Fronts}\label{sec:feasibility}

In our exploratory set of simulations, we have bracketed the range of
possibilities for K-H instabilities at sloshing cold fronts for our
two examples of low-mass and massive clusters. On the one hand,
the cold fronts in our inviscid simulations are very susceptible to
K-H instabilities, despite the presence of magnetic
fields (this is even more true if magnetic fields are very weak or absent, cf.~Section \ref{sec:nomag}, ZML11). There appear to be more instabilities in these simulations than we see at cold fronts in actual clusters (a quantitative comparison is forthcoming in our future work). On the other hand, an ICM with an isotropic Spitzer
viscosity exhibits no evidence of instabilities at all, which is
also contrary to some observations. Our more physically-motivated simulations, with
either Braginskii viscosity or a reduced isotropic Spitzer
viscosity, appear to qualitatively reproduce the observations in terms of
the degree of smoothness of the cold fronts. A quantitative comparison
with several well-observed cold fronts will be presented in a subsequent publication. 
However, from our present study, we infer that it may be difficult to
distinguish between an isotropic ``sub-Spitzer'' viscosity and
anisotropic Braginskii viscosity using X-ray observations alone. The degree
of suppression of K-H instabilities at cold front surfaces and the
effect on turbulence are both qualitatively similar. From a purely theoretical perspective, Braginskii viscosity is the preferred model because it is known that the ICM is magnetized, and the small Larmor radius of the ions will constrain the momentum transport accordingly.

Other avenues for constraining viscosity in the ICM
exist. \citet{don09} used Braginskii-MHD simulations of buoyant
bubbles in a cluster atmosphere to investigate their stability
properties. They found that bubble stability was impacted most by the
magnetic field geometry. In the core of the Perseus cluster, the
smooth shape of some of the H$\alpha$ filaments may indicate the flow around these filaments is laminar and significantly viscous \citep{fab03}. Further simulation studies of these systems and their observations need to be combined with studies of cold fronts to provide better constraints on ICM viscosity. 

%
%
\subsection{Origin of Sloshing-Driven Turbulence and Effects of
 Resolution}\label{sec:res}

The different prescriptions for viscosity in our simulations produce
significant differences in the amount of turbulence that is
produced. To more fully understand the origin of these differences, it
is instructive to examine the relevant scales for turbulence in our
simulations. The first is the scale of the sloshing motions
themselves, $L_0$, which at early times is roughly $L_0 \sim 50$~kpc,
increasing at later times to $L_0 \sim 200$~kpc. The second scale is
the scale at which the turbulent cascade is dissipated by viscosity,
$l_{\rm diss}$, assuming a roughly continuous generation of turbulence by the sloshing
motions as they expand. For dissipation due to Coulomb collisions, this scale depends on the Reynolds number of the plasma \citep[e.g.,][]{rod13a}:
\begin{eqnarray}
{\rm Re} &=& 7 f^{-1}\left(\frac{V_L}{300~{\rm
     km~s^{-1}}}\right)\left(\frac{L_0}{10~{\rm
     kpc}}\right) \\
\nonumber &\times& \left(\frac{n_{\rm th}}{3 \times 10^{-3}~{\rm
     cm^{-3}}}\right)\left(\frac{T}{3~{\rm keV}}\right),\label{eqn:reynolds}
\end{eqnarray}
where $V_L$ is the characteristic turbulent eddy velocity on the
length scale $L_0$, $n_{\rm th}$ is the thermal gas number density,
$T$ is the thermal gas temperature, and $f$ is the viscous suppression
factor. For Kolmogorov turbulence, $l_{\rm diss} \sim L_0({\rm
 Re})^{-3/4}$. For our ``Virgo'' setup, Re~$\sim 200$ and $l_{\rm diss}
\sim 2$~kpc, and for the ``AM06'' setup, Re~$\sim 100$ and $l_{\rm
 diss} \sim 3$~kpc, in the relevant region of the cold
fronts (assuming $f = 1$ and a value of $L_0 \sim 100$~kpc for the length scale of the
sloshing motions). Since these values are very close to the size of our smallest cells, the numerical dissipation dominates, since its effect sets in at a length scale
$\sim$$8\Delta{x}$ \citep{zuh13b}.

It therefore may seem surprising at first that the different
simulations have such wide differences in the amount of turbulence
that is produced, since all of the simulations have the same cell
size, and in any case the turbulent dissipation from viscosity does
not occur except at the smallest scales. However, the relevant scale
for generating turbulence, as well as for disrupting the cold front
surfaces, is not the dissipation scale of the
turbulent cascades, but rather it is the wavelength of the smallest
perturbations that will not be damped out by viscosity. In an
extensive set of plane-parallel simulations of the K-H instability,
\citet{rod13b} determined this scale to be
\begin{eqnarray}
l_{\rm crit} &=& {\rm 30~kpc}\left(\frac{\rm Re_{crit}}{30}\right)f\left(\frac{V_L}{400~{\rm
     km~s}^{-1}}\right)^{-1} \\
\nonumber &\times& \left(\frac{n_{\rm th}}{10^{-3}~{\rm cm}^{-3}}\right)^{-1}\left(\frac{T}{2.4~{\rm keV}}\right)^{5/2}
\end{eqnarray}
Taking our lead from \citet{rod13b}, we assume the critical Reynolds
number at which K-H instabilities develop is Re$_{\rm crit} \sim 30$. This
criterion yields roughly $l_{\rm crit} \sim 40$~kpc for the outermost cold fronts at late epochs (``Virgo'', $t = 2.7$~Gyr, ``AM06'', $t = 3.25$~Gyr) in both of our
models assuming full Spitzer viscosity ($f = 1$). For $f =
0.1$, $l_{\rm crit}$ is an order of magnitude smaller for each model,
and a wider range of perturbations that we can resolve in our
simulations can become unstable. For Braginskii viscosity, the
effective $f$ may be as high as 1/5 for an isotropically tangled field
or even smaller \citep{nul13}, depending on the local magnetic field geometry.

If we examine the cold fronts in the early stages of sloshing (upper
panels of Figures \ref{fig:virgo_temp} and \ref{fig:AM06_temp}), we find that in all
cases except the {\it isotropic}, $f=1$ simulations that K-H instabilities
develop very quickly. These instabilities generate 
turbulence that eventually develops within the sloshing region in
these simulations. In the {\it isotropic}, $f=1$ simulation, instabilities
never really develop in the first place, the flow is always quite
laminar, and as such there is not much turbulence at later
stages. Therefore, the most relevant scale that determines the
difference between our simulations is $l_{\rm crit}$, the scale of the
smallest perturbations that become unstable. This scale is
very well-resolved by our simulations if $f = 1$, $l_{\rm crit} \sim 
40~{\rm kpc}$. In the $f = 0.1$ case, $l_{\rm crit} \sim 4~{\rm kpc}$, and this
scale is resolved only marginally well (the $8\Delta x$ criterion is not fulfilled), so numerical resolution may start to have a significant effect on our results (see Section
\ref{sec:ICs}). 

%
%
\subsection{Impact of Unresolved Microphysics on Our Results}\label{sec:limitations}

When the pressure anisotropy violates the approximate inequalities
\begin{equation}\label{eqn:firehosemirror}
-\frac{B^2}{4\pi} \lesssim p_\perp - p_\parallel \lesssim \frac{B^2}{8\pi} ,
\end{equation}
a situation which is expected to occur readily in the ICM \citep{sch05,lyu07,kun11}, rapidly growing Larmor-scale instabilities (namely, the firehose instability on the left side of the equation and the mirror instability on the right side) are triggered and act to regulate the pressure anisotropy back to within its stability boundaries. 
The lack of finite-Larmor-radius effects in the Braginskii-MHD equations means that the fastest-growing firehose 
modes occur at arbitrarily small scales and that the mirror mode, an inherently kinetic instability, is not described properly. 
How these microscale instabilities regulate the pressure anisotropy and, in doing so, place constraints on the allowed transport 
of momentum and heat remains very much an open question, one that has received increased attention in recent years in both 
the astrophysical \citep[e.g.][]{sch06,sha06} and solar wind \citep[e.g.][]{gar01,bal09} communities. While it is perhaps premature to judge whether the qualitative 
evolution found by our Braginskii-MHD description of large-scale sloshing and K-H instabilities in the ICM is contingent upon formulating a rigorous description 
of the kinetic microphysics, we nevertheless feel obliged to speculate upon how such microphysics may impact our results. 

A key finding to emerge from (collisionless) kinetic simulations of the driven firehose instability \citep[e.g.][]{mat06,ht08,kun14} is that the collisionality 
of the plasma, supplemented by the anomalous scattering of particles off the microscale fluctuations, adjusts to maintain a marginally firehose-stable pressure anisotropy. 
In a weakly collisional plasma, this effectively reduces the parallel viscosity from $\sim$$v^2_{\rm th} / \nu_{\rm ii}$ to $\sim$$v^2_{\rm A} / S$, where $S$ is 
the shear frequency of the viscous-scale motions \citep{kun14,mog14}. If this behavior holds true in the ICM, the effective Reynolds number may be in fact larger than that given by Equation \ref{eqn:reynolds} and K-H instabilities may develop more easily. Moreover, with the pressure anisotropy microphysically pinned at the firehose 
stability threshold (the left inequality of Equation \ref{eqn:firehosemirror}), the resulting viscous stress would effectively cancel the magnetic tension -- another effect that may ease the development of K-H instabilities. 
The situation with the driven mirror instability is a bit less clear \citep[see][]{kun14,riq14}, in that the pressure anisotropy appears to be regulated not just by anomalous particle scattering but also by an increasing population of resonant particles becoming trapped in magnetic mirrors where the pressure is naturally less anisotropic. It is this trapped population that has been suspected of dramatically reducing the effective thermal conductivity of the ICM \citep{sch08}, an effect that would  
call into question the applicability of the simulations in Section \ref{sec:conduction} to real clusters.

Clearly, more work is needed not only to assess the effects of microphysical kinetic processes on the transport of heat and momentum, but to also converge on a set of 
well-posed MHD-like equations that can be profitably used to study the large-scale dynamics and thermodynamics of the ICM. For now, we take comfort in the fact that the cold fronts produced in our simulations are also the locations where the magnetic-field strength is very much increased, making the inequalities (Equation \ref{eqn:firehosemirror}) difficult to violate.

%
%
\subsection{Anisotropic Thermal Conduction and Cold Fronts}\label{sec:cond_disc}

We find in this work that the inclusion of Braginskii viscosity fails
to have any significant effect on the elimination of cold front temperature and density jumps by anisotropic thermal conduction, despite its effect to reduce the
tangling of magnetic field lines along cold front surfaces by K-H
instabilities. This is perhaps not surprising, given the value of the Prandtl number of the ICM:
\begin{equation}
{\rm Pr} \equiv \frac{\nu}{\chi} = 0.5 \,
\frac{\ln\Lambda_{\rm e}}{\ln\Lambda_{\rm i}}\left(\frac{2m_{\rm e}}{m_{\rm i}}\right)^{1/2} \simeq 0.02.
\end{equation}
This indicates that viscous forces operate on a much longer timescale than thermal conduction. Even if heat conduction is prevented from occurring directly across the cold front interface by a magnetized layer, the heating of the
cold side of the front still occurs, since the front is still surrounded on multiple sides in three dimensions by hot gas that it is not magnetically isolated from (cf.~Z13). 

%
%
\subsection{Comparison With Previous Work}\label{sec:comparison}

We have already noted the many points of comparison between our work
and that of \citet{rod13a}. In the limit that the magnetic field is very weak, modeling the ICM using hydrodynamics with a suppressed Spitzer viscosity would be a sufficient approximation to an MHD simulation with Braginskii viscosity to reproduce the cold fronts as we see them in observations. If the magnetic field is strong and stretched along the cold front surfaces, it will provide an additional source of suppression of K-H instabilities. Without an independent measurement of the magnetic field strength and direction, it would be difficult to determine which mechanism is largely responsible for determining the shape of the fronts. We investigate this matter further in Section \ref{sec:nomag}. 

The unmagnetized, reduced Spitzer viscosity approach is also limited when it comes to turbulence, since an isotropic viscosity will damp all modes of a turbulent cascade, whereas Braginskii viscosity will damp the magnetosonic modes only. In the sloshing cluster core, the dominant motions are solenoidal, and the resulting turbulent cascade is mainly Alfv\'{e}nic. In a major merger, with strong compressible turbulence, the turbulent cascade is likely to be dominated by magnetosonic waves. The details of the effect of the different prescriptions for viscosity on this latter scenario are not straightforward to
determine from the conclusions of this study, and requires further
work.  Though in our study the viscous dissipation scale was very
small in the cluster core, in a major merger with low densities and
high temperatures this scale could be much larger, and so the
differences between the two approaches at this scale would be more
prevalent. 

\citet{suz13} simulated the effect of isotropic and anisotropic
viscosity on a ``bullet'' of cold gas propagating through a hot,
magnetized medium. For the MHD simulations presented in their work,
they assumed a uniform magnetic field direction but varied its
direction with respect to the cold bullet's velocity (along the $x$-axis) between the different simulations. Not surprisingly, for isotropic Spitzer viscosity they
find that K-H instabilities are completely suppressed. For anisotropic
viscosity, the degree of suppression is highly dependent on the
magnetic field direction. For magnetic field lines perpendicular to
the direction of motion of the bullet, the cold bullet sweeps these lines up into a draping layer very quickly, with the velocity gradients perpendicular to the field lines. As a result, the K-H instability is only weakly suppressed. For magnetic
field lines inclined at a 45\degree angle to the $x$-axis, more
suppression occurs, due to the fact that the magnetic field is more
aligned with the velocity gradient initially. 

In this work, our initial conditions are more appropriate for
conditions in a relaxed cluster core with an initially tangled magnetic field. It is difficult to compare our results with theirs directly, 
due to the fact that their setup is more akin to a cold front produced by a major merger (such as in the ``Bullet Cluster'') than that of sloshing motions. 
However, our results are consonant with theirs in the sense
that Braginskii viscosity results in much less suppression of K-H
instabilities than does full isotropic viscosity, except in the case where there
is significant alignment of the velocity gradients with the magnetic
field.

%
%
\subsection{Unmagnetized Simulations}\label{sec:nomag}

\begin{figure*}
\begin{center}
\includegraphics[width=0.9\textwidth]{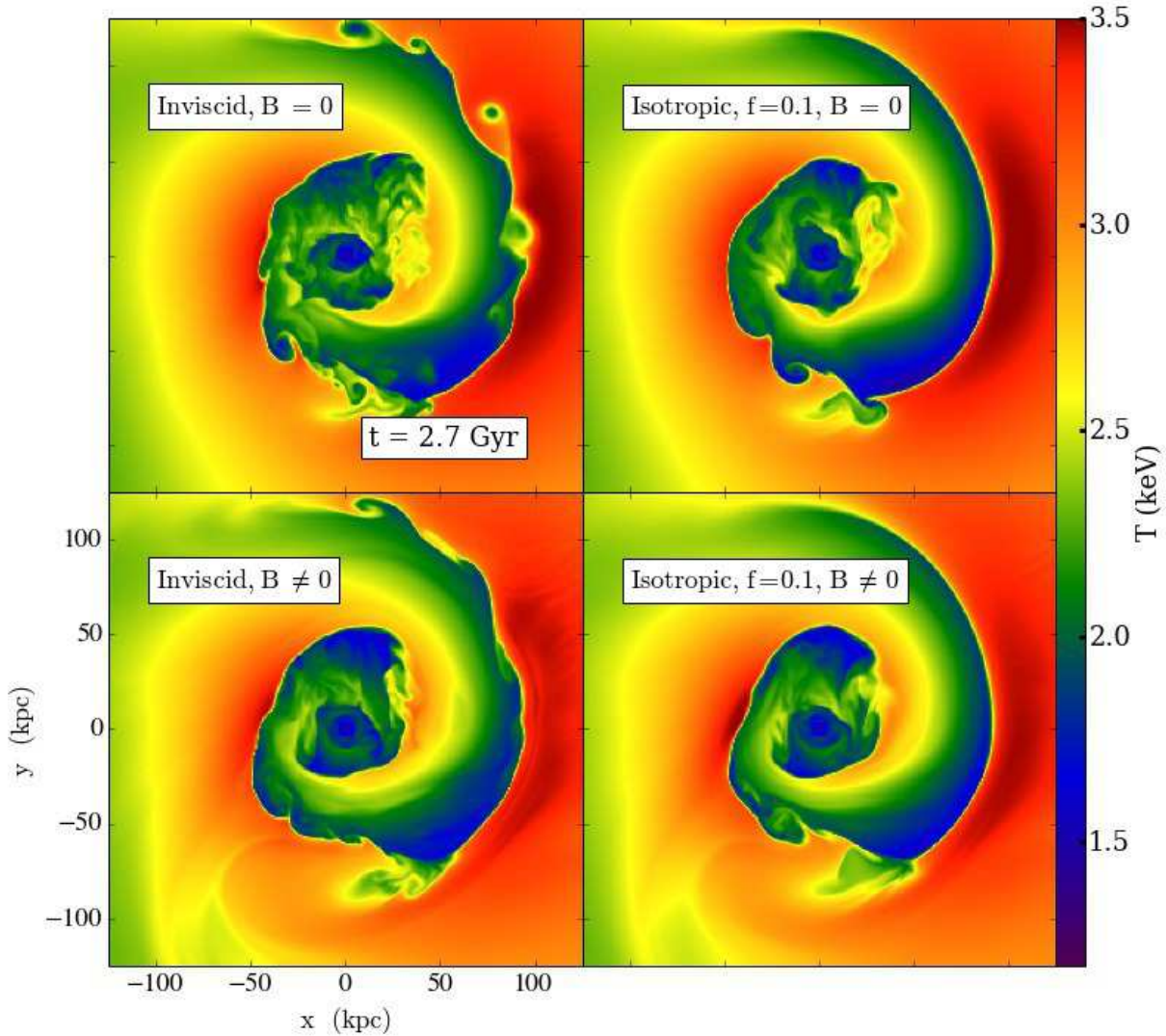}
\caption{Temperature slices in keV of the unmagnetized ``Virgo'' simulations at the epoch $t = 2.7$~Gyr, compared to the corresponding magnetized simulations.\label{fig:virgo_nomag_temp}}
\end{center}
\end{figure*}

\begin{figure*}
\begin{center}
\includegraphics[width=0.9\textwidth]{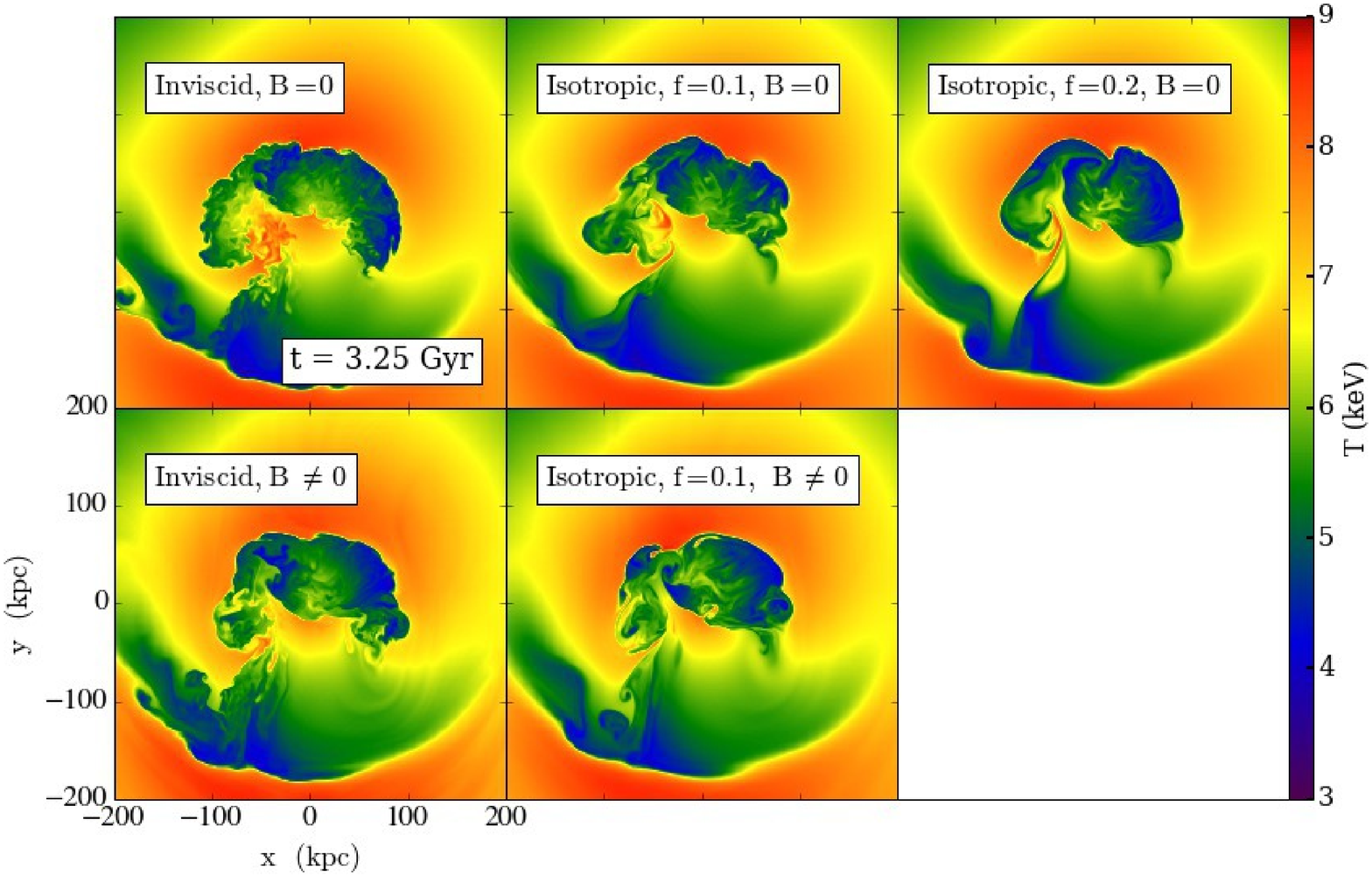}
\caption{Temperature slices in keV of the unmagnetized ``AM06'' simulations at the epoch $t = 3.25$~Gyr, compared to the corresponding magnetized simulations.\label{fig:AM06_nomag_temp}}
\end{center}
\end{figure*}

ZML11 showed that shear-amplified magnetic fields can suppress K-H instabilities in ICM cold fronts, even without viscosity, although this is highly dependent upon the strength of the initial magnetic field. The magnetic field in our simulations obviously plays a similar role, although determining whether its influence dominates over that of viscosity is complicated by the nonlinear nature of the problem. To date, most simulations of cold-front formation and evolution that included viscosity \citep[e.g.][]{zuh10,rod13a} had not included magnetic fields as well \citep[][is a notable exception]{suz13}. For completeness, we have performed unmagnetized versions of our ``Virgo'' and ``AM06'' simulations, both without viscosity and with an isotropic Spitzer viscosity reduced by $f=0.1$. We also performed an unmagnetized simulation of the ``AM06'' model with an isotropic Spitzer viscosity reduced by $f=0.2$. Sample temperature slices of these runs at different epochs, compared to their magnetized counterparts, are shown in Figures \ref{fig:virgo_nomag_temp} and \ref{fig:AM06_nomag_temp}. 

%
%

%

In the unmagnetized, inviscid simulations, more small-scale perturbations of the cold-front surface develop than in any of our magnetized simulations, since there is no mechanism (aside from the comparatively small numerical viscosity) to suppress them. More coherent large-scale instabilities are able to grow as well, which are evident in the temperature slices. 

In the unmagnetized simulations with $f = 0.1$ Spitzer isotropic viscosity, K-H instabilities are somewhat inhibited (especially the smallest-scale perturbations), but some large-scale features still persist. In the ``Virgo'' simulation (Fig.~\ref{fig:virgo_nomag_temp}), the cold fronts appear qualitatively similar to those in its magnetized $f = 0.1$ counterpart. This is not the case, however, in the ``AM06'' simulation (Fig.~\ref{fig:AM06_nomag_temp}), as its cold fronts appear more disrupted by K-H instabilities than its magnetized $f = 0.1$ counterpart. This is especially true at later epochs, on length scales of several tens of kpc. The additional ``AM06'' unmagnetized run with $f = 0.2$ shows better qualitative agreement with the magnetized $f = 0.1$ run. This demonstrates that even the somewhat weak initial field in these simulations ($\beta \sim 1000-1500$, c.f.~ZML11) can have a non-negligible effect on the appearance of the cold fronts, even in a viscous ICM.

These differences illustrate the difficulty in discerning the precise mechanism for K-H suppression at cold front surfaces from X-ray observations. In general, some combination of the effects of viscosity and amplified magnetic fields are responsible, but these do not combine in a linear way and the separation of these two effects is complicated by the fact that, while the thermal properties of the ICM plasma are well-constrained in the vicinity of a cold front, the local magnetic-field strength is generally unknown \citep[unless it can be constrained by other observations, e.g.~radio mini-halo emission, Faraday rotation, or discontinuities in the thermal pressure---for the latter see][]{rei12}. Unfortunately, these results indicate that there is not a general answer at this time as to whether the magnetic field or the viscosity is dominant in producing the observed smoothness of many cold fronts.

\section{Summary}\label{sec:summary}

We have performed a suite of exploratory MHD simulations of gas sloshing in a
galaxy cluster core using various prescriptions for viscosity, with
the aim of determining to what degree physically plausible models for
viscosity are capable of suppressing K-H instabilities at cold fronts. We also performed
simulations including viscosity and anisotropic thermal conduction to determine the
joint effect of these two microphysical processes on sloshing cold
fronts. Our main results are:
\begin{itemize}
\item We find that anisotropic Braginskii viscosity partially
 suppresses K-H instabilities at the surfaces of cold fronts and results in
 smoother fronts. However, it appears to be far less effective than isotropic viscosity, since the magnetic-field lines are oriented predominantly along the front surfaces (and therefore perpendicular to the velocity gradients).
\item The suppression of turbulent motions by a full isotropic
 Spitzer viscosity is too extreme to be consistent with current observational indications of K-H instabilities and
 turbulence in galaxy cluster cores. While a more quantitative comparison
 with real clusters is needed (and will be presented in future work), we have found that either Braginskii
 viscosity or a suppressed isotropic Spitzer viscosity is needed to at least qualitatively reproduce
 the appearance of X-ray clusters. 
\item Though magnetized Braginskii viscosity is a more accurate model for the 
 physics of the ICM, a suppressed isotropic Spitzer
 viscosity may provide a good qualitative approximation in terms of suppressing K-H instabilities at cold fronts. However, this conclusion depends on the magnetic-field strength and direction in the cold front region. 
\item The generation of cluster turbulence from sloshing motions is
 only somewhat inhibited by Braginskii viscosity, whereas it is
 completely inhibited by unsuppressed isotropic Spitzer viscosity. 
\item With or without Braginskii viscosity, and in cluster cores at
 different temperatures, an unsuppressed Spitzer conductivity along
 the field lines reduces density and temperature jumps at the cold fronts 
 to such a degree that they are inconsistent with observations. 
\item Unmagnetized simulations with viscosity may or may not give results that are similar to their magnetized counterparts, due to the nonlinear way in which the two effects combine to suppress K-H instabilities. 
\end{itemize}

There are several directions for further research in this area. Cold
fronts produced by major mergers, such as the bullet in the ``Bullet Cluster'' have different characteristics than sloshing cold fronts 
(e.g.~the fact that the location of the amplified magnetic field is on the hotter side of the front rather than the colder
side). Our results cannot be straightforwardly extended to this
scenario, and so simulations of this kind including the same physics
are needed. Other ICM features, such as active galactic nucleus  
bubbles and cold filaments, may provide complementary constraints on
the ICM viscosity. As the Braginskii-MHD equations are ill-posed at small scales 
and an adequate closure has yet to be developed, the study of microscale plasma instabilities and their effects on the macroscales 
may require changes to our model for magnetized viscosity. The 
effects of resistivity (and possibly magnetic reconnection) on gas sloshing and cold
fronts, particularly at the locations of amplified magnetic fields,
have yet to be investigated. Finally, direct comparisons between high-exposure
observations of cold fronts in real clusters and synthetic observations of simulated
clusters designed to mimic these systems would further strengthen our understanding of diffusive processes in the ICM. 

\acknowledgments
JAZ thanks Ian Parrish, Ralph Kraft, and Paul Nulsen for useful discussions. Calculations were performed using the computational resources of the Advanced Supercomputing Division at NASA/Ames Research Center. Analysis of the simulation data was carried out using the
AMR analysis and visualization toolset \code{yt} \citep{tur11}, which is
available for download at \code{\url{http://yt-project.org}}. Support for JAZ was provided by NASA though Astrophysics Theory Program Award Number 12-ATP12-0159. Support for MWK was provided by NASA through Einstein Postdoctoral Fellowship Award Number PF1-120084, issued by the {\it Chandra} X-ray Observatory Center, which is operated by the Smithsonian Astrophysical Observatory for and on behalf of NASA under contract NAS8-03060.

\appendix

\section{Resolution Test}\label{sec:res_test}

\begin{figure}
\begin{center}
\includegraphics[width=0.9\textwidth]{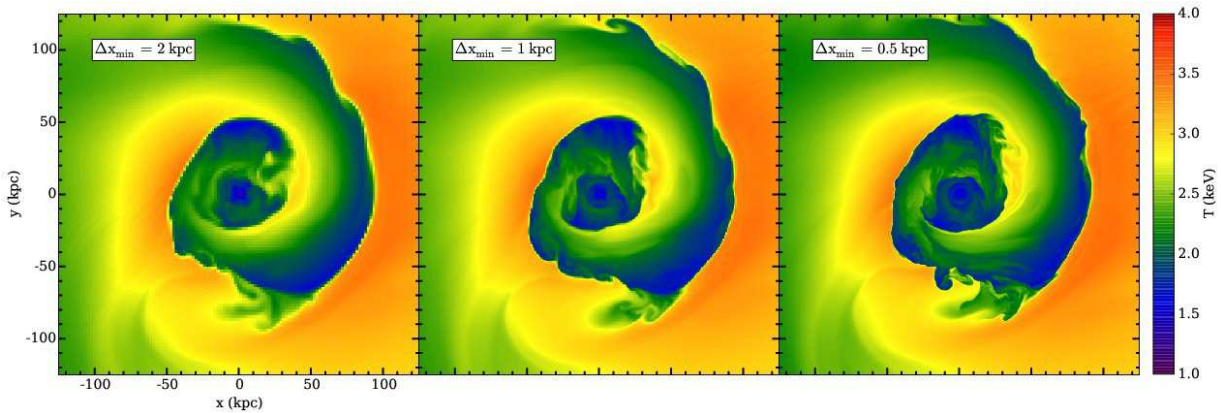}
\caption{Slices of temperature through the center of the simulation
 domain for inviscid ``Virgo'' simulations with varying resolution at
 time $t$ = 2.7~Gyr.\label{fig:diff_res}}
\end{center}
\end{figure}

We found in ZML11 that the ability of the magnetic field to suppress
K-H instabilities was somewhat resolution-dependent. To test the
dependence of our results on spatial resolution, we performed
simulations of our ``Virgo'' model at a finest resolution of $\Delta{x} = 2$~kpc and
$\Delta{x} = 0.5$~kpc, bracketing our default resolution of $\Delta{x}
= 1$~kpc. Figure \ref{fig:diff_res} shows slices of temperature
through the center of the simulation domain at time $t$ =
2.7~Gyr. Though the increase in resolution permits more small-scale
instabilities to develop, the overall behavior on larger scales
(structures of $\sim$tens of kpc) is the same. Since these are the
scales that X-ray telescopes will be able to observe, we conclude that
our default resolution of $\Delta{x} = 1$~kpc is more than adequate
for the purposes of this paper.

\section{Synthetic X-Ray Observations}\label{sec:xray_obs}

Our synthetic X-ray observations have been generated by a module derived
from the \code{PHOX} code \citep{bif12,bif13} that has been specifically
designed for use with \code{yt} \citep{tur11}. In the first step, we determine each cell's specific photon emissivity using an \code{APEC} \citep{smi01} model from its density and temperature and assumed metallicity. Using this emissivity as a distribution function for the photon energies, we generate photon samples from each cell assuming an
exposure time of $t_{\rm exp} = 300$~ks and collecting area $A_{\rm coll} =
6000$~cm$^{2}$. This ensures that we have a large number of photon
samples. In the second step, the photons are projected along a given
line of sight, and the energies are shifted by the line-of-sight
velocity and the cosmological redshift. A subset of the photons are chosen by adjusting the exposure time, accounting for Galactic absorption via a \code{thabs}
model, and assuming an energy-dependent effective area. For the latter we use the on-axis effective area curve of the ACIS-S3 chip on {\it Chandra}. Since we are not performing
spectral analysis, we have not applied any spectral responses. The simulated events are then binned into an image with pixels that correspond to the same size as
the finest SMR cells in the simulation ($\Delta{x} \approx 1$~kpc). Details on the
\code{yt} X-ray observation module may be found at \url{http://yt-project.org/doc/analyzing/analysis_modules/photon_simulator.html}.

%
%

\end{document}